\documentclass[12pt, a4paper]{article}

\usepackage{geometry}
\usepackage{authblk}
\usepackage{graphicx}
\usepackage{newtxtext}
\usepackage{newtxmath}
\usepackage{natbib}
\usepackage{hyperref}
\usepackage{xcolor}
\usepackage{amsmath}
\usepackage{setspace}
\usepackage[normalem]{ulem}
\usepackage{color,soul}

\usepackage{caption}
\hypersetup{
    colorlinks = true,
    urlcolor   = blue,
    citecolor  = black,
}

\newtheorem{corollary}{Corollary}

\newcommand{\RomanNumeralCaps}[1]
\linenumbers

\DeclareMathAlphabet{\mathsfbi}{OT1}{\sfdefault}{bx}{sl}
\DeclareMathVersion{sfletters}
\SetSymbolFont{letters}{sfletters}{OML}{ntxsfmi}{b}{it}

\makeatletter
\newcommand{\mathbfsbilow}[1]{%
  \text{\mathversion{sfletters}$\m@th#1$}%
}

\DeclareRobustCommand{\tensor}[1]{%
  \begingroup
  \ifcat\noexpand #1\relax
    \edef\greek@test{\detokenize{#1}}%
    \edef\greek@test{\expandafter\@cdr\greek@test\@nil}%
    \edef\greek@test{\expandafter\@car\greek@test\@nil}%
    \edef\x{\the\lccode\expandafter`\greek@test}%
    \edef\y{\number\expandafter`\greek@test}%
    \ifnum\x=\y\relax
      \mathbfsbilow{#1}%
    \else
      \mathsfbi{#1}%
    \fi
  \else
    \mathsfbi{#1}%
  \fi
  \endgroup
}
\makeatother

\newcommand{\norm}[1]{\left\lVert#1\right\rVert}

\newcommand{\norms}[1]{\tilde{\|}#1\tilde{\|}}

\newtheorem{theorem}{Theorem}
\newtheorem{remark}{Remark}

\newtheorem{app-remark}{Remark {C.}}
\newtheorem{app-lemma}{Lemma {C.}}
\newtheorem{app-corollary}{Corollary {C.}}
\newtheorem{app-definition}{Definition {C.}}
\newtheorem{app-proposition}{Proposition {C.}}
\newtheorem{app-theoremprf}{Theorem {C.}}
\newtheorem{app-corollaryprf}{Corollary {C.}}

\title{A kinematic-dynamic 3D model for density-driven ocean flows: Construction, global well-posedness, and dynamics}

\author[1]{Ori Saporta-Katz \thanks{Corresponding Author, ori.katz4@mail.huji.ac.il}}
\author[1,2,3]{Edriss S. Titi}
\author[4]{Hezi Gildor}
\author[1,5]{Vered Rom-Kedar}

\affil[1]{Department of Computer Science and Applied Mathematics, Weizmann Institute of Science, Rehovot  7610001, Israel}
\affil[2]{Department of Applied Mathematics and Theoretical Physics, University of Cambridge, Cambridge CB30WA, UK}
\affil[3]{Department of Mathematics, Texas A\& M University, College Station, TX 77843, USA}
\affil[4]{Institute of Earth Sciences, Hebrew University, Jerusalem 9190401, Israel}
\affil[5]{ The Estrin Family Chair of Computer Science and Applied Mathematics}

\date{}
\begin{document}
\setcounter{tocdepth}{1}
\maketitle

\newpage
\begin{abstract}
Differential buoyancy sources at an ocean surface may induce a density-driven flow that joins faster flow components to create a multi-scale, 3D flow.
Potential temperature and salinity are active tracers that determine the ocean's potential density: their distribution strongly affects the density-driven component while the overall flow affects their distribution.
We present a robust framework that allows one to study the effects of a general prescribed
3D flow on a density-driven velocity component through temperature and salinity transport,
by constructing a modular 3D model of intermediate complexity. The model contains an incompressible velocity that couples two advection-diffusion equations for the two tracers. Instead of solving the Navier-Stokes equations for the velocity, we consider a prescribed flow composed of several spatially predetermined modes.
One of these modes models the density-driven flow: its spatial form describes a density-driven flow structure
and its strength is determined dynamically by averaged density differences.
The other modes are completely predetermined, consisting of any incompressible, possibly unsteady, 3D flow, e.g. as determined by kinematic models, observations, or simulations. 
The result is a hybrid kinematic-dynamic model, formulated as a non-linear, weakly coupled system of two non-local PDEs.
We prove its well-posedness in the sense of Hadamard, and obtain {a priori} rigorous bounds regarding analytical solutions.
When the relevant Rayleigh number is small enough, we show, both rigorously and numerically, that for all initial conditions, the corresponding solutions converge to a unique steady state.
Motivated by the Atlantic Meridional Overturning Circulation, the model's relevance to oceanic systems is demonstrated by tuning the parameters to mimic the North Atlantic ocean. 
We show that in one limit the model may recover a simplified oceanic box model, including a bi-stable regime, and in another limit a kinematic model of oceanic chaotic advection, suggesting it can be utilized to study spatially dependent feedback processes in the ocean. 
\end{abstract}

%
\newpage
\tableofcontents

\section{Introduction}
\label{sec:intro}

 In various oceanic systems, significant horizontal density differences are induced by differential surface buoyancy sources, such as spatially varying heating, precipitation, evaporation, ice formation, and ice melting. These drive a sedate flow that leads, eventually, to the sinking of heavier water under lighter water. 
While transient sources allow the system to equilibrate through stratification and mixing, a continuous differential horizontal forcing may sustain a stable circulation.
One well-known geophysical example is the Atlantic Meridional Overturning Circulation (AMOC), an important component of Earth's climate system \citep{BuckleyMarshall2016, johnson2019recent}, induced in part by differential heating and freshwater sources between the equator and high latitudes \citep{FerreiraEtAl2019}.

%

{Density-driven flows have been studied qualitatively by dynamic box models, presented in the seminal work by Henry Stommel} \citep{Stommel1961} {as minimal models that capture some of the main qualitative features of ocean dynamics; see review by} \citet{DijkstraGhil2005} {for a survey and motivation.
Since their initial presentation, box models have been extended in,} to name a few, \citet{HuangEtAl1992, TzipermanEtAl1994, Cessi1994, GriffiesTziperman1995, GildorTziperman2001, PasqueroTziperman2004, calmanti2006north, AshkenazyTziperman2007wind, barham2019eddifying, an2021rate, BuddEtAl2021}.
In these models, a 2D ocean basin, forced from its surface border by temperature and precipitation/evaporation, is divided into instantaneously mixed boxes. The interbox transport scales like the average density differences between the boxes, as determined by the advected, forced, and mixed temperature and salinity tracers.
The scaling {can be} derived from a finite-difference approximation to the Boussinesq equations, and has been tested in some observational and numerical studies \citep{sijp2012key,butler2016reconstructing, cheng2018can}. The opposing effects of temperature and salinity on the density result in bi-stability and hysteresis; this theoretical prediction has been observed in complex models as well \citep{RahmstorfEtAl2005}, and has incited works regarding the stability of the AMOC (\citet{WeijerEtAl2019} and references therein).

Geophysical flows contain additional fast-varying large-scale phenomena that affect tracer transport. {To address  some of these effects, }box models have been extended to study effects of varying weather patterns \citep{Cessi1994, GriffiesTziperman1995, AshkenazyTziperman2007wind, barham2019eddifying, an2021rate, BuddEtAl2021} and simple wind-driven flows \citep{PasqueroTziperman2004, AshkenazyTziperman2007wind, barham2019eddifying}. These additions tend to stimulate dynamical tipping points between different modes, and enhance the variability of the overturning circulation.
In a different class of toy models, the transport in time-varying flows is commonly studied by kinematic models. These models illustrate that tracer transport is greatly affected by  chaotic advection, i.e. chaotic Lagrangian transport of passive scalars in a (generally non-turbulent) prescribed flow \citep{Aref1984, KoshelPrants2006, ArefEtAl2017, Ghil2017}.
A canonical example is the oscillating double-gyre kinematic flow model, where a laminar flow that models the oceanic double-gyre flow with a strong seasonal variability leads to chaotic advection in the ocean and non-trivial transport statistics \citep{YangLiu1997, YangLiu1994, Yang1996, KoshelPrants2006, AharonEtAl2012, Ghil2017}.
However, employed to provide a qualitative understanding of physical processes associated with passive fluid mixing and transport, such studies do not include any feedback mechanism. {Note that the box models and the kinematics models have an important common feature: they both use a prescribed spatial form of the velocity field to study transport phenomena.}

%
%
%
%

Here we present, explore, and analyze a novel kinematic-dynamic 3D model, that couples idealized box models with kinematic transport models that may exhibit chaotic advection.
The model consists of  two advection-diffusion equations for the temperature and salinity, coupled by a velocity field that has an externally driven component and an internal component with an amplitude that is determined, like in traditional box models, by averaged density differences between different regions of the basin.
Thus, our model isolates the effect of a prescribed, general, time-dependent, 3D velocity mode on a density-driven velocity component, 
and specifically on its strength, stability and variability.
The distributions of temperature and salinity in the basin are affected by the overall flow, while determining the density and thus affecting the density-driven velocity component.
%
The model is the natural extension of 2D box models into a 3D setting that takes into account additional flows and innerbox density variations: {each box is not fully mixed instantaneously, as in most box models. Our model allows a more natural mixing resulting from advection, diffusion, and sources. Although these innerbox density variations do not impact the strength of the dynamical velocity component (it is a result of density averages inside each box as in regular box models), they do impact the quantity of tracers that is transported between the boxes at each time step. Thus, indirectly, innerbox density variations impact the temporal dependence of the average density in each box, and, correspondingly, impact the strength of the overturning velocity component.}
Adjusting the parameters appropriately, the model recovers various box-model versions, as we show both rigorously and numerically.
Despite its apparent simplicity, the model can be tuned to imitate the current-day North Atlantic ocean, including semi-realistic results of temperature, salinity, and AMOC strength.
This kinematic-dynamic framework can easily be modified to a large variety of settings: different domains, spherical geometry, various types of boundary conditions, or more complicated scaling laws.

The model is a non-local, non-linear, coupled system of PDEs, for which basic properties such as the very existence of solutions are not obvious.
A desired property of a PDE model is global well-posedness in the sense of Hadamard, defined as the existence and uniqueness of solutions to the system for all times, along with a smooth dependence on the problem's data \citep{hadamard1902problems}. This property has been shown to be satisfied in some oceanic and atmospheric models: see, e.g., the survey papers 
\citet{temam2005some,Li2018} and references therein.
Since explicit solutions to nonlinear evolution equations are generally inaccessible, and these models are probed by approximated numerical solutions, such a proof strengthens the physical viability of employed models.
We address this issue rigorously, proving the global well-posedness of our model in the sense of Hadamard: if the boundary conditions satisfy a compatibility condition, we prove that the corresponding initial-boundary-value problem has a unique strong solution for all times, along with a smooth dependence of solutions on the problem data.
The proof reveals several bounds on averages of solutions. In the case of time-independent sources and boundary conditions, we also calculate a bound for the Rayleigh number of the problem below which the system has a globally stable unique steady state solution, and above which the system may exhibit 
nontrivial long-time dynamics. 
For example, in our simulations, when using time-independent forcing, we observe bi-stability when the  Rayleigh number is large enough; see section \ref{sec:box}.
Moreover, regardless of the size of the Rayleigh number, one can show  that
the infinite-dimensional dynamical system generated by this model is dissipative and possesses
a finite-dimensional non-empty global attractor, a subject that we postpone to future study. 

The structure of the work is as follows.
In section \ref{sec:modelderivation} we present the model formulation.
We describe the relevance of its numerical solutions to the North Atlantic ocean given an appropriate tuning of the parameters in section \ref{sec:NA}.
In section \ref{sec:box} we show that our model is a natural generalization of the box model scheme, and in particular we illustrate that it is mathematically equivalent to the popular $2\times2$ box model for certain values of the parameters.
Rigorous mathematical analysis of our model - a proof of the system's well-posedness, in the sense of Hadamard, as well as some useful bounds - is detailed in section \ref{sec:hadamard}.
We discuss the results and outline future directions in section \ref{sec:theend}.

\section{Model construction}\label{sec:modelderivation}
\begin{figure}[h!]
\centering
\includegraphics[width = 120mm]{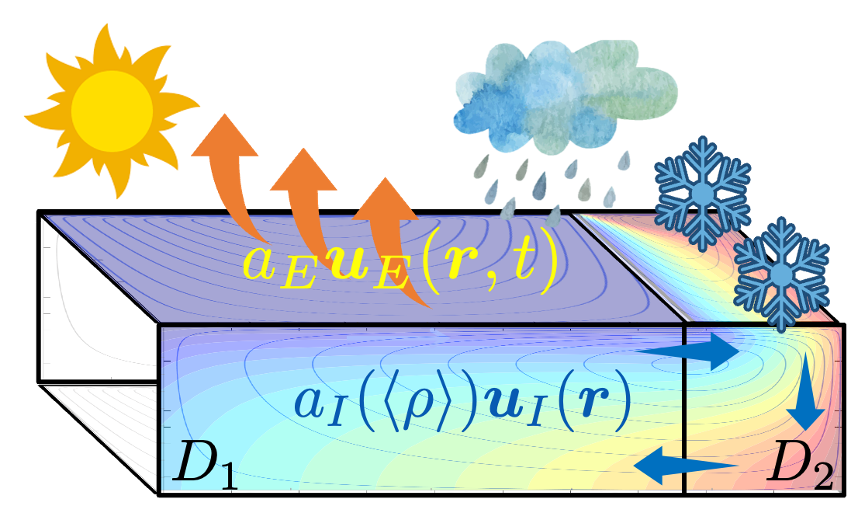}
\caption{
A cartoon of the kinematic-dynamic model.
A general velocity with two components, $a_E \boldsymbol{u}_E(\boldsymbol{r},t)$ and $a_I(\left<\rho\right>) \boldsymbol{u}_I(\boldsymbol{r})$, couples the advection-diffusion equations for temperature $T$ and salinity $S$ with mixed Robin-Neumann surface boundary conditions that model heating, cooling, evaporation and precipitation. Together $T$ and $S$ determine the density $\rho$, illustrated by the color variations in the figure. In turn, the difference between the density averages on the two predetermined boxes $D_1$ and $D_2$ determine $a_I(\left<\rho\right>)$.}
\label{fig:modelcartoon}
\end{figure}

The evolution of the temperature $T(\boldsymbol{r},t)$ and salinity $S(\boldsymbol{r},t)$  is modelled by the coupled advection-diffusion equations with sources, and subject to the relevant physical boundary conditions, as described below,
\begin{gather}
\partial_t T(\boldsymbol{r},t) + \boldsymbol{u}(\boldsymbol{r}, t; \rho(T,S)) \boldsymbol{\cdot} \boldsymbol{ \nabla} T(\boldsymbol{r},t) = \boldsymbol{ \nabla} \boldsymbol{\cdot} (\tensor{\kappa} \boldsymbol{ \nabla} T(\boldsymbol{r},t)) + f_T(\boldsymbol{r},t), 
\label{eq:Tpde}
\\
\partial_t S(\boldsymbol{r},t) + \boldsymbol{u}(\boldsymbol{r}, t; \rho(T,S)) \boldsymbol{\cdot} \boldsymbol{ \nabla} S(\boldsymbol{r},t) = \boldsymbol{ \nabla} \boldsymbol{\cdot} (\tensor{\kappa} \boldsymbol{ \nabla} S(\boldsymbol{r},t)) + f_S(\boldsymbol{r},t),
\label{eq:Spde}
\end{gather}
inside a closed rectangular basin $\boldsymbol{r} \in \Omega =  (0, L_x)\times (0, L_y) \times (0, L_z \subset \mathbb{R}^3$. We define the aspect ratio $A = L_z/L_x$, and consider the realistic case of  $L_x \sim L_y$. For large scale ocean phenomena, $A\sim 10^{-3}$, similar to a sheet of paper. This emphasizes how close to 2D are large-scale oceanic flows. Nonetheless, understanding the ocean requires a 3D approach; for our purposes studying density-driven flows, taking into account the vertical direction is imperative.
We use the linearized equation of state for the density, 
\begin{equation}
\rho(T,S) = \rho_0(1-\alpha T + \beta S),
\label{eq:EOS}
\end{equation}
where $\alpha$ is the thermal expansion coefficient and $\beta$ is the haline contraction coefficient. We note that we define $T$ and $S$ as anomalies around $\rho_0$; in ambient ocean conditions, the density $\rho_0 \sim 1026 kg/m^3$ corresponds to a temperature of $15^\circ C$ and a salinity of $35 psu$. The respective coefficients are of the order $\alpha \sim 10^{-4} K^{-1}$ and $\beta \sim 10^{-4} psu^{-1}$.

Equations \eqref{eq:Tpde}\ and \eqref{eq:Spde} are coupled only via their common advecting velocity field $\boldsymbol{u}(\boldsymbol{r}, t; \rho(T,S))$, rendering the tracers active as described in the introduction.
Instead of coupling the equations to the incompressible Navier-Stokes (NS) equations, our approach employs a significant simplification: Assume we know the spatial form of a certain large-scale, basin-wide solution $\boldsymbol{u}$, e.g.~based on observations of the actual velocity field in the ocean or by numerical simulations. We assume that this solution
can be written as a sum of several modes, and that only some of its modes' amplitudes are 
 determined dynamically by the density distribution. The phenomenological motivation for this approach is that
the velocity field in the ocean is built from several components; some are mainly driven by external sources, for example wind stress and tides, and some are dominantly driven by internal stresses deriving from density inhomogeneities, as discussed above. 
While the NS are nonlinear, one hopes that such a division can be justified by separation of temporal or spatial scales.

Our model example will be a general oceanic velocity that is decomposed into two components: an externally-forced velocity mode that does not depend on density, and
an internally-forced velocity mode that does.
We further assume that density inhomogeneities affect the strength, but not the form, of the internally forced velocity field components.
Finally, we consider an internal velocity strength that depends on spatial averages of the density, denoted in general $\left<\rho\right>$, and not on pointwise density values.
Thus, the internally-forced velocity field has the following structure: $\sum_j a_I^j(\left<\rho\right>) \boldsymbol{u}_I^j(\boldsymbol{r})$, where $I$ signifies internal.
Taking a first-order approximation we consider one such internally forced mode:
\begin{equation}
\boldsymbol{u}(\boldsymbol{r},t;\left<\rho\right>) = a_I(\left<\rho\right>) \boldsymbol{u}_I(\boldsymbol{r})
+ a_E \boldsymbol{u}_E ( \boldsymbol{r},t),
\label{eq:velfield}
\end{equation}
where $E$ signifies external, and $a_E$ is some constant. $a_I$ and $a_E$ have units of velocity, while $\boldsymbol{u}_I$ and $\boldsymbol{u}_E$ are dimensionless, and incompressible by construction:
\begin{equation}
\boldsymbol{ \nabla}\boldsymbol{\cdot} \boldsymbol{u}_I = \boldsymbol{ \nabla}\boldsymbol{\cdot} \boldsymbol{u}_E = 0.
\label{eq:velincompressible}
\end{equation}
Due to the incompressibility, the horizontal velocity components scale as $1/A$ larger than the vertical components.
We consider a flow enclosed in the domain:
defining $\partial\Omega$ as the boundary of the domain, and $\boldsymbol{\boldsymbol{\hat{n}}}(q)$ as the unit normal vector at $q\in\partial\Omega$ pointing  outwards of the domain, we demand
an impermeability (no-normal flow) boundary condition
\begin{equation}
\boldsymbol{u}_I|_{\partial\Omega} \boldsymbol{\cdot} \boldsymbol{\boldsymbol{\hat{n}}} = \boldsymbol{u}_E|_{\partial\Omega} \boldsymbol{\cdot} \boldsymbol{\boldsymbol{\hat{n}}} \equiv 0.
\label{eq:velclosed}
\end{equation}
Motivated by both models and observations 
\citep{Stommel1961, TzipermanEtAl1994, GildorTziperman2001, mullarney2007role, sijp2012key,butler2016reconstructing, cheng2018can},
the internal strength parameter $a_I(\left<\rho\right>)$ is proportional to the average density difference between two different regions of the basin,
$D_1, D_2 \subset \Omega$ with $D_1 \cup D_2 = \Omega$.
Thus, the formula for $a_I$ is
\begin{equation}
a_I(\left<\rho\right>) = \Gamma 
\left(\left< \rho \right>_2 - \left< \rho \right>_1\right)
=
\Gamma \rho_0 \left(
-\alpha (\left< T \right>_2 - \left< T \right>_1) + \beta (\left< S \right>_2 - \left< S \right>_1) \right),
\label{eq:aE}
\end{equation}
where $\Gamma$ is a proportionality constant with units of velocity over density,
$\left< T \right>_{i} = \frac{1}{|D_i|} \int_{D_i} T dV$, $i = 1,2$,
and
$\left< S\right>_{i} = \frac{1}{|D_i|} \int_{D_i} S dV$, $i = 1,2$.
This coupling provides a natural temporal separation between the two velocity components - the changes in the internal flow amplitude are governed by spatial averages and thus have less time fluctuations than the external flow.
Regarding the external flow, $a_E \boldsymbol{u}_E(\boldsymbol{r},t)$ can be any large-scale incompressible flow that does not depend on the tracers $T$ and $S$, e.g. a prescribed wind-driven surface flow, tidally driven flow,  or any kinematic flow model.

The source terms $f_T$ and $f_S$ may be set to quantify sources of heat and salt in the defined oceanic basin, such as exchange flow with marginal seas, river runoff, sea-ice formation and melting, or volcanic activity.
Motivated by oceanographic models, we consider a diffusion coefficient matrix $\tensor{\kappa}$ that parameterizes background turbulence and small-scale eddy flow processes as an effective diffusion (eddy diffusivity, see \citet{MajdaKramer1999}), rendering it orders of magnitude larger than molecular diffusivity. The diffusion constants are the same for both $T$ and $S$, but differ between the horizontal ($\hat{x},\hat{y}$) directions and the vertical ($\hat{z}$) direction, as isopycnal mixing is generally stronger than diapycnal mixing \citep{GentMcwilliams1990, gargett1984vertical} (isopycnals are approximated as horizontal):
\begin{equation}
\tensor{\kappa} =
\left(
\begin{matrix}
\kappa_H & 0 & 0 \\
0 & \kappa_H & 0 \\
0 & 0 & \kappa_V
\end{matrix}
\right), \; \; \kappa_V \ll \kappa_H.
\label{eq:kappamatrix}
\end{equation}
The relevant orders of magnitude for the ocean are $\kappa_H \sim 10^3 m^2/sec$, $\kappa_V \sim 10^{-4} m^2/sec$ \citep{GentMcwilliams1990, gargett1984vertical, MajdaKramer1999}.

For a full formulation of the problem at hand, boundary conditions must be specified. In the oceanic basin, the air-sea interface is its most significant boundary in terms of heat and freshwater forcing. 
The surface heat flux is related to the atmosphere-ocean temperature difference \citep{Haney1971};
a warmer ocean surface will release heat to the atmosphere, cooling the ocean (and making it
denser) while warming the atmosphere. The surface
freshwater flux is related to the evaporation and precipitation rates such that net evaporation will result in higher
surface salinity and thus in denser water. There is no direct feedback from ocean salinity on the atmosphere.
Thus, it is common to use the so-called mixed boundary conditions for the surface \citep{Haney1971, TzipermanEtAl1994}: a Robin, also known as a relaxation, boundary condition for the
temperature, in which the amount of heat flux depends on the air-sea temperature difference; and a Neumann boundary condition for the salinity. 
Neglecting (for now) geothermal heating and exchange flow with marginal seas, we consider a zero  flux boundary condition for the sides and bottom of the box. Defining $z=L_z$ as the domain's surface boundary (air-sea interface), the appropriate boundary conditions are thus written as
\begin{equation}
(\tensor{\kappa} \boldsymbol{ \nabla} T) \boldsymbol{\cdot} \boldsymbol{\boldsymbol{\boldsymbol{\hat{n}}}} = \begin{cases}
g_A^T (T^*(x,y,t) - T) & \hbox{if}\; \; z=L_z
\\
 0 & \hbox{else}
\end{cases}
,
\label{eq:BCT}
\end{equation}
\begin{equation}
(\tensor{\kappa} \boldsymbol{ \nabla} S) \boldsymbol{\cdot} \boldsymbol{\boldsymbol{\hat{n}}} = \begin{cases}
g_A^S S^*(x,y,t) & \hbox{if}\; \; z=L_z
\\
 0 & \hbox{else}
\end{cases}
.
\label{eq:BCS}
\end{equation}
$g_A^T$, $g_A^S$ are the effective rates of convective heat and mass transfer at the boundary, respectively, with units of velocity. 
 $T^*(x,y,t)$ and $S^*(x,y,t)$ are the temperature and salinity atmospheric sources (based, e.g.~on observations), and are predetermined.
Compatibility conditions on the box boundaries imply that at the box edges the normal derivatives of $T^*$, $S^*$ must vanish (see {Appendix} \ref{appC}), namely, that there is no flux to the "shore".
We note that the salinity forcing function $S^*(x,y,t)$ may be negative or positive, where a positive (negative) forcing signifies more (less) evaporation than precipitation. In order for the overall salinity to remain constant, we demand $\int_0^{L_x} \int_0^{L_y} S^*(x,y,t) dx dy = 0$ and $\int_\Omega f_S(\boldsymbol{r},t) dV = 0$.

The kinematic-dynamic model, depicted in Figure \ref{fig:modelcartoon}, is fully described by equations \eqref{eq:Tpde} - \eqref{eq:BCS}: a system of modified advection-diffusion equations for temperature $T$ and salinity $S$ with source terms, \eqref{eq:Tpde} and \eqref{eq:Spde}, that are weakly coupled by averages with a partially kinematic and partially dynamic incompressible velocity field \eqref{eq:velfield}. The system is subject to mixed Robin-Neumann boundary conditions for the temperature and salinity, \eqref{eq:BCT} and \eqref{eq:BCS}. The diffusion matrix \eqref{eq:kappamatrix}, parameterizing the eddy diffusivity, is diagonal, and the overall salinity is conserved in the basin throughout the evolution.
The dynamic coefficient of the internal velocity, $a_I(\left<\rho\right>)$, provides a natural observable of the dynamics. It reveals when the density-induced velocity component stabilizes to a steady state and when, as parameters are changed, the steady state bifurcates (and the solutions become bi-stable, oscillatory or, possibly,  chaotic).

\subsection{Non-dimensionalization and rescaling}\setcounter{tocdepth}{0}\label{subsection2.1}
%
System  \eqref{eq:Tpde} - \eqref{eq:BCS} has multiple natural timescales, associated with the horizontal and vertical eddy diffusion, the velocity field and the surface sources. 
Rescaling the problem to contain dimensionless parameters that demonstrate the ratios between these, we define the basic timescale $\tau \equiv L_z^2/\kappa_V$, describing the time it would take a tracer to cross the domain from surface to bottom via vertical eddy diffusion only.
We note that $\tau$ takes into account both molecular diffusion and eddy mixing  on the small-scale (e.g. due to breaking of internal waves) that is parameterized by the vertical  eddy diffusivity $\kappa_V$ and is not included in $\boldsymbol{u}$, and therefore is an insightful timescale for the advection-diffusion equation.
For typical oceanic values of $L_z \sim 4\times 10^3 m$ and $\kappa_V \sim 10^{-4} m^2/sec$, the timescale is of the order $\tau \sim 5\times 10^3 yrs$.
We further define the temperature and salinity scales as the maximal differences in the surface sources,
$ T^*_{\Delta} = \max T^* - \min T^*$ and
$ S^*_{\Delta} = \max S^* - \min S^*$, with the additional conditions $ T^*_{\Delta} \neq 0$, $ S^*_{\Delta} \neq 0$ to ensure horizontal density diferences.

We perform a non-dimensionalization of the variables
$t$, $x$, $y$, $z$, $T$, and $S$
by
$\tau$, $L_x$, $L_y$, $L_z$, $T^*_{\Delta}$, and $S^*_{\Delta}$, respectively.
Notice that the rescaling differs between the spatial dimensions, such that the rescaled domain is the symmetric cube $(0,1)^3$.
Correspondingly, we rescale the
dimensional functions $T^*, S^*, f_T, f_S$ 
by $T^*_{\Delta}, S^*_{\Delta}, T^*_{\Delta}/\tau, S^*_{\Delta}/\tau$.
Likewise, the
 dimensional parameters of $\tensor{\kappa}$ are replaced by
 $\delta_x \equiv  \frac{\kappa_H/L_x^2}{\kappa_V/L_z^2}$ in the $\hat{x}$ component, 
 $\delta_y \equiv  \frac{\kappa_H/L_y^2}{\kappa_V/L_z^2}$ in the $\hat{y}$ component, and $1$ in the $\hat{z}$ component; we further define $\delta \equiv \min\{\delta_x, \delta_y\}$.
Note that due to stratification strongly limiting cross-isopycnal flow, and due to the almost-2D nature of oceanic domains, 
 it may occur that $\delta_x \sim \delta_y \sim 1$.
 $g_A^T$ and $g_A^S$ are replaced by $Nu \equiv \frac{g_A^T}{\kappa_V/L_z}$, the Nusselt number for heat transfer at the boundary, and
$Sh \equiv  \frac{g_A^S}{\kappa_V/L_z}$, the Sherwood number for mass transfer at the boundary, respectively.
The $\hat{x}$, $\hat{y}$ and $\hat{z}$ components of the velocity field $\boldsymbol{u}$, as defined in equation \eqref{eq:velfield}, are rescaled, respectively, by $L_x/\tau$, $L_y/\tau$, and $L_z/\tau$.
Thus, by defining the P\'eclet number related to the external velocity field as $Pe \equiv \frac{a_E}{\kappa_V/L_z}$; 
the thermal Rayleigh number as $Ra^T \equiv \frac{\Gamma \rho_0 \alpha T^*_{\Delta}}{\kappa_V/L_z}$;
the salinity Rayleigh number as $Ra^S \equiv \frac{\Gamma \rho_0 \beta S^*_{\Delta}}{\kappa_V/L_z}$;
and the density stability ratio as $R_\rho \equiv \frac{\beta  S^*_{\Delta}}{\alpha T^*_{\Delta}} = Ra^T/Ra^S$; the dimensionless, rescaled velocity field is given by
\begin{equation}
\begin{aligned}
&\boldsymbol{u}_{rs} = Pe \boldsymbol{u}_{E, rs} + Ra^T a_{I, rs} \boldsymbol{u}_{I, rs},
\\
&a_{I, rs}( \left<T\right>,  \left<S\right>) = \frac{a_I( \left<\rho\right>)}{\Gamma \rho_0  \alpha  T^*_{\Delta}} = 
-\frac{
 \left<T\right>_2 - \left<T\right>_1}
{T^*_\Delta}
 +
 R_\rho
\frac{
 \left<S\right>_2 - \left<S\right>_1}
{S^*_\Delta},
\end{aligned}
\end{equation}
where $rs$ signifies rescaled.
All rescaled components of the velocites  $ \boldsymbol{u}_{E, rs}$ and $ \boldsymbol{u}_{I, rs}$ are now of order $1$, and defined in the  symmetric cube $(0,1)^3$.
The difference in timescales between the externally-driven velocity $ \boldsymbol{u}_{E, rs}$ and the density-driven velocity $ \boldsymbol{u}_{I, rs}$ is controlled by $Pe$ and $Ra^T$, such that
 $1$ is the timescale of eddy diffusion from surface to bottom, $1/Ra^T$ is the timescale of a vertical circulation around the basin, and $1/Pe$ is the timescale of externally-driven velocity contributions; typically, $1 \gg 1/Ra^T \gg 1/Pe$.
Summarizing, $Ra^T$ is the tuning parameter of the rescaled problem (through the free parameter $\Gamma$), and the rescaled variable $a_{I, rs}( \left<T\right>,  \left<S\right>)$ is the dynamic observable of interest that results from the model.
Hereafter we consider the nondimensional problem (and drop the $rs$ subscript), unless indicated otherwise.

\subsection{Simulating the model}
{Numerical solutions of the model are obtained step-wise: the density difference between the boxes at a given time-step, derived from the temperature and salinity, is used to calculate the amplitude of the internal velocity field, and this updated velocity field is used to advance the temperature and salinity to the next time step according to their respective equations} \eqref{eq:Tpde} and \eqref{eq:Spde}.
This 3D, optionally time-varying, non-linear, coupled PDE system,
with multiple time-scales, requires a customized numerical scheme, which we have implemented in Matlab. The scheme employs a finite volume integration scheme, which allows for a differential box size and has good conservation properties. The diffusion and source terms are straightforward to simulate using
standard discretization methods. For the advection term, in order to
retain numerical stability on the one hand while reducing numerical diffusion on the other hand, we use the stable flux-limiter advection scheme, combining the {low-order} upwind scheme with the {high-order} Lax-Wendroff scheme, weighted with a Sweby flux limiter with $\beta = 1.5$ \citep{sweby1984high}.
The spatial and temporal grid sizes were determined to be below the Courant-Friedrichs-Lewy (CFL) stability condition \citep{courant1967partial}, and fine enough to resolve the small structure elements of the velocity field. The final grids we used were checked for sensitivity of results to grid size changes.
The simulations shown here are run with a given parameter set until a quasi-stable equilibrium state/periodic state is reached; results shown are of the equilibria solutions.
The datasets generated and analyzed during the current study are available from the corresponding author on reasonable request.

\section {The North Atlantic ocean's large-scale flow}\label{sec:NA}

\begin{figure}[h]
\centering
\includegraphics[width = 180mm]{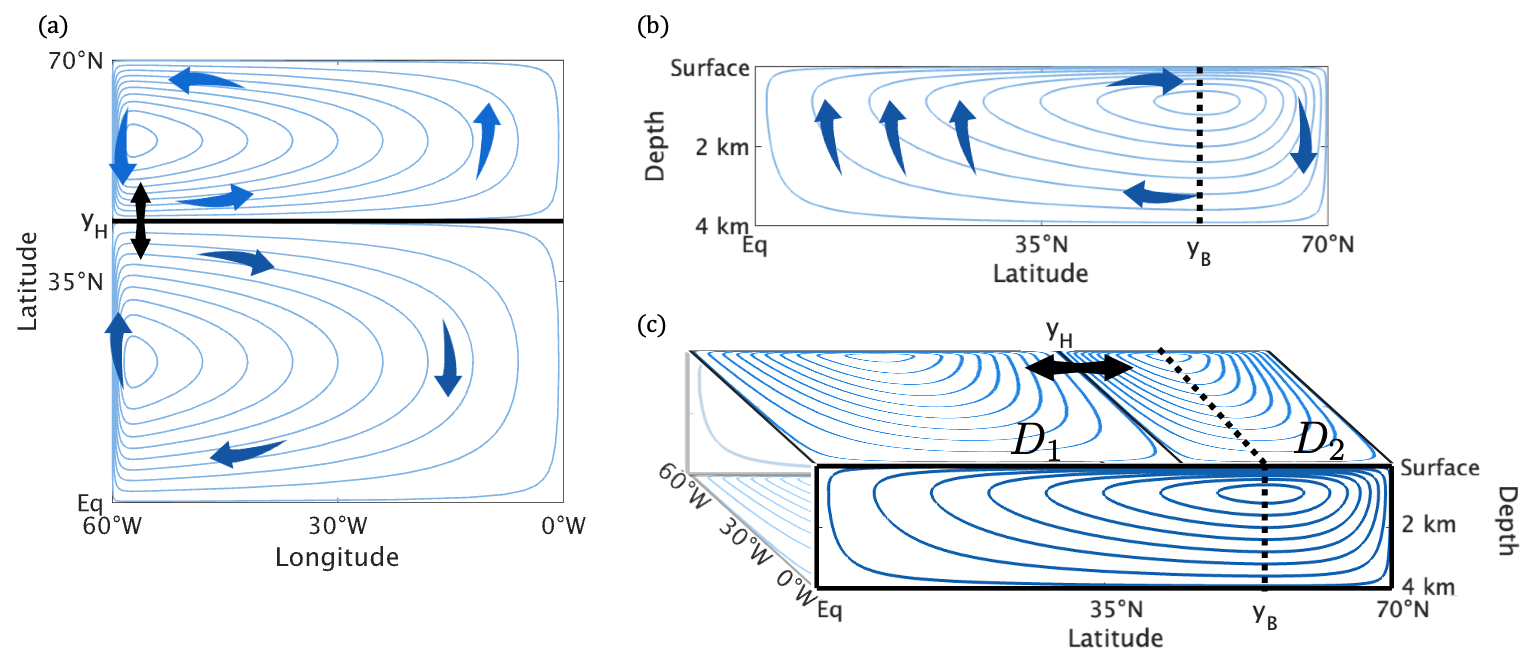}
\caption{
Flow field structure for the North Atlantic. (a) Surface flow ${\boldsymbol{u}}_E(\boldsymbol{r},t)$, with a penetration depth of $\sim 600 m$. $y_H$ is the inter-gyre demarcation line's position, and may be time-varying to model its seasonal north-south oscillations \citep{FrankignoulEtAl2001}.
(b) Zonally averaged overturning flow ${\boldsymbol{u}}_I(\boldsymbol{r})$.
Downwelling occurs north of $y_B = 55^\circ$N for the North Atlantic. $y_B$ is also the border between the south box $D_1$ and the north box $D_2$.
 (c) The composite 3D flow. The difference between the average densities in $D_1$ and $D_2$ determines the strength (and direction) of the overturning flow illustrated in (b).}
\label{fig:kinematicFlow}
\end{figure}

The North Atlantic's large-scale flow is built mainly from two velocity modes with a clear time-scale separation between them: a wind-driven double-gyre flow with a timescale of $\sim 5$ years \citep{vallis2017atmospheric}, 
and the AMOC, with a timescale of $\sim 1000$ years \citep{Ghil2017, johnson2019recent}.
The AMOC, unique to the Atlantic Ocean under present-day climate \citep{FerreiraEtAl2019}, transports approximately $18$ Sv (Sverdrup, equals $10^6$m$^3$/sec) of warm surface waters from the tropics towards the North Pole, sinks in the northern latitudes and flows southwards in the abyss.
The circulation is driven and maintained by a combination of wind-driven upwelling in the Antarctic region, upwelling throughout the ocean, and an anomalously high salinity in the northern latitudes causing a sinking of salty cold surface waters upon winter cooling \citep{FerreiraEtAl2019}. The high salinity is commonly attributed to a combination of factors, including an abundance of salty sources, atmospheric forcing, and the salt-advection feedback loop: the AMOC sustains itself by advecting high-salinity surface waters from equatorial latitudes to the north. 
Evaluating the relative contribution of each of these factors to the structure, strength and variability of the AMOC is important in order to understand its (somewhat debated) stability in current, past and projected climates (see \citet{WeijerEtAl2019},  and references therein). 

The extent of the effect of the feedback loop is related to the amount of salty waters that actually reach the northern latitudes. This depends on the time-dependent pathway statistics, affected both by the overturning circulation itself, and by chaotic advection resulting from the large-scale double-gyre surface flow of the basin, along with the other large-scale flow components in the basin, since the actual transport pathways differ from the mean flow \citep{Aref1984}.
Thus, the AMOC should be sensitive to density variations resulting from inhomogeneous external fluxes of salinity and temperature, that are transported in an interplay between diffusion, resulting in local mixing, and 3D chaotic advection, resulting in stirring \citep{Aref1984, brett2019competition}.
The extent of this sensitivity is unknown \citep{WeijerEtAl2019}.

To apply the kinematic-dynamic model to the North Atlantic flow,
consider an idealized rectangular basin with a constant depth of $4\times10^3 m$ and straight edges, and a horizontal extent of order $L \sim \mathcal{O}(10^6 m)$ that approximates the longitudinal region $[0^\circ W, \; 60^\circ W]$ and the latitudinal region $[0^\circ N, \; 70^\circ N]$.
See {Appendix} \ref{appB} for a comprehensive list of the parameters we used for the North Atlantic simulations.
In this work, we neglect the spherical geometry of the North Atlantic (equirectangular projection), however it is easy to see that the same framework can be naturally extended to a spherical basin.

The two major contributions to the large-scale flow in the basin are the horizontal wind-driven surface flow in the horizontal (${x},{y}$) direction, representing the North Atlantic sub-tropical and sub-polar gyres, described by the external velocity field ${\boldsymbol{u}}_E(\boldsymbol{r},t)$;
and the vertical density-driven overturning flow in the latitudinal-abyssal ($\hat{y},\hat{z}$) direction, described by ${\boldsymbol{u}}_I(\boldsymbol{r})$.
The surface double-gyre flow ${\boldsymbol{u}}_E$ we use is a solution to the Sverdrup balance with a boundary \citep{YangLiu1994}, see Figure~\ref{fig:kinematicFlow}(a). It exhibits a westward-biased asymmetry that models the western boundary flow, and a finite penetration depth of a few hundred meters, corresponding to the approximate measured depth of the oceanic thermocline. In some of the simulations, we also consider the effect of a seasonal north-south variation of the demarcation line between the gyres with an amplitude of approximately $1^\circ$ in latitude, with the north-most position obtained in the fall season \citep{FrankignoulEtAl2001}.
The overturning flow ${\boldsymbol{u}}_I$ is set as a zonally ($\hat{x}$) independent gyre in the $\hat{y}-\hat{z}$ plane, with a northward branch above the thermocline, a downwelling branch around the northern basin border, a southward branch below the thermocline and semi-uniformly upwelling south of the downwelling region. To model this flow, we use a similar functional form as the surface flow, see Figure~\ref{fig:kinematicFlow}(b).
The composite 3D flow is depicted in Figure~\ref{fig:kinematicFlow}(c).
%
 The exact functions we use for these flows are presented in {Appendix} \ref{appA}, and the parameter values that mimic the North Atlantic flow are presented in {Appendix} \ref{appB}.
%
%
By construction, both flow vector field components $ {\boldsymbol{u}}_E$ and $ {\boldsymbol{u}}_I$ are of order $1$ and non-dimensional, as described in section \ref{subsection2.1}.
In the composite flow, they are multiplied by their (dimensionless) respective strength coefficients $Pe$ and $Ra^T a_I( \left<T\right>,  \left<S\right>)$.
$Pe \sim 10^6$ corresponds to realistic sub-tropical gyre velocities with a circulation time of approximately 5 years, and $Ra^T \sim 10^3$ corresponds to realistic AMOC velocities with a circulation time of approximately $1000$ years and an AMOC strength of approximately $18$ Sv.
The vertical strength parameter $a_I( \left<T\right>,  \left<S\right>)$, of order $1$ due to the rescaling, is determined dynamically by the average density difference between the north part $D_2$ and the south part $D_1$ of the basin, as described in section \ref{sec:modelderivation}. The north-south box border is chosen according to the overturning gyre's center, where the flow switches from upwelling to downwelling, see Figure \ref{fig:kinematicFlow}(c).

We note that the AMOC does not have a southern border at the equator; although the transport loop does indeed attain partial closure to the north of the equator by upwelling, most of the southbound transport branch continues all the way south to Antarctica, where it participates in a complicated interplay with the Antarctic Circumpolar Current and eventually returns as a northbound branch \citep{johnson2019recent}.
We follow previous studies that consider a similar configuration: \citet{
Stommel1961, TzipermanEtAl1994, Cessi1994}, and others. Yet, our model could take this into account by inserting a source term at the southern border that parameterizes the appropriate inputs and outputs of temperature and salinity through the equatorial basin wall; alternatively, the velocity field ${\boldsymbol{u}}_I$ itself could have an open southern border. These options are left for exploration in later works.

%

\begin{figure}[h]
\centering
\includegraphics[width = 180mm]{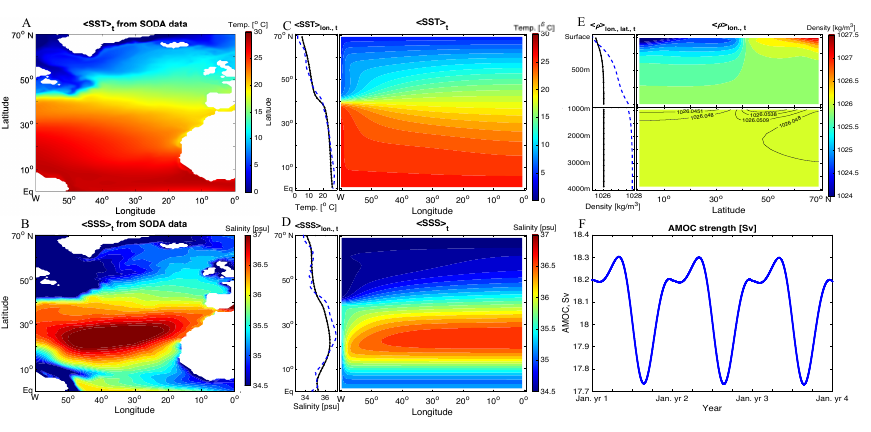}
\caption{
Simulation results of Atlantic Ocean model, in comparison with realistic reanalysis experimental data.
(A-B) SODA3.4.2 reanalysis experiment data \citep{carton2008reanalysis}, averaged over 1980-2017: (A) depicts sea-surface temperature (SST), and (B) depicts sea-surface salinity (SSS).
(C-F) Simulation results: (C) SST [$^\circ C$], (D) SSS [$psu$],  (E) zonally-averaged density [$kg/m^3$], and (F) simulation result of the AMOC strength [Sverdrup (Sv)], proportional to $a_I(\left<\rho\right>)$ of the model.
The black curves on the left of (C) and (D) are
longitudinal averages of the colored data.
The black curve on the left of (E) is a zonal average of the colored data.
The blue dashed curves are the corresponding averages from the SODA3.4.2 data.}
\label{fig:realistic}
\end{figure}

We used 
the SODA3.4.2 reanalysis experiment \citep{carton2008reanalysis} to tune the surface forcing, extracting the climatological average from 1980 to 2017 and deriving the zonally averaged data of the sea-surface temperature (SST) and salinity (SSS). This realistic data was used as a restoring (Robin) boundary condition for both $T$ and $S$. After approximately 1000 simulation years, the system reached a quasi stable time-periodic state. Then, we switched the salinity restoring force with a corresponding constant flux forcing (Neumann) boundary condition, a common procedure for tuning realistic systems with mixed Robin-Neumann boundary conditions, e.g.~\citet{TzipermanEtAl1994}.
Tuning the overturning strength parameter $\Gamma$, the system settled at a reasonable distribution of sea-surface temperature (SST), sea-surface salinity (SSS), zonally averaged density (except at northern-most latitudes, where there are density inversions), and AMOC overturning strength, see Figure~\ref{fig:realistic}. The tracer fronts observed between the two gyres are due to the velocity field, and are clearly apparent also in the realistic observations. In section \ref{sec:theend} we discuss similarities and differences between the model results and the realistic data.

\section{Relation to simplified box models}\label{sec:box}

\begin{figure}
\centering
\includegraphics[width = 170mm]{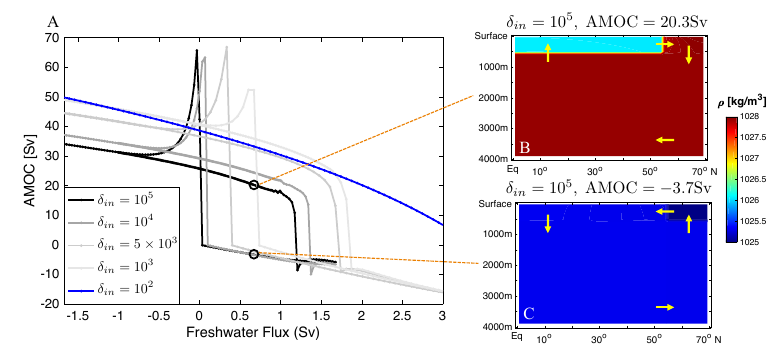}
\caption{
2D autonomous model, $Pe = 0$. (A) Bi-stability of steady state solutions as a function of freshwater flux (FWF, see main text).
$\delta_{in}$ is the ratio between inter-box and inner-box diffusion.
A hysteresis loop exists for $\delta_{in} \geq 10^3$ ($Ra^T \geq 2\times 10^3$). As $\delta_{in}$ (and $Ra^T$) increases, the hysteresis loop moves toward higher FWF, and the positive overturning strength grows. At $\delta_{in} = 100$ ($Ra^T = 200$) bistability is lost and a single steady state appears for a given forcing value (Theorem \ref{theorem:2}(iii)).
(B, C) Density for two steady states in the bistable regime, with $FWF = 0.67$ Sv, $Ra^T = 2\times 10^5$. (B) Overturning is $20.3$ Sv, (C)  overturning is $-3.7$ Sv.}
\label{fig:hysteresisBox}
\end{figure}



Our system, equations \eqref{eq:Tpde} - \eqref{eq:BCS}, is a natural extension of the $2\times 2$ box model presented in \citet{TzipermanEtAl1994, HuangEtAl1992}, extended to take into account density perturbations resulting from 3D advection, diffusion and forcing. Under the assumptions of inner-box fast mixing, and inter-box interactions resulting only from advection, and using ${\boldsymbol{u}}_I$ from the motivating example (Figure~\ref{fig:kinematicFlow}B) as the internal velocity in equation~\eqref{eq:velfield}, the system reduces to the four-box model.
To show this, we divide the basin into four boxes $B_{ij}$, $i,j=1,2$ (see Figure~\ref{fig:hysteresisBox}), and define the instantaneous box averages $T_{ij}(t) \equiv \frac{1}{|B_{ij}|} \int_{B_{ij}} T(\boldsymbol{r},t) dV$, and a function describing the density perturbations from the box averages, $T'(\boldsymbol{r},t)$, such that $T(\boldsymbol{r},t) = T_{ij}(t) + T'(\boldsymbol{r},t) $ for $ \boldsymbol{r} \in B_{ij}$,
and repeat the same for the salinity.
While $T(\boldsymbol{r},t)$ and $S(\boldsymbol{r},t)$ are smooth, $T'(\boldsymbol{r},t)$ and $S'(\boldsymbol{r},t)$ are only piecewise-smooth.
The fast mixing assumption is equivalent to assuming $T'(\boldsymbol{r},t)$ and $S'(\boldsymbol{r},t)$ are small at all times; immediate inner-box mixing corresponds to $T'(\boldsymbol{r},t) =  S'(\boldsymbol{r},t) \equiv 0$.
Employing these definitions, we integrate equation \eqref{eq:Tpde}  over $B_{ij}$, setting $f_T = 0$, and divide by its volume.
Using Gauss theorem and the upwind advection scheme, we obtain the following equations:
\begin{equation}
\begin{aligned}
\label{eq:4boxEOM}
&\dot{T}_{11} 
=
\frac{Nu}{1-z_B}(T^*_{South} - T_{11})
+ \frac{1}{y_B} \left( T_{21} - T_{11}\right) V
+
h_{11}(\delta_y, T')
\\ 
&\dot{T}_{12} 
=
\frac{Nu}{1-z_B}(T^*_{North} - T_{12})
+ \frac{1}{1-y_B} \left( T_{11} - T_{12}\right) V
+
h_{12} (\delta_y, T')
\\ 
&\dot{T}_{21} 
=
\frac{1}{y_B} \left( T_{22} - T_{21}\right) V
+
h_{21} (\delta_y, T')
\\ 
&\dot{T}_{22} 
=
\frac{1}{1-y_B} \left( T_{12} - T_{22}\right) V
+
h_{22} (\delta_y, T')
\end{aligned}
\end{equation}
where $T^*_{South} = \frac{1}{y_B} \int_0^1 dx \int_0^{y_B}dy \; T^*(x,y)$,
$T^*_{North} = \frac{1}{1-y_B} \int_0^1 dx \int_{y_B}^1 dy \; T^*(x,y)$ are the average surface fluxes in the south and north, correspondingly; 
and $V = \frac{1}{1-z_B}\int_0^1 dx \int_{z_B}^1 dz \; v|_{y=y_B}$ is the average velocity between the two top boxes.
The averaged velocity $V$ is determined as explained in section \ref{sec:modelderivation}, where $D_1 = B_{11} \cup B_{21}$, and $D_2 = B_{21} \cup B_{22}$: 
\begin{equation*}
\begin{aligned}
&V = Ra^T a_I(T_{ij}, S_{ij}) V_I ; \; \;V_I = \frac{1}{1-z_B}\int_0^1 dx \int_{z_B}^1 dz \;  {\boldsymbol{u}}_I \boldsymbol{\cdot}\hat{y}|_{y=y_B},
\\
&a_I(T_{ij}, S_{ij}) = 
-\frac{1}{T^*_\Delta} \left(\frac{\sum_{j=1}^2 B_{j2} T_{j2}}{\sum_{j=1}^2 B_{j2}} 
- 
\frac{\sum_{j=1}^2 B_{j1} T_{j1}}{\sum_{j=1}^2 B_{j1}} \right)
+ 
\frac{R_\rho}{S^*_\Delta} \left(\frac{\sum_{j=1}^2 B_{j2} S_{j2}}{\sum_{j=1}^2 B_{j2}} 
- 
\frac{\sum_{j=1}^2 B_{j1} S_{j1}}{\sum_{j=1}^2 B_{j1}} \right),
\end{aligned}
\end{equation*}
Note that $V_I$ is
a constant determined by the box division and the chosen overturning velocity field form ${\boldsymbol{u}}_I$.
In equations \eqref{eq:4boxEOM}, we assume $ a_I>0$, and use the upwind scheme: since the transport between the boxes is unidirectional by construction of the boxes, we assume the velocity is in a thermally dominant mode, i.e. from $B_{11}$ to $B_{12}$ and so on.
Thus, if $a_I<0$ the overturning flow switches direction and, just like in the 4-box model, the velocity terms switch signs and the advected values change accordingly. 

The remaining terms are contained in $h_{ij}(\delta_y, T')$, including the diffusion between the boxes and the advection of $T'$, which is the deviation in each box from the mean value inside the box. $h_{ij}$ can be calculated in a straightforward manner.
%
The equations for $S_{ij}$ are similar, except for the surface boundary conditions which have a constant flux for the salinity. 
If $T'$, $S'$ and inter-box diffusion are neglected, $h_{ij} \equiv 0$ and the equation set  \eqref{eq:4boxEOM} attains closure; along with the corresponding salinity equations, these are exactly the 2$\times$2-box equations, with known steady-state solutions \citep{HuangEtAl1992, TzipermanEtAl1994}. 
Using similar methods, it is easy to see that with this framework we can reproduce various previously-studied box models with any number of boxes, with a  controllable amount of inter-box diffusion and inner-box inhomogeneities: all that is required is a tailoring of a suitable internal velocity field $\boldsymbol{u}_I$ with transport in the desired direction between the various boxes.

To reproduce the $2\times 2$ box model limit, we shut off the external velocity field (i.e. $Pe=0$), and introduce a spatially dependent diffusion matrix with high diffusion inside each box $\tensor{\kappa}_{in}$, and small diffusion between the boxes $\tensor{\kappa}_{out}$. The results are summarized in Figure~\ref{fig:hysteresisBox}, and the parameters used are summarized in {Appendix} \ref{appB}.
As commonly performed with AMOC stability studies, we performed a simple hosing experiment: allowing the system to equilibrate, we slowly increased and then decreased the northern freshwater flux, defined as defined as $FWF [\hbox{Sv}]= 
L_x L_y  g_A^S \frac{\left<S^*\right>_{{y>y_B}}}{35.5 \hbox{psu}} / 10^6$.
We also define $\delta_{in}^i = \frac{\kappa_{in}^i/L_i^2}{\kappa_{out}^z/L_z^2}$, $i = x,y,z$, and the problem is set up such that $\delta_{in}^x = \delta_{in}^y = \delta_{in}^z \equiv \delta_{in}$. We note that the thermal Rayleigh number $Ra^T$ scales like $\delta_{in}$.
We performed the same hosing experiment on varying values of  $\delta_{in}$ (and thus, of  $Ra^T$). As $Ra^T$ decreases, the hysteresis loop shifts and shrinks until bi-stability is apparently lost,  as expected from Theorem \ref{theorem:2}(iii) presented in the next section.

\begin{figure}
\centering
\includegraphics[width = 170mm]{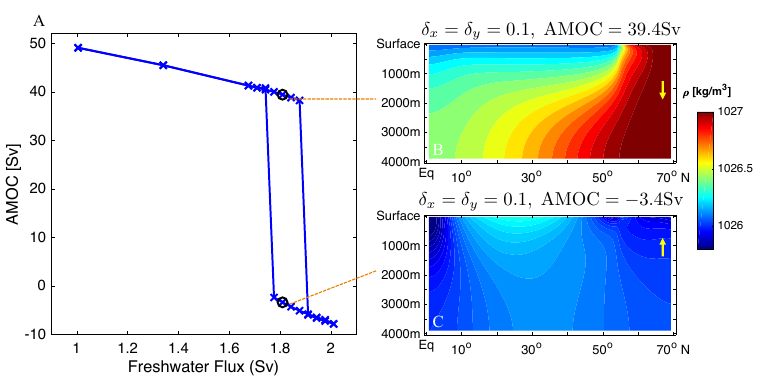}
\caption{
3D autonomous model. (A) Bi-stability of steady state solutions as a function of freshwater flux (FWF) with realistic eddy diffusivity. The mixing between the north and south boxes is reduced by the horizontal gyre flow. In all data points, the diffusion is the same, with $\delta_x = \delta_y = 0.1$.
For a narrow range of FWF, bistability is observed. 
(B, C) Density for two steady states in the bistable regime, with $FWF = 1.8$ Sv. (B) Overturning is $39.4$ Sv, (C) overturning is $-3.4$ Sv.}
\label{fig:hysteresis3D}
\end{figure}

As stated previously, even when the diffusion is homogenous in the full domain, there may be cases where neglecting $T'$ and $S'$ is justified, for example if there are barriers for mixing between the boxes, induced perhaps by localized coherent structures, that would render the eddy diffusivity between boxes effectively negligible. 
To demonstrate this idea, we performed an autonomous 3D experiment with homogenous eddy diffusivity throughout the basin. We
aligned the inter-gyre demarcation line of the horizontal velocity field with the horizontal barrier between the boxes $y_B$. Thus, each box is relatively mixed due to the combination of velocity fields and vertical diffusion, while interbox mixing is minimized due to the velocity field structure.
The results are summarized in Figure \ref{fig:hysteresis3D}, and the parameters used are summarized in {Appendix} \ref{appB}.
As in the 2D experiment, we varied the $FWF$ parameter continuously and plotted the resulting steady-state AMOC strength. The resulting hysteresis loop and steady states bear a qualitative resemblance to those obtained in the 2D experiment for relatively high values of inter-box diffusion; see Figure \ref{fig:hysteresis3D}.

As in previous studies of hysteresis AMOC loops in GCM models \citep{RahmstorfEtAl2005},
 finding these hysteresis loops requires some tweaking of parameters. Thus, some of 
the parameters of Figures \ref{fig:hysteresisBox} and \ref{fig:hysteresis3D} differ from those of Figure \ref{fig:realistic} (see Table \ref{table:new}). 
In particular, notice that the spatial dependence of the velocity field changes considerably the bi-stable range.  
A more comprehensive study of the range of realistic geophysical settings in which bi-stability 
is possible in our model (possibly including unsteady forcing) and its relation to corresponding parameter ranges in box models 
and in full GCM models is left for future research.
%
%
%
%
\section{Well-posedness of the model}\label{sec:hadamard}

It is well known that the linear evolution problem for the inhomogeneous 3D advection-diffusion equation subject to  Neumann or Robin boundary conditions, has a unique solution if the initial values, boundary values and {coefficients} satisfy some smoothness properties (see, e.g.,  \citet{Nittka2014}, and references therein). 
On the other hand, the 3D Boussinesq system 
describing ocean flow is still lacking a proof of existence of global, smooth solutions. 
Our system \eqref{eq:Tpde} - \eqref{eq:BCS}, is a non-trivial extension of the regular advection-diffusion equation due to its non-local coupling. 
In this section we present the theorem that guarantees 
the global, for all times and initial data, existence and uniqueness of well-behaving solutions.
Furthermore, we find several {a priori} bounds on the temperature and salinity functions, their derivatives, and the coupling term $a_I$ itself, that may be useful for numerical purposes, allowing some distinction between numerical errors and real features.
Moreover, in the case of time independent sources and boundary data,
we prove that there is a bound on the thermal and salinity Rayleigh numbers with respect to a function of the other dimensionless parameters of the problem, such that if the former is large enough with respect to the latter,
for all initial conditions the corresponding solutions will eventually converge to a single, stable steady-state solution.
Finally, relatively simple generalizations of the system are subject to similar proofs, and are quite wide-spread in various applications (see discussion in section \ref{sec:theend}).

In the next part we use the following notation:
For an appropriate {function} $Q(
{\boldsymbol{r}})$, we denote
\begin{itemize}
\item 
$\norm{Q} \equiv \left(\int_{{\Omega}} Q({\boldsymbol{r}})^2 dV\right)^{1/2} $, the $L^2$ norm of $Q$ on the domain $\Omega$;
\item 
$\norms{Q} \equiv \left(\iint Q\left({x},{y},{z}=1\right)^2 d{x} d{y}\right)^{1/2}$, the $L^2$ norm of $Q$ on the surface part of the boundary of $\Omega$, defined as ${z}=1$ in the rescaled problem. 
\end{itemize}

The main claim we prove regards the well-posedness in the sense of Hadamard of system \eqref{eq:Tpde}-\eqref{eq:BCS}, in the case of time-independent forcing and external velocity. 
We note that the proof can be easily extended to time-dependent forcing and external velocity, as long as the relevant spatial norms are bounded in time: see Remark \ref{remark:2}.
The exact mathematical formulation of the problem and the theorems, and the definitions of strong and classical $L^2$ solutions, along with an exact list of the regularity requirements on the system parameters and functions and the full proof of the theorem appear in {Appendix} \ref{appC}; here we present the main ideas.

\begin{theorem}
\label{theorem}
Given initial conditions $T_0, S_0$ that satisfy assumptions \eqref{eq:a}:
\begin{enumerate}
 \item The evolution problem described by equations \eqref{eq:Tpde} - \eqref{eq:BCS} has a unique, global-in-time strong solution, denoted $( T, S)$.
 \item This solution depends continuously on initial conditions, boundary conditions and sources in the sense described by equation \eqref{eq:continuousdependence}.
\item For every time $\bar{\mathcal{{T}}}>0$, the rescaled, non-dimensional solution satisfies the following bounds:
\begin{equation}
\begin{aligned}
\label{eq:bounds}
&\sup_{t\in [0,\bar{\mathcal{{T}}]}}{\norm{T}^2(t)} \leq C_1^T; \; \;
\int_0^{\bar{{\mathcal{T}}}}  \norm{\tensor{\kappa}^{1/2} \boldsymbol{ \nabla} T}^2(t) dt
\leq
C_2^T + {\bar{\mathcal{T}}} C_3^T;
\\
&\sup_{t\in [0,\bar{\mathcal{{T}}]}}{\norm{S}^2(t)} \leq C_1^S; \; \;
\int_0^{\bar{\mathcal{T}}}  \norm{\tensor{\kappa}^{1/2} \boldsymbol{ \nabla} S}^2(t) dt
\leq
C_2^S + {\bar{\mathcal{T}}} C_3^S;
\end{aligned}
\end{equation}
where
\begin{equation}
\begin{aligned}
\label{eq:Cs}
C_1^T &\equiv \norm{T_0}^2
+ \frac{2 Nu}{\min\{ Nu, 1\}(1 - a)} \norms{T^*}^2
 +
\frac{4}{\min\{ Nu^2, 1\} a (1 -  a)} \norm{f_T}^2; \\
C_2^T &\equiv \frac{1}{2}\norm{T_0}^2; \\
C_3^T &\equiv \frac{\min\{ Nu, 1\}}{8} \norm{T_0}^2
+ Nu \norms{T^*}^2
+ \frac{4}{\min\{ Nu, 1\}} \norm{f_T}^2;
\\
C_1^S &\equiv
\norm{S_0}^2 + 
\frac{0.27}{b} \frac{Sh^2}{\min\{1,\delta^2\}}  \norms{S^*}^2
+
\frac{0.06}{\left(\frac{1}{2}-b\right)}\frac{1}{\min\{1,\delta^2\}}
\norm{f_S}^2
;
\\
C_2^S &\equiv 
1.6 \min\{\delta, 1\}  \norm{S_0}^2;\\
C_3^S &\equiv 
1.6 \min\{\delta, 1\}   \norm{S_0}^2
+
\frac{6 \; Sh^2}{\min\{\delta, 1\}  }\norms{S^*}^2
+
\frac{3.9 \; Sh}{\min\{\delta^2, 1\}}\norm{f_S}^2
,
\end{aligned}
\end{equation}
for $a = \left({1 + \sqrt{1 + 2 Nu \frac{\norms{T^*}^2}{\norm{f_T}^2}}}\right)^{-1}$ and
$b = \frac{1}{2}{\left(1+
\frac{1}{ Sh \sqrt{2+\pi} }
\frac{\norm{f_S}}
{\norms{S^*}}
\right)}^{-1}.$
\item If, in addition, the initial conditions satisfy  \eqref{eq:b}, namely, if they satisfy the boundary conditions and are sufficiently regular,
then the problem
has a unique, global-in-time classical $L^2$-solution.
\end{enumerate}
\end{theorem}

Here we present the proof outline.
The proof of existence relies on the concept of iteration. As a first step, given initial conditions $(T_0(\boldsymbol{r}), S_0(\boldsymbol{r}))$, we construct a sequence of solutions $\{(T_n(\boldsymbol{r},t), S_n(\boldsymbol{r},t))\}_{n=1}^\infty$ to the iteratively-defined, linear set of advection-diffusion equations
\begin{equation}
\partial_t T_{n} - \boldsymbol{ \nabla}\boldsymbol{\cdot}(\tensor{\kappa}\boldsymbol{ \nabla} T_{n}) + (\boldsymbol{u}_{n-1}\boldsymbol{\cdot\nabla})T_{n} = f_T; 
\; \; 
\partial_t S_{n} - \boldsymbol{ \nabla}\boldsymbol{\cdot}(\tensor{\kappa}\boldsymbol{ \nabla} S_{n}) + (\boldsymbol{u}_{n-1}\boldsymbol{\cdot\nabla})S_{n} = f_S,
\end{equation}
where
$\boldsymbol{u}_{n-1} \equiv \boldsymbol{u}(\boldsymbol{r}; S_{n-1},T_{n-1})$ is determined from the previous step, according to equation \eqref{eq:velfield}.
The above system is subject to the boundary conditions of the original system, \eqref{eq:BCT} and \eqref{eq:BCS}, with $T_n$ and $S_n$ in place of $T$ and $S$, respectively.
 Note that these are linear PDEs subject to mixed Robin and Neumann boundary conditions, and the existence and uniqueness of a solution $(T_n, S_n)$ is guaranteed in \citet{Nittka2014}.
 This solution is then proven to satisfy certain energy estimates from which we deduce bounds on the relevant norms of the solutions, that are independent of $n$. Using these bounds, we show that this iterative sequence is a Cauchy sequence in the $L^\infty([0,\tau];L^2(\Omega))$ topology, converging strongly to functions that solve the PDEs and satisfy the boundary conditions for short times. 
We can show that the relevant norms of the solutions remain finite on the maximal interval of existence, and this guarantees the global existence of solutions.
 Next, the idea of a maximal interval of existence is used in order to  show that these solutions are in fact global in time. Uniqueness of these solutions is shown along with smooth dependence on initial and boundary conditions by bounding the difference between two solutions with a bound that goes to zero when the difference in their initial and boundary conditions goes to zero. 

We emphasize that the bounds in Theorem \ref{theorem} are generally true for temperature and salinity functions that satisfy equations \eqref{eq:Tpde} - 
\eqref{eq:BCS}, with any velocity field that is incompressible and non-penetrating, irrespective of the method through which the velocity field is achieved.

A straightforward conclusion from the proof of Theorem \ref{theorem} is that the dynamical overturning strength parameter $a_I$ is bounded:

\begin{corollary}
\label{cor:1}
For a given rescaled, non-dimensional solution $\left(T,S\right)$ of problem \eqref{eq:Tpde} - \eqref{eq:BCS}, the dynamic weight of the velocity function, $a_I(\left<T\right>, \left<S\right>)$, is bounded at all times:
\begin{equation}
|a_I(\left<T\right>, \left<S\right>)| \leq \frac{\sqrt{{C_1^T}} + R_\rho \sqrt{ C_1^S}}
{\min\{ |D_1|, |D_2|\} },
\end{equation}
where the values of $C_1^T$ and $C_1^S$ are given in \eqref{eq:Cs}.
\end{corollary}

Finally, we discuss steady-state solutions of the problem.
We show that for any choice of time-independent sources and boundary data, parameters and external velocity functions, there exists a steady state solution to the system. Note that
for a general choice of parameters, multiple steady-state solutions to our nonlinear problem are expected to coexist, as we show numerically in Section \ref{sec:box}.
%
However, as is expected in advection-diffusion-type problems, 
when the system is not vigorously forced with respect to its dissipation, the steady-state solution \textit{is} unique; furthermore, solutions to the time-dependent problem converge to this unique steady state solution as $t \rightarrow \infty$.
These results are summarized in the following theorem:

\begin{theorem}
\label{theorem:2}
Suppose that the sources and the boundary data are time-independent. Then:
\begin{enumerate}
\item The nonlinear steady-state problem corresponding to
\eqref{eq:Tpde} - \eqref{eq:BCS}
has a weak steady-state solution, denoted $(T,S)$.
\item All the rescaled, non-dimensional steady-state solutions satisfy the following bounds:
\begin{equation}
\begin{aligned}
\label{eq:ssH1bounds}
\norm{T}^2 \leq C_4^T&, 
\; \; 
\norm{\tensor{\kappa}^{1/2} \boldsymbol{ \nabla} T}^2 \leq C_5^T,
\; \; 
\norm{S}^2 \leq C_4^S,
\; \; 
\norm{\tensor{\kappa}^{1/2} \boldsymbol{ \nabla} S}^2 \leq C_5^S;
\end{aligned}
\end{equation}
where
\begin{equation}
\begin{aligned}
\label{eq:ssH1Csbody}
&C_4^T = 
\frac{1}{16\min\{Nu,1\}^2} \norm{f_T}^2 
+
 \frac{Nu}{4 \min\{Nu,1\}} \norms{T^*}^2,
\; \;
C_5^T =
\frac{1}{4 \min\{Nu,1\}} \norm{f_T}^2 
+
 \frac{2 Nu}{3} \norms{T^*}^2,
\\
&C_4^S =
\frac{1}{\epsilon_1} \norms{S^*}^2
+
\frac{1}{\epsilon_2} \norm{f_S}^2,
\; \;
C_5^S = 
\frac{1}
{\epsilon_1} \norms{S^*}^2
+
\frac{1}{\epsilon_3} \norm{f_S}^2.
\end{aligned}
\end{equation}
The constants are given by:

$\epsilon_1 = 2 \min\{\delta,1\}/Sh^2$,
$\epsilon_2 = 4 \min\{\delta,1\} \left(3 \pi/8 -1\right)$,
$\epsilon_3 = 2 \left( \min\{\delta,1\} (\pi-2)+1 \right)$.
\item
Let $T_0, S_0$ be initial conditions that satisfy assumptions \eqref{eq:a}, and let $(\tilde{T},\tilde{S})$ be a global strong solution to equations \eqref{eq:Tpde} - \eqref{eq:BCS}, as established in Theorem \ref{theoreminproof}.
If the following condition is satisfied by the problem parameters:
\begin{equation}
\label{eq:rayleighbound}
{\max\{Ra^T, Ra^S\}}
\leq
\min \{ |D_1|, |D_2| \}
\frac
{
{\min\{\delta, 1\}}
\min \{ 4 \pi \delta, {Nu}, {1} \}
}{
8 \max \left\{ 
C_5^T, 
C_5^S\right\}
}
,
\end{equation}
where $C_5^T$, $C_5^S$ are given in equation \eqref{eq:ssH1Csbody},
then $(\tilde{T},\tilde{S})$ converges to a unique steady-state solution as $t\rightarrow \infty$.
\end{enumerate}
\end{theorem}

Theorem \ref{theorem:2}(iii) proves that the bistability of the system, a robust feature of known models of density-driven flows with both salinity and temperature forcing, is lost if the minimal diffusion in the domain is large enough with respect to the advection. This ratio is represented by $Ra^T/\min\{\delta,1\}$.
Since the salinity Rayleigh number is $Ra^S = Ra^T R_\rho$,  Theorem \ref{theorem:2}(iii) provides a bound on both the thermal and the salinity Rayleigh numbers.
In the special case of no external sources, $f_T = f_S = 0$, equation \eqref{eq:rayleighbound} simplifies; for example, we can immediately conclude that the bound on both $Ra^T$ and $Ra^S$ is smaller than both
${1}/{\norms{T^*}^2}$ and 
${1}/{Sh^2 \norms{S^*}^2}$.

We note that Theorems \ref{theorem} and \ref{theorem:2} and Corollary \ref{cor:1} provide 
{a priori}
rigorous bounds that are not in general tight, since in their formulation, the ``worst case scenario" of the most extreme functions allowed by the problem formulation is considered.
This is illustrated by Corollary \ref{cor:1}, where the bound for $a_I(\left<T\right>, \left<S\right>)$ does not depend on gradients or differences of $T^*$ and $S^*$, although it is obvious that if these are constant in space, there will be no overturning circulation after sufficient time.
Also, one may calculate the numerical values of the bounds for the parameters used in the simulations. For example, Corollary \ref{cor:1} promises that the overturning strength of the steady state solution in Figure~\ref{fig:hysteresis3D} for freshwater forcing $1.8$ Sv is smaller than $900$ Sv - this is indeed the case. However, this bound is not very useful, as the asymptotic overturning strength is found numerically to be smaller than $25$ Sv.
On the other hand, qualitatively the statements are sound. For example, the transition promised by Theorem \ref{theorem:2}(iii) between a unique steady state and non-trivial dynamics is indeed observed  in the simulations: While for a large Rayleigh number, bistable solutions are found, when the minimal diffusion in the basin is large enough with respect to the advection coefficient as encapsulated by a small enough Rayleigh number, bistability is lost and all solutions seem to converge to a unique steady state; see Figures \ref{fig:hysteresisBox} and \ref{fig:hysteresis3D}.

\begin{remark}{Generalizations.}\label{remark:1}

The proof presented here can be generalized to the following cases:
\begin{itemize}
\item Any number of boxes $n\in \mathbb{N}$, and any linear dependence of the coupling on the tracer functions $T$ and $S$, which can be different in each box. Thus, the internal velocity mode's coupling strength parameter $a_I$ can take the general form:
\begin{equation}
\label{eq:5.8}
a_I(\left<T\right>,\left<S\right>) = \sum_{j=1}^n \alpha_j \frac{\int_{D_j^T}  T dV /T^*_\Delta}{ D_j^T} 
+
\beta_j \frac{\int_{D_j^S}  S dV /S^*_\Delta }{ D_j^S} ,
\end{equation}
where $D_j^T, D_j^S \subset \Omega$; $\alpha_j, \beta_j \in \mathbb{R}$, $j = 1,...,n$.
One could further allow $\alpha_j$, $\beta_j$ to be spatially dependent, as long as they are bounded.
The variable values of $\alpha_j$ and $\beta_j$ can be used to better approximate the nonlinear equation of state \citep{gill2016atmosphere}.
\item Any number of density-driven velocity modes $m\in\mathbb{N}$, resulting in a composite velocity field of the form 
\begin{equation}
\begin{aligned}
&\boldsymbol{u}(\boldsymbol{r},t;T,S) = Pe \boldsymbol{u}_E(\boldsymbol{r},t) + Ra^T \sum_{k=1}^m a_I^k(\left<T\right>,\left<S\right>) \boldsymbol{u}_I^k(\boldsymbol{r}),
\end{aligned}
\end{equation}
where $a_I^k$, $k=1,...,m$, are as in equation \eqref{eq:5.8}.
Such a generalization may be used to examine the balance between several competing effects, possibly leading to nontrivial temporal competition between the different modes.
\item Different geometries, domain shapes, etc., as long as the velocity modes do not exit the boundaries, i.e. their normal component to the boundaries of the domain is 0. Also, the work can be generalized to spherical geometry by inserting the appropriate curvature parameters.
\item Spatially dependent values of $\tensor{\kappa}$, provided they are bounded from below by a positive constant.
\item A different combination of boundary conditions - Robin-Robin or Neumann-Neumann.
\item Non-autonomous systems - we expand on these in the following remarks.
\end{itemize}
\end{remark}
\begin{remark}{Non-autonomous systems.}\label{remark:2}

In the case of a non-autonomous system, with the time-dependence arising from the external velocity field $\boldsymbol{u}_E$, the source terms $f_T, f_S$ and/or the boundary forcing terms $T^*, S^*$, Theorem \ref{theorem} is still valid using the same proof. The only modifications would be in the bounds, which would need to always take into account a global-in-time bound for each time-dependent quantity. Similarly, Corollary \ref{corproof:1}
is still valid, with the same modifications to the bounds.
\end{remark}

\begin{remark}{Periodically non-autonomous systems.} \label{remark:3}

In the case of time-dependent sources and boundary data with a period $\mathcal{T}$, it is possible to follow similar arguments as in Theorem \ref{theorem} to show that a time-periodic set of solutions exists for any parameters. The structure of this proof would be to create a time-$\mathcal{T}$ Poincar\'e map of the original PDEs. Since the equations are parabolic, they have a smoothing effect that would allow one to show that the map's embedding is compact by the Rellich lemma. The Schauder-Tychonoff fixed point theorem would then guarantee the existence of a fixed point for the time-$\mathcal{T}$ map. Since the same steps can be followed for any starting point, this would prove the existence of a time-$\mathcal{T}$-periodic solution.
Furthermore, if the forcing and boundary data are weak with respect to the diffusion, it should be possible to show in this case that a time-periodic solution is stable in much the same way as we prove Theorem \ref{theorem:2}(iii)
in this work.
\end{remark}

Indeed, we see numerical evidence of Remark \ref{remark:3}. For example, in Figure~\ref{fig:realistic}, a time-periodic forcing with a yearly period in the velocity field and surface forcing results in a stable time-periodic solution, with the same period.

\section{Discussion and conclusions}\label{sec:theend}

We have presented a novel phenomenological kinematic-dynamic model of the interaction between a 3D time-dependent kinematic flow with a density-driven component and the density function in a closed basin.
The model, depicted in Figure \ref{fig:modelcartoon}, is an extended version of the advection-diffusion equations for temperature and salinity, with  the velocity field serving as an integral coupling term. Additionally, the model allows one to include realistic sources for temperature and salinity.
The coupling renders the equation set non-linear and non-local. We have proven here, using an iteration scheme and energy estimates, that the model is well-posed in the sense of Hadamard (Theorem \ref{theorem}). We have shown that when the sources and the boundary data are time-independent, for a small enough Rayleigh number, all solutions will converge to a unique steady state (Theorem \ref{theorem:2}). Though the analytical bounds we obtain are not tight, this qualitative behavior does appear in numerical simulations of the system (Figure \ref{fig:hysteresisBox}).
Furthermore, the theorems hold for a larger class of velocities, coupling and boundary conditions, as described in Remark \ref{remark:1}. 
In addition, our model bears resemblance to non-local continuum models used to model swarm dynamics and cell migration \citep{mogilner1999non,li2019global}. Assuming the advection field is incompressible, proving well-posedness of such models should follow the same framework as presented here.

The presented model is an intermediate model between the fully coupled NS equations and the uncoupled advection-diffusion equation which is used in kinematic models. Thus, it may allow, in the future, to produce additional simplifications by which Lagrangian trajectories produced by chaotic advection can serve as the active force that changes the internal velocity field component.  

The model is relevant for any incompressible flow with a density-driven component. Our motivating example  has been the North Atlantic large scale flow, built from rapid externally driven modes and a slow density-driven overturning mode (Figure \ref{fig:kinematicFlow}). Indeed, we have shown that, given realistic parameters, the model can produce a stable periodic state with an overturning strength, and temperature and salinity functions, that resemble realistic and more complex simulations of the North Atlantic ocean in current climate conditions (Figure \ref{fig:realistic}). 
The major discrepancy between our results and realistic distributions of temperature and salinity, respectively,   is the density inversions we obtain in the northern latitudes.
These density inversions are to be expected, as we do not have a mechanism that mixes inverted water columns in the model, such as convective adjustment  \citep{gough1997isopycnal}. In the actual North Atlantic ocean, density inversions appear only in the wintertime, as a result of intense cooling events, upon which convection ensues \citep{MarshallSchott1999}. Introducing a mechanism for resolving the density inversions in our model will solve this discrepancy. There are a few options for such solutions, including convective-adjustment correction schemes \citep{gough1997isopycnal}, a local increase of the vertical diffusion coefficient, or a transient, localized kinematic flow that mixes the column through chaotic advection. We are currently exploring the different options; in later works we will present a model that includes density-inversion corrections. 
 
 Another realistic effect that is left for a future {study is} the Mediterranean outflow water's (MOW) role in the variability of the AMOC. The Mediterranean, through the Strait of Gibraltar ($36^\circ$N, $5.7^\circ$W),
inserts approximately 1 Sv of warm, salty and dense water that sinks to a depth of 1000 meters, and spreads westwards, all the way to the eastern coasts of America.
 The various paths of these salty dense waters to the abyss and their dominance in driving the AMOC can be examined by our model.
The effect of a modelled MOW on the asymptotic solutions is interesting in the oceanographic context, as it is still under debate to what extent the MOW influences the AMOC strength in current climate conditions \citep{Reid1978, mccartney2001origin, LozierStewart2008, Ivanovic2014}.

\section*{Acknowledgements}
OSK acknowledges the support of
a research grant from the Yotam Project and the Weizmann Institute Sustainability and Energy Research Initiative; and the support of the S\'ephora Berrebi Scholarship in Mathematics.
VRK and OSK acknowledge the support of the Israel Science Foundation, Grant 1208/16.
EST is grateful to support in part  by the Einstein Stiftung/Foundation - Berlin, through the Einstein Visiting Fellow Program.
HG acknowledges the support of the Vigevani Research Project Prize.

\appendix

\section{Velocity field of the North Atlantic ocean - kinematic model}\label{appA}

Here we present the rescaled version of the velocity fields $\boldsymbol{u}_{E,rs}$ and $\boldsymbol{u}_{I,rs}$ that we use in this work to mimic the North Atlantic ocean's large-scale flow, where  $\boldsymbol{u}_{E,rs}$ describes the wind-driven surface flow, and $\boldsymbol{u}_{I,rs}$ describes the density-driven overturning flow.
We emphasize that the choice of this velocity field is for the numerical part and not for the general analysis part: the general analysis holds for any velocity fields that satisfy the minimal requirements of incompressibility and impermeability.
Thus, the domain we consider is $\Omega = (0,1)^3 \subset \mathbb{R}^3$, and the velocity fields' components are of the order $1$.
In the following, we neglect the $rs$ notation for simplicity of notation.

%
%

The flow is incompressible, and is derived from a vector streamfunction.
To model the wind-driven surface flow, we build a streamfunction component with a ``hill'' and ``valley''
along the latitudinal component $\hat{y}$, and an asymmetric westward-biased ``hill'' along the longitudinal component $\hat{x}$.
The function we use along the $\hat{y}$ component is a simple combination of sine functions,
$\psi_y^H = \sin \pi y  \sin \pi(y-y_H(t))$,
where $y_H(t)$ is the demarcation line between the gyres. Considering a yearly periodicity, we set
$y_H(t) = \bar{y}_H + \tilde{y}_H \sin \pi t {\tau}/{year}$,
where $t=0$ corresponds to January, and $\tau = L_z^2/\kappa_V$ is the typical timescale of the rescaling (see subsection \ref{subsection2.1}).
The time dependency of the demarcation line causes chaotic
mixing between the two gyres, as shown in \cite{YangLiu1994, AharonEtAl2012}.

The function we use along the $\hat{x}$ component is an approximated solution to the Sverdrup equation with a narrow
boundary layer in a wide basin \citep{YangLiu1994}:

\begin{equation*}
\psi_x^H = \frac{l_x}{m_x} (1 - e^{-x/l_x})(x-1),
\end{equation*}

where $m_x$ is a normalizing factor, $m_x = \max_{x} \frac{l_x}{m_x} (1 - e^{-x/l_x})(x-1)$, so that the peak of the ``hill'' is at the value 1. $l_x$ is the westward asymmetry parameter, and determines the width of the western boundary current and the Gulf Stream. At its limits, $l_x\rightarrow 0$ corresponds to a width of 0 and an infinite velocity at $x=0$, while $l_x\rightarrow\infty$ corresponds to a symmetric function around $x=0.5$.

This construction is similar to that used in \cite{YangLiu1994}, in which they considered
a two-dimensional flow built from two asymmetric gyres of a similar functional form with an oscillating
demarcation line between gyres. For a given time $t$, the Eulerian streamlines are closed contours with a westward
bias. The time dependancy of $\psi_y$ causes the Lagrangian trajectories to be chaotic, resulting in transport between the
two gyres.

To these wind-driven features we add the influence of the penetration depth, set as the depth of the thermocline $H_{TC}$. Thus the $z$-dependence of the flow is given by an exponential decay of the horizontal velocity with depth,
$\psi_z^H = 1 - 
\left(
1 + 
\exp - ({1-z})
/
H_{TC}\right)^{-1}$.
The overall streamfunction component $\psi_H$ is the product of these {contributions:}
\begin{equation*}
\psi_H(\boldsymbol{r}, t) = \psi_x^H(x) \psi_y^H(y, t) \psi_z^H(z),
\end{equation*}
and the horizontal component of the velocity field is derived in the common {method:}
\begin{equation*}
\boldsymbol{u}_E \equiv (u_E, v_E, w_E)\; ; \; \; u_E = \partial_y \psi_H, \; v_E = - \partial_x \psi_H, \; w_E = 0.
\end{equation*}
In the full, 3D velocity field, $\boldsymbol{u}_E$ is multiplied by the strength parameter $Pe$.
Present day measurements evaluate the Gulf Stream width at around 150km, and its maximal velocity at around
$2.5$m/s. The average demarcation line position is around $40^\circ N$, and its oscillation is seasonal with an amplitude of approximately $2^\circ$ latitude. The thermocline depth is around 500 meters. In order to reproduce these numbers, the relevant parameters are set to be of the order
$Pe \sim 10^6$,$\bar{y}_H = 0.57$, $\tilde{y}_H = 0.05$, $H_{TC} = 0.1$, $l_x = 0.01$.

To model the vertical overturning flow, we build a zonally symmetric streamfunction component with an asymmetric ``hill''
along the depth component $\hat{z}$, and an asymmetric ``hill'' along the latitudinal component $\hat{y}$. This formulation creates an overturning flow with north-bound transport above the thermocline, a steep downwelling regime along the northern border of the basin, south-bound flow below the thermocline and semi-uniform upwelling south of the downwelling regime.
The asymmetric functions we use are of the same form as $\psi_x^H${:}
\begin{equation*}
\psi_z^V = \frac{l_z}{m_z} \left(1 - e^{\frac{z-1}{l_z}}\right)  z,
\; \; \psi_y^V = \frac{l_y}{m_y} \left(1 - e^{-\frac{y}{l_y}}\right)  (1-y),
\end{equation*}
where $m_z$, $m_y$ are rescaling parameters as in $\psi_H$.
The overall streamfunction is given {by}
\begin{equation*}
\psi_V(\boldsymbol{r}) = \psi_z^V(z)  \psi_y^V(y),
\end{equation*}
 and the velocity is derived in the common {form:}
\begin{equation*}
\boldsymbol{u}_I \equiv (u_I, v_I, w_I) \; ; \; \; \; u_I = 0, \; v_I = \partial_z \psi_V, \;  w_I = - \partial_y \psi_V.
\end{equation*}
For a downwelling regime between $55^\circ N$ and $70^\circ N$, the $\hat{y}$-direction asymmetry parameter is set to $l_y = 0.1$. For a northbound transport extending to a depth of 1000 meters, the $\hat{z}$-direction asymmetry parameter is set to $l_z = 0.1$.
In this case, the maximal velocity is obtained at the downwelling branch.
The full velocity field is given by a weighted sum of the individual contributions{:}
\begin{equation*}
\boldsymbol{u}(\boldsymbol{r},t) = Pe \boldsymbol{u}_E(\boldsymbol{r}, t) + Ra^T a_I(\left<\rho\right>) \boldsymbol{u}_I(\boldsymbol{r}).
\end{equation*}

\section{Model parameters}\label{appB}
In Table \ref{table:new}, we summarize the model parameters used in the numerical experiments shown in figures \ref{fig:realistic} (realistic experiment), \ref{fig:hysteresisBox} (2D autonomous experiment) and \ref{fig:hysteresis3D} (3D autonomous experiment).
To obtain the bi-stability seen in Figures \ref{fig:hysteresisBox} and \ref{fig:hysteresis3D}, the dimensions of the top box $z_B$ and the effective rates of convective heat and mass transfer at the boundary, $g_A^T$ and $g_A^S$ respectively, had to be tweaked, as is common in box model experiments.
The rest of the changes in parameters between the experiments are motivated in section \ref{sec:box}.
Overall, the physical control parameters we changed between the three experiments are the internal strength parameter $a_I$'s proportionality constant $\Gamma$, the eddy diffusivity $\tensor{\kappa}$, the temperature and salinity surface functions $T^*$ and $S^*$, the strength of the external velocity field $a_E$, the intergyre mean demarcation line $\bar{y}_H$ and oscillation amplitude $\tilde{y}_H$, the depth box boundary $z_B$ and the depth asymmetry parameter $l_z$.
Correspondingly, the rescaling parameters $\tau = L_z^2/\kappa_V$, $T^*_\Delta = \max T^* - \min T^* $, and $S^*_\Delta = \max S^* - \min S^*$ change between the experiments, as do the non-dimensional parameters $\delta_x = \frac{\kappa_H/L_x^2}{\kappa_V/L_z^2}$, 
$\delta_y = \frac{\kappa_H/L_y^2}{\kappa_V/L_z^2}$, 
$Sh = \frac{g_A^S}{\kappa_V/L_z}$,
$Pe = \frac{a_E}{\kappa_V/L_z}$,
$Ra^T = \frac{\Gamma \rho_0 \alpha T_\Delta^*}{\kappa_V/L_z}$,
and $R_\rho = \frac{\alpha T^*_\Delta}{\beta S^*_\Delta}$.

%
%

%
%
%
%
%
%
%
\begin{table}
  \begin{center}
\def~{\hphantom{0}}
  \begin{tabular}{llccc}
  \textbf{Parameters}  & \textbf{Parameter description} & \textbf{Realistic} & \textbf{2D aut.} & \textbf{3D aut.} 
                  \\
 $[$units$]$ &  &  {Figure}~\ref{fig:realistic}  & Figure~\ref{fig:hysteresisBox} &  Figure~\ref{fig:hysteresis3D}\\
                   \hline
                   \\
  \textbf{Rescaling} & & &
  \\
  $L_x$ [m] & Longitudinal length  ($0 - 60^\circ W$) & $4 \times 10^6$ m  & $4 \times 10^6$ m & $4 \times 10^6$ m \\
  $L_y$ [m] & Latitudinal length  ($Eq - 70^\circ N$) & $7.7 \times 10^6$ m & $7.7 \times 10^6$ m & $7.7 \times 10^6$ m \\
  $L_z$ [m] & Depth  (AMOC depth) & $4 \times 10^3$ m & $4 \times 10^3$ m & $4 \times 10^3$ m \\
 $\tau$ [yrs]  & Vertical eddy diffusivity timescale & $2\times 10^3$ yr & $10^3-10^6$ yr & $100$ yr \\
$T^*_\Delta$ [$^\circ$C]  & Temperature scale ($\max T^* - \min T^*$) & $30^\circ$C & $18^\circ$C & $18^\circ$C \\
$S^*_\Delta$ [psu]  & Salinity scale ($\max S^* - \min S^*$) & $17$ psu & $0.1-5$ psu & $0.1 - 5$ psu \\
& & & &\\
\textbf{Diffusion} & & &
\\
$\delta_x$ & Rescaled longitudinal diffusion & $10$ & $10^2-10^5$
& $0.1$
\\
$\delta_y$ &  Rescaled latitudinal diffusion & $10$ & $10^2-10^5$
& $0.1$
\\
& & & &\\
\textbf{Sources} & & &
\\
$Nu$ & Nusselt number& $140$ & $140$
& $140$\\
$Sh$ & Sherwood number & $20$ & $10-10^4$ & $1$
\\
$f_T$ & Bulk temperature source & $0$ & $0$ &$0$ 
\\
$f_S$ & Bulk salinity source & $0$ & $0$ &$0$ 
\\
& & & &\\
\textbf{External velocity} &\textbf{(wind-driven)} & &
\\
$Pe$ & P\'eclet number  & $1.6\times 10^6$ & $0$
& $5\times 10^4$\\
$\bar{y}_H$  & Intergyre demarcation line (Figure \ref{fig:kinematicFlow})  & $0.57$ & - & $0.78$
\\
$\tilde{y}_H$  & Intergyre oscillation amplitude & $0.02$ & -
& $0$ \\
$l_x$ & Longitudinal asymmetry parameter & $0.01$ & -
& $0.01$ \\
$H_{TC}$ [m] & Penetration depth & $600$ m & - & $600$ m
\\
& & & &\\
\textbf{Internal velocity} &\textbf{(density-driven)} & &
\\
$Ra^T$ & Thermal Rayleigh number & 
$1.8\times 10^3$ & 
$2\times 10^2 - 2 \times 10^5$ & $20$
\\
$R_\rho$ & Density stability ratio & $5.8$ & $1-50$ & $1-50$
\\
$y_B$ & Latitudinal box boundary (Figure \ref{fig:kinematicFlow}) & $0.78$ & $0.78$ & $0.78$
\\
$z_B$ & Depth box boundary (Figure \ref{fig:kinematicFlow}) & $0.75$ & $0.85$ & $0.85$
\\
$l_y$ & Latitudinal asymmetry parameter & $0.1$ & $0.1$ & $0.1$
\\
$l_z$ & Depth asymmetry parameter & $0.1$ & $0.04$ & $0.04$
\\
\hline \end{tabular}
  \caption{Parameters used in the simulations presented in this work. Aut. = Autonomous.}
  \label{table:new}
  \end{center}
\end{table}

\section{Proof of global well-posedness of the model and corollaries}\label{appC}

\subsection{Problem formulation}
Let $\Omega = (0,1)^3$ be a box domain describing a (rescaled)  oceanic basin, $\boldsymbol{r} = (x,y,z) \in \Omega$ a general point in the domain, $q \in \partial\Omega$ a general point on the boundary of the domain, and define $\boldsymbol{\hat{n}}(q)$ as the unit vector pointing outwards from the boundary of the domain. We further define $\sigma_1 \equiv (x,y,z=1) \subset \partial\Omega$ as the upper surface of the ocean, and $\sigma_0 \equiv (x,y,z=0) \subset \partial\Omega$ as the ocean bottom.
We divide the domain into two subdomains $D_1, D_2 \subset \Omega$, $D_1 \cup D_2 = \Omega$, and define for any function $\phi \in L^2(\Omega)$ the averages
$\left< \phi \right>_{i} = \frac{1}{D_i} \int_{D_i} \phi dV$,
for $i = 1,2$. As a shorthand, we define the vector $\left< \phi \right> \equiv (\left< \phi \right>_1, \left< \phi \right>_2)\in\mathbb{R}^2$.

Let $\tensor{\kappa}$, $\Gamma$, $\alpha$, $\beta$, $a_E$, $\boldsymbol{u}_E$, $\boldsymbol{u}_I$, $f_T$, $f_S$, $g_A^T$, $g_A^S$, $T^*$, $S^*$, $T_0$, and $S_0$ be given, and satisfy the following assumptions:
\begin{equation}
\begin{aligned}
\label{eq:a}
\begin{cases}
&
(a1) \; \; \tensor{\kappa} =
\left(
\begin{matrix}
\kappa_x & 0 & 0 \\
0 & \kappa_y & 0 \\
0 & 0 & \kappa_z
\end{matrix}
\right);
\; \; \kappa_x, \kappa_y, \kappa_z \; \; positive \; constants,
\\
&
(a2) \; \; \Gamma, \alpha, \beta, a_E, g_A^T, g_A^S \; \; positive \; constants,
\\
&
(a3) \; \; \boldsymbol{u}_E, \boldsymbol{u}_I \in (L^\infty(\boldsymbol{r}))^3,
\\
&
(a4) \; \; \boldsymbol{ \nabla} \boldsymbol{\cdot}\boldsymbol{u}_E(\boldsymbol{r})  = \boldsymbol{ \nabla} \boldsymbol{\cdot}\boldsymbol{u}_I(\boldsymbol{r}) = 0,
\\
&
(a5) \; \; \boldsymbol{u}_E \boldsymbol{\cdot}\boldsymbol{\hat{n}}(q) = \boldsymbol{u}_I \boldsymbol{\cdot}\boldsymbol{\hat{n}}(q) = 0,
\\
&
(a6) \; \; f_T, f_S \in L^2(\Omega); \int_\Omega f_S dV = 0,
\\
&
(a7) \; \; T^*, S^* \in H^1(\sigma_1); \int_{\sigma_1} S^* dx dy =  0,
\\
&
(a8) \; \; \partial_x T^*(x,y)|_{x=0,1} = \partial_y T^*(x,y)|_{y=0,1} = 0,
\\
&
(a9) \; \; \partial_x S^*(x,y)|_{x=0,1} = \partial_y S^*(x,y)|_{y=0,1} = 0,
\\
&
(a10) \; \; T_0, S_0 \in L^2(\Omega).
\\
&
(a11) \; \; \int_\Omega S_0 dV = 0.
\end{cases}
\end{aligned}
\end{equation}
Then we consider the following nonlinear evolution problem for a temperature $T$ and a salinity $S$: 
\begin{equation}
\begin{aligned}
\label{eq:P}
&\begin{cases}
(P1) \; \; \partial_t T(t,\boldsymbol{r}) - \boldsymbol{ \nabla}\boldsymbol{\cdot}(\tensor{\kappa}\boldsymbol{ \nabla} T(t,\boldsymbol{r}) )
+ (\boldsymbol{u}(\boldsymbol{r}; \left<T\right>, \left<S\right>)\boldsymbol{\cdot}\boldsymbol{ \nabla})T(t,\boldsymbol{r}) 
= f_T(\boldsymbol{r}), & t > 0 \; , \; \;  \boldsymbol{r}\in\Omega
\\
(P2) \; \; \partial_t S(t,\boldsymbol{r}) - \boldsymbol{ \nabla}\boldsymbol{\cdot}(\tensor{\kappa}\boldsymbol{ \nabla} S(t,\boldsymbol{r})) +
 (\boldsymbol{u}(\boldsymbol{r}; \left<T\right>, \left<S\right>)\boldsymbol{\boldsymbol{\cdot\nabla}})S(t,\boldsymbol{r}) = f_S(\boldsymbol{r}),&  t > 0 \; , \; \;  \boldsymbol{r}\in\Omega
\\
\\
(P3) \; \; a_I(\left<T\right>, \left<S\right>) = \Gamma  \left(-\alpha(
\left<T\right>_{2}(t) - \left<T\right>_{1}(t))
+
\beta(\left<S\right>_{2}(t) - \left<S\right>_{1}(t))
\right),
& t>0
\\
(P4) \; \; \boldsymbol{u}(\boldsymbol{r};
 \left<T\right>, \left<S\right>) = a_E \boldsymbol{u}_E(\boldsymbol{r}) 
+ a_I(\left<T\right>, \left<S\right>) \boldsymbol{u}_I(\boldsymbol{r}), & t>0 \; , \; \; \boldsymbol{r}\in\Omega
\\
\\
(P5) \; \; (\tensor{\kappa} \boldsymbol{ \nabla} T(t,q)) \boldsymbol{\cdot}\boldsymbol{\hat{n}}(q)
=
\begin{cases}
g_A^T (T^*(x,y) - T(t,q)) & \hbox{if}\; \; q \in \sigma_1\\
 0 & \hbox{else}
\end{cases}
,
& t > 0 \; , \; \;  q\in\partial\Omega
\\
(P6) \; \; (\tensor{\kappa} \boldsymbol{ \nabla} S(t,q)) \boldsymbol{\cdot}\boldsymbol{\hat{n}}(q)
=
\begin{cases}
g_A^S S^*(x,y) & \hbox{if}\; \; q \in \sigma_1\\
 0 & \hbox{else}
\end{cases}
,
& t > 0 \; , \; \;  q\in\partial\Omega
\\
\\
(P7) \; \; T(t=0,\boldsymbol{r}) = T_0(\boldsymbol{r}), & \boldsymbol{r}\in\Omega
\\
(P8) \; \; S(t=0,\boldsymbol{r}) = S_0(\boldsymbol{r}). & \boldsymbol{r}\in\Omega
\end{cases}
\end{aligned}
\end{equation}
Note that assumptions $(a8)$ and $(a9)$ are compatibility conditions for the boundary.
The coupled system $(P1)-(P8)$ equipped with 
 assumptions \eqref{eq:a} is denoted by 
$\left(P_{T_{0}, S_{0}, T^*, S^*, f_{T}, f_{S}}\right)$.
Let 
$\lambda = \frac{1}{4} \min \{ g_A^T, \kappa_z\}$ and 
$\kappa_{min} = \min \{ \kappa_x, \kappa_y, \kappa_z \}$; $\lambda$ and $\kappa_{min}$ are strictly positive from assumption (a2).
Throughout the text, $\tensor{\kappa}$, $\Gamma$, $\alpha$, $\beta$, $a_E$, $\boldsymbol{u}_E$, $\boldsymbol{u}_I$ $g_A^T$, $g_A^S$, $\lambda$ and $\kappa_{min}$ are called the problem parameters;
$f_T$, $f_S$ are called the source functions;
$T^*$, $S^*$ are called the boundary functions,
and
$T_0$, $S_0$ are the initial conditions.

\begin{app-remark}
Integrating equation $(P2)$ over the domain, using the relevant boundary conditions, one can show that $\int_\Omega S(t,\boldsymbol{r}) dV = Constant$. 
\end{app-remark}
Hence, by virtue of (a11) without loss of generality and for simplicity we consider solutions with $\int_\Omega S(t,\boldsymbol{r}) dV = 0$. Suppose $W \subset L^1(\Omega)$, then we denote by
$\dot{W} = \{ f \in W : \int_\Omega f dV = 0\}$.

Here we follow closely the presentation in  \citet{Nittka2014}, and correspondingly we define three notions of solutions to $\left(P_{T_{0}, S_{0}, T^*, S^*, f_{T}, f_{S}}\right)$.
For $\mathcal{T}>0$, we denote the Banach spaces:
$W_T = C([0,{\mathcal{T}}]; L^2(\Omega)) \cap L^2(0,{\mathcal{T}}; H^1(\Omega))$,
$W_S = C([0,{\mathcal{T}}]; \dot{L}^2(\Omega)) \cap L^2(0,{\mathcal{T}}; \dot{H}^1(\Omega))$.

\begin{app-definition}
Let ${\mathcal{T}}>0$. $( T, S ) \in  W_T \times W_S$ is a \textit{weak solution 
on $[0,{\mathcal{T}}]$} of $\left(P_{T_{0}, S_{0}, T^*, S^*, f_{T}, f_{S}}\right)$
 if, for all test functions $\psi \in H^1([0,{\mathcal{T}}]; H^1(\Omega))$ with $\psi({\mathcal{T}}) = 0$, the following holds:
\begin{equation}
\label{eq:weakTSnonlinear}
\begin{aligned}
&- \int_0^{\mathcal{T}} \int_\Omega T(t) \partial_t \psi(t) dV dt
+ \int_0^{\mathcal{T}} \left( 
\int_\Omega (\tensor{\kappa} \boldsymbol{ \nabla} T(t) - \boldsymbol{u} T(t)) \boldsymbol{\cdot}\boldsymbol{ \nabla} \psi(t) dV +
\int_{{\sigma_1}} g_A^T T(t) \psi(t) dx dy
\right)dt
\\
&=
\int_\Omega T_0 \psi(0) dV + \int_0^{\mathcal{T}} \int_\Omega f_T \psi(t) dV dt
+ 
\int_0^{\mathcal{T}} \int_{{\sigma_1}} g_A^T T^*(x,y) \psi(t) dx dy dt;
\\
\\
&- \int_0^{\mathcal{T}} \int_\Omega S(t) \partial_t \psi(t) dV dt
+ \int_0^{\mathcal{T}} 
\int_\Omega (\tensor{\kappa} \boldsymbol{ \nabla} S(t) - \boldsymbol{u} S(t)) \boldsymbol{\cdot}\boldsymbol{ \nabla} \psi (t) dV dt
\\
&=
\int_\Omega S_0 \psi(0) dV + \int_0^{\mathcal{T}} \int_\Omega f_S \psi(t) dV dt
+ 
\int_0^{\mathcal{T}} \int_{{\sigma_1}} g_A^S S^*(x,y) \psi(t) dx dy dt,
\end{aligned}
\end{equation}
where $\boldsymbol{u} = \boldsymbol{u}(\boldsymbol{r}; \left< T\right>, \left< S\right>)$ is given by equation $(P4)$ in \eqref{eq:P}.
\end{app-definition}

\begin{app-remark}
If a weak solution exists, then from \eqref{eq:weakTSnonlinear} one can prove that $(T(t=0),S(t=0)) = (T_0, S_0)$.
\end{app-remark}


\begin{app-definition}
Let ${\mathcal{T}}>0$. $( T, S)$ is a  \textit{strong solution
on $[0,{\mathcal{T}}]$} of $\left(P_{T_{0}, S_{0}, T^*, S^*, f_{T}, f_{S}}\right)$
if it is a weak solution, and $( T, S) \in H^1([0,{\mathcal{T}}]; L^2(\Omega))\times H^1([0,{\mathcal{T}}]; \dot{L}^2(\Omega))$.
\end{app-definition}

\begin{app-definition}
Let $\mathcal{T}>0$. 
$( T, S)$ is a \textit{classical} $L^2$\textit{-solution
on} $[0,{\mathcal{T}}]$ of
$\left(P_{T_{0}, S_{0}, T^*, S^*, f_{T}, f_{S}}\right)$
if:
\begin{enumerate}

\item $( T, S) \in  (C^1([0,\mathcal{T}]; L^2(\Omega)) \cap C([0,\mathcal{T}]; H^1(\Omega))) \;
\times
\;
(C^1([0,\mathcal{T}]; \dot{L}^2(\Omega)) \cap C([0,\mathcal{T}]; \dot{H}^1(\Omega))) $

\item $\boldsymbol{ \nabla} \boldsymbol{\cdot}(\tensor{\kappa} \boldsymbol{ \nabla} T),
\boldsymbol{ \nabla} \boldsymbol{\cdot}(\tensor{\kappa} \boldsymbol{ \nabla} S) \in C([0,\mathcal{T}]; L^2(\Omega))$

\item 
$(\tensor{\kappa} \boldsymbol{ \nabla} T) \boldsymbol{\cdot}\boldsymbol{\hat{n}}|_{\partial \Omega},
(\tensor{\kappa} \boldsymbol{ \nabla} S) \boldsymbol{\cdot}\boldsymbol{\hat{n}}|_{\partial \Omega}
\in C([0,\mathcal{T}]; L^2(\partial\Omega))$

\item $(T(t=0), S(t=0)) =( T_0, S_0 )$

\item $( T, S)$ satisfies equations $(P1)$, $(P2)$ 
in the sense of $C([0,\mathcal{T}]; L^2(\Omega))$ and equations
$(P5)$, $(P6)$ in the sense of  $C([0,\mathcal{T}]; L^2(\partial\Omega))$.
\end{enumerate}
\end{app-definition}

As we shall see in Theorem {C.}\ref{theoreminproof}, for classical $L^2$-solutions the initial conditions will need to satisfy, in addition to the relevant assumptions in \eqref{eq:a}, the following compatibility assumptions:
\begin{equation}
\begin{aligned}
\label{eq:b}
\begin{cases}
&
(b1) \; \; T_0, S_0 \in H^4(\Omega)
\\
&
(b2) \; \; (\tensor{\kappa} \boldsymbol{ \nabla} T_0(q)) \boldsymbol{\cdot}\boldsymbol{\hat{n}}(q)
=
\begin{cases}
g_A^T (T^*(x,y) - T_0(q)) & \hbox{if}\; \; q \in \sigma_1\\
 0 & \hbox{else}
\end{cases}
\\
&
(b3) \; \; (\tensor{\kappa} \boldsymbol{ \nabla} S_0(q)) \boldsymbol{\cdot}\boldsymbol{\hat{n}}(q)
=
\begin{cases}
g_A^S S^*(x,y) & \hbox{if}\; \; q \in \sigma_1\\
 0 & \hbox{else}
\end{cases}
\end{cases}
.
\end{aligned}
\end{equation}

We further define the notion of temporally global solutions as follows:
\begin{app-definition}
$(T, S)$ is a \textit{ global weak/strong/classical solution}
of $\left(P_{T_{0}, S_{0}, T^*, S^*, f_{T}, f_{S}}\right)$
if, for any $\mathcal{T}>0$, it is a weak/strong/classical solution on $[0,\mathcal{T}]$, respectively.
\end{app-definition}

\begin{app-theoremprf}
\label{theoreminproof}
Let $\tensor{\kappa}$, $\Gamma$, $\alpha$, $\beta$, $a_E$, $\boldsymbol{u}_E$, $\boldsymbol{u}_I$, $f_T$, $f_S$, $g_A^T$, $g_A^S$, $T^*$, $S^*$, $T_0$, and $S_0$ be given as in \eqref{eq:a}. Then:
\begin{enumerate}
  \item 
 The nonlinear problem $\left(P_{T_{0}, S_{0}, T^*, S^*, f_{T}, f_{S}}\right)$  has
  a unique, global strong solution $( T, S)$.
  \item This solution depends continuously on initial conditions, boundary conditions and sources 
  in the sense described by equation \eqref{eq:continuousdependence}, below.
  \item For every $\mathcal{\bar{T}}>0$, 
the solution satisfies the following bounds:
\begin{equation}
\begin{aligned}
\label{eq:bounds}
&
\sup_{t\in[0,\bar{\mathcal{T}}]}\norm{T}^2(t) \leq C_1^T; \; \;
\int_0^{\mathcal{\bar{T}}}  \norm{\tensor{\kappa}^{1/2} \boldsymbol{ \nabla} T}^2(t) dt
\leq
C_2^T + {\mathcal{\bar{T}}} C_3^T;
\\
&
\sup_{t\in[0,\bar{\mathcal{T}}]}\norm{S}^2(t) \leq C_1^S; \; \;
\int_0^{\mathcal{\bar{T}}}  \norm{\tensor{\kappa}^{1/2} \boldsymbol{ \nabla} S}^2(t) dt
\leq
C_2^S + {\mathcal{\bar{T}}} C_3^S;
\end{aligned}
\end{equation}
where
\begin{equation}
\begin{aligned}
\label{eq:Cs}
C_1^T &\equiv \norm{T_0}^2
+ \frac{g_A^T}{2\lambda(1 - a)} \norms{T^*}^2
 +
\frac{1}{4\lambda^2 a (1 -  a)} \norm{f_T}^2; \\
C_2^T &\equiv \frac{1}{2}\norm{T_0}^2; \; \;
C_3^T \equiv \frac{\lambda}{2} \norm{T_0}^2
+ g_A^T \norms{T^*}^2
+ \frac{1}{\lambda} \norm{f_T}^2;
\\
C_1^S &\equiv
\norm{S_0}^2 + \frac{1}{2 \pi \kappa_{min}^2}
\left(
 \frac{1+\frac{2}{\pi}}{b} \left(g_A^S\right)^2  \norms{S^*}^2
+
\frac{1}{\pi (\frac{1}{2} - b)} \norm{f_S}^2
\right);
\\
C_2^S &\equiv \frac{\pi \kappa_{min}}{2}  \norm{S_0}^2; \; \;
C_3^S \equiv 
\frac{\pi \kappa_{min}}{2}  \norm{S_0}^2
+
6 \frac{g_A^S}{\kappa_{min}}
\left(
g_A^S \norms{S^*}^2
+
\frac{2}{\pi \kappa_{min}} \norm{f_S}^2
\right),
\end{aligned}
\end{equation}
 for $a = \left({1 + \sqrt{1 + 2 g_A^T \frac{\norms{T^*}^2}{\norm{f_T}^2}}}\right)^{-1}$ and
$b = \frac{1}{2}{\left(1+
\frac{1}{ g_A^S \sqrt{2+\pi} }
\frac{\norm{f_S}}
{\norms{S^*}}
\right)}^{-1}.$
  \item If, in addition, the initial conditions satisfy \eqref{eq:b}
then
$\left(P_{T_{0}, S_{0}, T^*, S^*, f_{T}, f_{S}}\right)$
has a unique, global classical $L^2$-solution.  
\end{enumerate}
\end{app-theoremprf}

\subsection{Proof of Theorem {C.}\ref{theoreminproof}}
\subsubsection{Construction of an iterative sequence of approximate solutions}
\label{sec:C21}
The proof will consist of several major steps, which we will describe in the next subsections.

Let us start by constructing an iterative scheme, that will result in a sequence of solutions to an iteratively defined linear problem that approximates the nonlinear problem 
$\left(P_{T_{0}, S_{0}, T^*, S^*, f_{T}, f_{S}}\right)$.
To this end, let $\mathcal{T}>0$, and let $T_m, S_m \in C([0,\mathcal{T}]; L^2(\Omega))$ be given. Then we present the following linear problem, defined for $t \in (0,\mathcal{T}]$:
\begin{equation}
\label{eq:Plinear}
\begin{cases}
(P1^l) \; \; \partial_t T(t,\boldsymbol{r}) - \boldsymbol{ \nabla}\boldsymbol{\cdot}(\tensor{\kappa}\boldsymbol{ \nabla} T(t,\boldsymbol{r}) )
+ (\boldsymbol{u}_m(\boldsymbol{r}, \left<T_m\right>, \left<S_m\right>)\boldsymbol{\cdot\nabla})T(t,\boldsymbol{r}) 
= f_T(\boldsymbol{r}), & t > 0 \; , \; \;  \boldsymbol{r}\in\Omega
\\
(P2^l) \; \; \partial_t S(t,\boldsymbol{r}) - \boldsymbol{ \nabla}\boldsymbol{\cdot}(\tensor{\kappa}\boldsymbol{ \nabla} S(t,\boldsymbol{r})) +
 (\boldsymbol{u}_m(\boldsymbol{r}; \left<T_m\right>, \left<S_m\right>)\boldsymbol{\cdot\nabla})S(t,\boldsymbol{r}) = f_S(\boldsymbol{r}),&  t > 0 \; , \; \;  \boldsymbol{r}\in\Omega
\\
\\
(P3^l) \; \; a_I(\left<T_m\right>, \left<S_m\right>) = \Gamma  \left(-\alpha(
\left<T_m\right>_{2}- \left<T_m\right>_{1})
+
\beta(\left<S_m\right>_{2} - \left<S_m\right>_{1})
\right),
& t>0
\\
(P4^l) \; \; \boldsymbol{u}_m(\boldsymbol{r};
 \left<T_m\right>, \left<S_m\right>) = a_E \boldsymbol{u}_E(\boldsymbol{r}) 
+ a_I(\left<T_m\right>, \left<S_m\right>) \boldsymbol{u}_I(\boldsymbol{r}), & t>0 \; , \; \; \boldsymbol{r}\in\Omega
\\
\\
(P5^l) \; \; (\tensor{\kappa} \boldsymbol{ \nabla} T(t,q)) \boldsymbol{\cdot}\boldsymbol{\hat{n}}(q)
=
\begin{cases}
g_A^T (T^*(x,y) - T(t,q)) & \hbox{if}\; \; q \in \sigma_1\\
 0 & \hbox{else}
\end{cases}
,
& t > 0 \; , \; \;  q\in\partial\Omega
\\
(P6^l) \; \; (\tensor{\kappa} \boldsymbol{ \nabla} S(t,q)) \boldsymbol{\cdot}\boldsymbol{\hat{n}}(q)
=
\begin{cases}
g_A^S S^*(x,y) & \hbox{if}\; \; q \in \sigma_1\\
 0 & \hbox{else}
\end{cases}
,
& t > 0 \; , \; \;  q\in\partial\Omega
\\
\\
(P7^l) \; \; T(t=0,\boldsymbol{r}) = T_0(\boldsymbol{r}), & \boldsymbol{r}\in\Omega
\\
(P8^l) \; \; S(t=0,\boldsymbol{r}) = S_0(\boldsymbol{r}). & \boldsymbol{r}\in\Omega
\end{cases}
\end{equation}
We call equations $(P1^l)$-$(P8^l)$ equipped with assumptions \eqref{eq:a} as 
$\left(P^{linear}_{T_0, S_0, T_m, S_m} \right)$.
Note that in this notation we explicitly specify only the initial conditions $T_0, S_0$ and the functions $T_m, S_m \in C([0,\mathcal{T}]; L^2(\Omega))$; the problem parameters, boundary functions, and source functions are the same as are given above in the theorem formulation.
We remark that this problem is almost identical to the nonlinear problem, except that the velocity field is predetermined by the given functions $T_m, S_m$. Hence, this problem is indeed linear.
In fact, one can decouple $\left(P^{linear}_{T_0, S_0, T_m, S_m} \right)$ into two separate linear problems:
equations $(P1^l), (P3^l), (P5^l)$ for $T(t,\boldsymbol{r})$, and equations $(P2^l), (P4^l), (P6^l)$ for $S(t,\boldsymbol{r})$.

Following exactly as before, we define three notions of solutions to $\left(P^{linear}_{T_0, S_0, T_m, S_m} \right)$:

\begin{app-definition}
Let $\mathcal{T}>0$ and $T_m, S_m \in C([0,\mathcal{T}]; L^2(\Omega))$ be given.
$(T,S) \in W_T \times W_S$ is called a \textit{weak solution 
on $[0,\mathcal{T}]$} of the linear problem $\left(P^{linear}_{T_0, S_0, T_m, S_m} \right)$ 
 if, for all test functions $\psi \in H^1([0,\mathcal{T}]; H^1(\Omega))$ with $\psi(\mathcal{T}) = 0$, the following holds:
\begin{equation}
\label{eq:weakT}
\begin{aligned}
&- \int_0^\mathcal{T} \int_\Omega T(t) \partial_t \psi(t) dV dt
+ \int_0^\mathcal{T} \left( 
\int_\Omega (\tensor{\kappa} \boldsymbol{ \nabla} T(t) - \boldsymbol{u}_m \; T(t)) \boldsymbol{\cdot}\boldsymbol{ \nabla} \psi(t) dV +
\int_{{\sigma_1}} g_A^T T(t) \psi(t) dx dy
\right)dt
\\
&=
\int_\Omega T_0 \psi(0) dV + \int_0^\mathcal{T} \int_\Omega f_T \psi(t) dV dt
+ 
\int_0^\mathcal{T} \int_{{\sigma_1}} g_A^T T^*(x,y) \psi(t) dx dy dt;
\end{aligned}
\end{equation}
\begin{equation}
\label{eq:weakS}
\begin{aligned}
&- \int_0^\mathcal{T} \int_\Omega S(t) \partial_t \psi(t) dV dt
+ \int_0^\mathcal{T} 
\int_\Omega (\tensor{\kappa} \boldsymbol{ \nabla} S(t) - \boldsymbol{u}_m  \; S(t)) \boldsymbol{\cdot}\boldsymbol{ \nabla} \psi (t) dV dt
\\
&=
\int_\Omega S_0 \psi(0) dV + \int_0^\mathcal{T} \int_\Omega f_S \psi(t) dV dt
+ 
\int_0^\mathcal{T} \int_{{\sigma_1}} g_A^S S^*( x,y) \psi(t) dx dy dt.
\end{aligned}
\end{equation}
\end{app-definition}

\begin{app-definition}
Let $\mathcal{T}>0$, $T_m, S_m \in C([0,\mathcal{T}]; L^2(\Omega))$ be given. $( T, S) $ is a \textit{strong solution 
on $[0,\mathcal{T}]$} 
of the linear problem
$\left(P^{linear}_{T_0, S_0, T_m, S_m} \right)$ 
if it is a weak solution on $[0,\mathcal{T}]$, and
\begin{equation*}
T, S \in H^1([0,\mathcal{T}]; L^2(\Omega)).
\end{equation*}
\end{app-definition}

\begin{app-definition}
Let $\mathcal{T}>0$, $T_m, S_m \in C([0,\mathcal{T}]; L^2(\Omega))$ be given. $(T,S)$ is a
\textit{classical $L^2$-solution 
on $[0,\mathcal{T}]$} 
of the linear problem 
$\left(P^{linear}_{T_0, S_0, T_m, S_m} \right)$ 
if:
\begin{enumerate}

\item $( T, S) \in  C^1([0,\mathcal{T}]; L^2(\Omega)) \cap C([0,\mathcal{T}]; H^1(\Omega))
\times
C^1([0,\mathcal{T}]; \dot{L}^2(\Omega)) \cap C([0,\mathcal{T}]; \dot{H}^1(\Omega))$

\item $\{ \boldsymbol{ \nabla} \boldsymbol{\cdot}(\tensor{\kappa} \boldsymbol{ \nabla} T),
\boldsymbol{ \nabla} \boldsymbol{\cdot}(\tensor{\kappa} \boldsymbol{ \nabla} S) \} \in (C((0,\mathcal{T}]; L^2(\Omega)))^2$

\item 
$\{ (\tensor{\kappa} \boldsymbol{ \nabla} T) \boldsymbol{\cdot}\boldsymbol{\hat{n}}|_{\partial \Omega},
(\tensor{\kappa} \boldsymbol{ \nabla} S) \boldsymbol{\cdot}\boldsymbol{\hat{n}}|_{\partial \Omega} \}
\in (C([0,\mathcal{T}]; L^2(\partial\Omega)))^2$

\item $(T(t=0), S(t=0)) =(T_0, S_0)$

\item $T,S$ satisfy equations $(P1^l, P2^l, P5^l, P6^l)$
as functions in $C([0,\mathcal{T}]; L^2(\Omega))$, $C([0,\mathcal{T}]; L^2(\partial\Omega))$.
In particular, thanks to (iii) above, 
$(\partial_t T, \partial_t S) \in \left( C((0,\mathcal{T}]; L^2(\Omega))\right)^2$
\end{enumerate}
\end{app-definition}
In particular, a classical $L^2$-solution is a strong solution.

\begin{app-proposition}\label{prop:1}
Let $\mathcal{T}>0$, $T_m, S_m \in C([0,\mathcal{T}]; L^2(\Omega))$ be given. Then:
\begin{enumerate}
\item $\left(P^{linear}_{T_0, S_0, T_m, S_m} \right)$ 
has a unique strong solution on $[0,\mathcal{T}]$.
\item
If, in addition, the initial conditions $T_0, S_0$ satisfy assumptions \eqref{eq:b},
then 
$\left(P^{linear}_{T_0, S_0, T_m, S_m} \right)$ 
has a unique classical $L^2$-solution on $[0,\mathcal{T}]$.
\end{enumerate}
\end{app-proposition}

{Proof:}
\begin{enumerate}
\item
Existence and uniqueness of weak solutions of the linear problem are guaranteed directly by Theorem 2.11 in \cite{Nittka2014}.
Moreover, existence and uniqueness of strong solutions are given by Remark 2.15 in \cite{Nittka2014}. We note that in order to use Remark 2.15 in \citet{Nittka2014} to prove the existence of a strong solution, one may construct a function $G\in H^1([0,\mathcal{T}]; H^2(\Omega))$ whose restriction to the boundary satisfies the boundary conditions. In our problem setting, due to the compatibility conditions $(a8), (a9)$ in \eqref{eq:a} this task can be easily achieved by setting $G(x,y,z,t) = T^*(x,y,t)$ for $T$, and $G(x,y,z,t) = g_A^S S^*(x,y,t) z^2/2$ for $S$, where $T^*$ and $S^*$ are lifted from the boundary to the full domain and are constant with respect to the $z$ variable. Due to the trace theorem, $G$ has the required regularity, i.e. $G \in H^1([0,\mathcal{T}]; H^2(\Omega))$.
\item Since the boundary condition functions $T^*$, $S^*$, and the source functions $f_T$, $f_S$, are time-independent,
this follows directly from Proposition 2.7(b) in \cite{Nittka2014}.
\end{enumerate}
$\blacksquare$

Equipped with the above, we can now construct a sequence of strong solutions to an iteratively defined linear problem.
\begin{enumerate}
\item 
We initialize the sequence with $(T_0, S_0)$.
Note that these are constant functions in time, and therefore by Proposition {C.}\ref{prop:1} the problem $\left(P^{linear}_{T_0, S_0,T_0, S_0} \right)$ has a unique strong solution, that we denote $(T_1, S_1)$.
Then $T_{1}, S_{1} \in C([0,\mathcal{T}]; L^2(\Omega))$ from the definition of a strong solution to the linear problem, and thus, by Proposition {C.}\ref{prop:1} ,
$\left(P^{linear}_{T_0, S_0,T_{1}, S_{1}} \right)$ has a strong solution, that we denote $( T_{2}, S_{2})$.

\item Let $n> 1$, and assume $(T_{n}, S_{n})$ is a strong solution of
$\left(P^{linear}_{T_0, S_0,T_{n-1}, S_{n-1}} \right)$ for given $T_{n-1}, S_{n-1} \in C([0,\mathcal{T}]; L^2(\Omega))$.
Then $T_{n}, S_{n} \in C([0,\mathcal{T}]; L^2(\Omega))$ from the definition of a strong solution to the linear problem, and thus, by Proposition {C.}\ref{prop:1},
$\left(P^{linear}_{T_0, S_0,T_{n}, S_{n}} \right)$ has a strong solution, that we denote $( T_{n+1}, S_{n+1})$.
\end{enumerate}
By repeating Step (ii) iteratively we construct a sequence $\{(T_n, S_n)\}_{n=1}^\infty$ of strong solutions to the corresponding sequence of linear problems
 $\left\{ \left(P^{linear}_{T_0, S_0,T_{n-1}, S_{n-1}} \right)\right\}_{n=1}^{\infty}$ for $n\geq 1$.
 Under the additional assumptions \eqref{eq:b}, by the same induction steps, the sequence 
  $\{(T_n, S_n)\}_{n=1}^\infty$ is a sequence of classical $L^2$-solutions to $\left\{ \left(P^{linear}_{T_0, S_0,T_{n-1}, S_{n-1}} \right)\right\}_{n=1}^{\infty}$.

\subsubsection{Global uniform bounds on the sequence of approximate solutions}

Now that we have established a sequence of approximate solutions $\{(T_n, S_n)\}_{n=1}^\infty$, we will use energy estimates to establish some uniform in $n$ estimates for $\{(T_n, S_n)\}_{n=1}^\infty$.
To this end, let $\mathcal{T} > 0$, and $n\geq 1$. 
{In the next lemma, we extend the space of test functions for weak solutions of $\left(P^{linear}_{T_0, S_0, T_{n-1}, S_{n-1}} \right)$.}

\begin{app-lemma}
\label{lemma:1}
{Let $(T_n, S_n)$ be a weak solution of $\left(P^{linear}_{T_0, S_0, T_{n-1}, S_{n-1}} \right)$.
Then for every} $\phi\in H^1([0,\mathcal{T}]; L^2(\Omega))\cap L^2(0,\mathcal{T}; H^1(\Omega))$ {with} $\phi(\mathcal{T}) = 0$, $(T_n, S_n)$ 
 {satisfies equations} \eqref{eq:weakT}, \eqref{eq:weakS} {with $\phi$ in place of a test function $\psi \in H^1([0,\mathcal{T}]; H^1(\Omega))$.}
\end{app-lemma}

{Proof:} Since $H^1(\Omega)$ is dense in $L^2(\Omega)$, there exists a sequence $\{ \phi^m \}_{m=1}^\infty$, $\phi^m \in H^1([0,\mathcal{T}]; H^1(\Omega))$ with $\phi^m(\mathcal{T}) = 0$, converging to $\phi$ in $H^1([0,\mathcal{T}]; L^2(\Omega))\cap L^2(0,\mathcal{T}; H^1(\Omega))$.
Thus, it is easy to see that equation \eqref{eq:weakT} with $\psi = \phi^m$ converges, term by term, to the same equation with $\phi$ in place of $\psi$ as $m\rightarrow\infty$.  
{$\blacksquare$}


{In the next lemma, we further extend the space of test functions for weak solutions of $\left(P^{linear}_{T_0, S_0, T_{n-1}, S_{n-1}} \right)$, namely by no longer requesting the condition $\psi(\mathcal{T}) = 0$.}

\begin{app-lemma}
\label{lemma:2}
{Let $(T_n, S_n)$ be a weak solution of 
$\left(P^{linear}_{T_0, S_0, T_{n-1}, S_{n-1}} \right)$.
Then for every} $\psi \in H^1([0,\mathcal{T}]; L^2(\Omega)) \cap L^2(0,\mathcal{T}; H^1(\Omega))$,
{the following holds:}
\begin{equation}
\label{eq:weakT2}
\begin{aligned}
&\int_\Omega T_n(\mathcal{T}) \psi(\mathcal{T}) dV
- \int_0^\mathcal{T} \int_\Omega T_n(t) \partial_t \psi(t) dV dt
+ \int_0^\mathcal{T} \left( 
\int_\Omega (\tensor{\kappa} \boldsymbol{ \nabla} T_n(t) -
\boldsymbol{u}_{n-1} T_n(t)) \boldsymbol{\cdot}\boldsymbol{ \nabla} \psi(t) dV + \right. \\
&\left. \int_{{\sigma_1}} g_A^T T_n(t) \psi(t) dx dy
\right)dt
=
\int_\Omega T_0 \psi(0) dV + \int_0^\mathcal{T} 
\left( \int_\Omega f_T \psi(t) dV
+ 
\int_{{\sigma_1}} g_A^T T^*(x,y) \psi(t) dx dy \right)dt.
\end{aligned}
\end{equation}
\begin{equation}
\label{eq:weakS2}
\begin{aligned}
\int_\Omega S_n(\mathcal{T}) \psi(\mathcal{T}) dV
- \int_0^\mathcal{T} \int_\Omega &S_n(t) \partial_t \psi(t) dV dt
+ \int_0^\mathcal{T}  
\int_\Omega (\tensor{\kappa} \boldsymbol{ \nabla} S_n(t) -
\boldsymbol{u}_{n-1} S_n(t)) \boldsymbol{\cdot}\boldsymbol{ \nabla} \psi(t) dV dt\\
&=
\int_\Omega S_0 \psi(0) dV + \int_0^\mathcal{T} 
\left( \int_\Omega f_S \psi(t) dV
+ 
\int_{{\sigma_1}} g_A^S S^*(x,y) \psi(t) dx dy \right)dt.
\end{aligned}
\end{equation}
\end{app-lemma}

{Proof:} Set $\tilde{\mathcal{T}} = \mathcal{T} + 1$.
Let $(\tilde{T}_n,\tilde{S}_n) $ be a strong solution of 
$\left(P^{linear}_{T_0, S_0, T_{n-1}, S_{n-1}} \right)$
 on $[0,\tilde{\mathcal{T}}]$. 
Define
$\phi(t) = \begin{cases}
\psi(t) \; \; t \in [0,\mathcal{T}]
\\
 \psi(\mathcal{T}) \varphi(t) \; \; t \in (\mathcal{T},\tilde{\mathcal{T}}]
\end{cases},$
where $\varphi(\mathcal{T}) = 1$, and $\varphi$ is a smooth function decreasing to $0$ over the interval $[\mathcal{T}, \tilde{\mathcal{T}}]$.
Then  $\phi$ satisfies the conditions for a test function in the weak solution according to Lemma {C.}\ref{lemma:1}, i.e. $(\tilde{T}_n, \tilde{S}_n)$ satisfy equations \eqref{eq:weakT}, \eqref{eq:weakS} with $\phi$ instead of $\psi$. Also, $(\tilde{T}_n,\tilde{S}_n)$ agrees with $(T_n, S_n)$ on $[0,\mathcal{T}]$ from uniqueness of the strong solution.

Set $\mathcal{T}$ as the new initial time, and let
 $\{\tilde{\tilde{T}}_n, \tilde{\tilde{S}}_n\}$
 be the solution of 
 $\left(P^{linear}_{\tilde{T}_n(\mathcal{T}), 
 \tilde{S}_n(\mathcal{T}),
  T_{n-1}, S_{n-1}} \right)$.
  If the sources and boundary functions are not autonomous, shift their time parameter corresppondingly.
Then $(\tilde{\tilde{T}}_n,\tilde{\tilde{S}}_n)$ agrees with $(\tilde{T}_n,\tilde{S}_n)$ on $[\mathcal{T}, \tilde{\mathcal{T}}]$ from uniqueness of the strong solution, and therefore the following is satisfied:
\begin{equation}
\begin{aligned}
\label{eq:lemma2weakT1}
&
- \int_\mathcal{T}^{\tilde{\mathcal{T}}} \int_\Omega 
\tilde{T}_n(t) \partial_t \phi(t) dV dt
+ \int_\mathcal{T}^{\tilde{\mathcal{T}}} \left( 
\int_\Omega (\tensor{\kappa} \boldsymbol{ \nabla} \tilde{T}_n(t) -
\boldsymbol{u}_{n-1} \tilde{T}_n(t)) \boldsymbol{\cdot}\boldsymbol{ \nabla} \phi(t) dV + \int_{{\sigma_1}} g_A^T \tilde{T}_n(t) \phi(t) dx dy
\right)dt
\\
&=
\int_\Omega \tilde{T}(\mathcal{T}) \phi(\mathcal{T}) dV + 
\int_\mathcal{T}^{\tilde{\mathcal{T}}}
\left( \int_\Omega f_T \phi(t) dV
+ 
\int_{{\sigma_1}} g_A^T T^*(t,x,y) \phi(t) dx dy \right)dt.
\end{aligned}
\end{equation}
On the other hand, $\tilde{T}_n$ satisfies
\begin{equation}
\begin{aligned}
\label{eq:lemma2weakT2}
&
- \int_0^{\tilde{\mathcal{T}}} \int_\Omega 
\tilde{T}_n(t) \partial_t \phi(t) dV dt
+ \int_0^{\tilde{\mathcal{T}}} \left( 
\int_\Omega (\tensor{\kappa} \boldsymbol{ \nabla} \tilde{T}_n(t) -
\boldsymbol{u}_{n-1} \tilde{T}_n(t)) \boldsymbol{\cdot}\boldsymbol{ \nabla} \phi(t) dV +
 \int_{{\sigma_1}} g_A^T \tilde{T}_n(t) \phi(t) dx dy
\right)dt
\\
&=
\int_\Omega T_0 \phi(0) dV + 
\int_0^{\tilde{\mathcal{T}}}
\left( \int_\Omega f_T \phi(t) dV
+ 
 \int_{{\sigma_1}} g_A^T T^*(t,x,y) \phi(t) dx dy \right)dt.
\end{aligned}
\end{equation}
Subtracting \eqref{eq:lemma2weakT1} from \eqref{eq:lemma2weakT2}, and separating the time integral to two parts, $\int_0^{\tilde{\mathcal{T}}} = \int_0^{\mathcal{T}} + \int_{\mathcal{T}}^{\tilde{\mathcal{T}}}$, we are left with:
\begin{equation*}
\begin{aligned}
&
\int_\Omega \tilde{T}_n(\mathcal{T}) \phi(\mathcal{T}) dV 
- \int_0^{\mathcal{T}} \int_\Omega 
\tilde{T}_n(t) \partial_t \phi(t) dV dt
+ \int_0^{\mathcal{T}} \left( 
\int_\Omega (\tensor{\kappa} \boldsymbol{ \nabla} \tilde{T}_n(t) -
\boldsymbol{u}_{n-1} \tilde{T}_n(t)) \boldsymbol{\cdot}\boldsymbol{ \nabla} \phi(t) dV \right.
\\
&\left. + \int_{{\sigma_1}} g_A^T \tilde{T}_n(t) \phi(t) dx dy
\right)dt
=
\int_\Omega T_0 \phi(0) dV + 
\int_0^{{\mathcal{T}}}
\left( \int_\Omega f_T \phi(t) dV
+ 
\int_{{\sigma_1}} g_A^T T^*(t,x,y) \phi(t) dx dy \right)dt.
\end{aligned}
\end{equation*}
Now, since $\tilde{T}_n(t) = T_n(t)$  on $[0,\mathcal{T}]$ and $\phi(t) = \psi(t)$ on $[0,\mathcal{T}]$, this equals
\begin{equation*}
\begin{aligned}
&
\int_\Omega {T}_n(\mathcal{T}) \psi(\mathcal{T}) dV 
- \int_0^{\mathcal{T}} \int_\Omega 
{T}_n(t) \partial_t \psi(t) dV dt
+ \int_0^{\mathcal{T}} \left( 
\int_\Omega (\tensor{\kappa} \boldsymbol{ \nabla} {T}_n(t) -
\boldsymbol{u}_{n-1} {T}_n(t)) \boldsymbol{\cdot}\boldsymbol{ \nabla} \psi(t) dV \right.
\\
&\left. + \int_{{\sigma_1}} g_A^T {T}_n(t) \psi(t) dx dy
\right)dt
=
\int_\Omega T_0 \psi(0) dV + 
\int_0^{{\mathcal{T}}}
\left( \int_\Omega f_T \psi(t) dV
+ 
\int_{{\sigma_1}} g_A^T T^*(t,x,y) \psi(t) dx dy \right)dt.
\end{aligned}
\end{equation*}
The same steps may be followed for $S_n$.
{$\blacksquare$}

\begin{app-remark}
\label{remark:lemma2}
Following similar arguments as in Lemma {C.}\ref{lemma:1} and Lemma {C.}\ref{lemma:2} above, one can show that a weak solution $(T,S)$ of the nonlinear problem $\left(P_{T_{0}, S_{0}, T^*, S^*, f_{T}, f_{S}}\right)$ satisfies equations analogous to equations \eqref{eq:weakT2} and \eqref{eq:weakS2}, with $T$ in place of $T_n$, $S$ in place of $S_n$, and, in place of $\boldsymbol{u}_{n-1}$, $\boldsymbol{u} = \boldsymbol{u}(\boldsymbol{r}; \left< T \right>, \left< S \right>)$ as in equation $(P4)$ in \eqref{eq:P} .
\end{app-remark}

Next we present a few additional lemmas that will be useful for proving our global bounds.

\begin{app-lemma}{(Poincar{\'e} inequality)}
\label{lemma:3}
{For a function $T \in H^1(\Omega)$, the following inequality holds:}
\begin{equation*}
\frac{g_A^T}{2} \norms{T}^2
 + \norm{\tensor{\kappa}^{1/2} \boldsymbol{ \nabla} T}^2
 \geq \lambda \norm{T}^2.
\end{equation*}
\end{app-lemma}

\begin{app-lemma}
\label{lemma:4}
{
Let $S \in H^1(\Omega)$. For any $\epsilon>0$,
\begin{equation*}
\begin{aligned}
\norms{S}^2
\leq
\left(1+\frac{1}{\epsilon}\right) \norm{S}^2 
+ \frac{\epsilon}{\kappa_{min} } 
\norm{\tensor{\kappa}^{1/2} \boldsymbol{ \nabla} S}^2.
\end{aligned}
\end{equation*}
}
\end{app-lemma}
\begin{app-remark}
Lemma {C.}\ref{lemma:4} is a version of the trace theorem customized for this problem.
\end{app-remark}

\begin{app-lemma}
\label{lemma:5}
Let $S \in H^1(\Omega)$. Then
\begin{equation*}
\norm{S - \left< S\right>_{\Omega}}^2 \leq \frac{1}{\pi \kappa_{min}} \norm{\tensor{\kappa}^{1/2}\boldsymbol{ \nabla} S}^2
\;
\hbox{ where  }
\left< S\right>_\Omega \equiv \frac{1}{|\Omega|} \int S dV.
 \end{equation*}
\end{app-lemma}

Proofs of these useful Lemmas {C.}\ref{lemma:3}-\ref{lemma:5} are left to the last section \ref{sec:lemmaproofs}.
Now, we can prove several global bounds on the strong solutions $(T_n, S_n)$ of 
$\left(P^{linear}_{T_0, S_0, T_{n-1}, S_{n-1}} \right)$, that we have established in section \ref{sec:C21}.

\begin{app-proposition}\label{prop:2}
{
Suppose 
$(T_n, S_n)$ is a strong solution on $[0,\mathcal{T}]$ of $\left(P^{linear}_{T_0, S_0, T_{n-1}, S_{n-1}} \right)$, as established in section  \ref{sec:C21}. Then $T_n$, $S_n$ satisfy the following bounds, that are independent of $n$:}
\begin{enumerate}

\item $\sup_{t\in[0, \mathcal{T}]} \norm{T_n}^2(t) \leq C_1^T$;

\item
$\int_0^{\mathcal{T}}  \norm{\tensor{\kappa}^{1/2} \boldsymbol{ \nabla} T_n}^2(t) dt
\leq
C_2^T + {\mathcal{T}} C_3^T$;

\item $\sup_{t\in[0, \mathcal{T}]}\norm{S_n}^2(t) \leq C_1^S$;

\item
$\int_0^{\mathcal{T}}  \norm{\tensor{\kappa}^{1/2} \boldsymbol{ \nabla} S_n}^2(t) dt
\leq
C_2^S + {\mathcal{T}} C_3^S$;
\end{enumerate}
where $C_i^j$ for $i\in \{1,2,3\} $, $j\in \{T,S\}$ are given explicitly in equation \eqref{eq:Cs}.
\end{app-proposition}

{Proof:}

 {(i)
 \underline{Uniform in $n$, $L^\infty([0,\mathcal{T}]; L^2(\Omega))$ bounds on $T_n$:}}

According to Lemma {C.}\ref{lemma:2}, $T_n$ can be used as a test function in equation \eqref{eq:weakT2} in place of $\psi$, and thus satisfies:
\begin{equation*}
\begin{aligned}
&\int_\Omega (T_n({\mathcal{T}}))^2 dV
- \int_0^{\mathcal{T}} \int_\Omega T_n(t) \partial_t T_n(t) dV dt
+ \int_0^{\mathcal{T}} \left( 
\int_\Omega (\tensor{\kappa} \boldsymbol{ \nabla} T_n(t) +
\boldsymbol{u}_{n-1} T_n(t)) \boldsymbol{\cdot}\boldsymbol{ \nabla} T_n(t) dV + \right. \\
&\left. \int_{{\sigma_1}} g_A^T (T_n(t))^2 dx dy
\right)dt
=
\int_\Omega T_0^2 dV + \int_0^{\mathcal{T}} 
\left( \int_\Omega f_T T_n(t) dV
+ 
\int_{{\sigma_1}} g_A^T T^*(t,x,y) T_n(t) dx dy \right)dt.
\end{aligned}
\end{equation*}
Therefore, since $T_n \in L^2(0,\mathcal{T}; H^1(\Omega)) \cap C([0,\mathcal{T}]; L^2(\Omega))$, one can employ assumptions $(a4), (a5)$ along with a generalized version of the divergence theorem \citep{constantin1988navier} to obtain:
\begin{equation*}
\int_0^{\mathcal{T}} \left(
\frac{1}{2}\frac{d}{dt} \norm{T_n}^2
 + g_A^T \norms{T_n}^2 
 + \norm{\tensor{\kappa}^{1/2} \boldsymbol{ \nabla} T_n}^2 \right) dt
 = \int_0^{\mathcal{T}} \left(
 g_A^T\int_{{\sigma_1}}{T_n T^*(x,y) d \sigma}
 +
 \int_\Omega{T_n f_T} \right) dt.
\end{equation*}
Thus, the integrands of time are equal almost everywhere. The next steps will be performed inside the integrands, and eventually will be integrated over time again.
We use the Cauchy-Schwarz and Young inequalities to obtain:
\begin{equation*}
\begin{aligned}
\frac{1}{2}\frac{d}{dt} \norm{ T_n}^2
 + g_A^T \norms{T_n}^2
 + \norm{\tensor{\kappa}^{1/2} \boldsymbol{ \nabla} T_n}^2
&\leq
  g_A^T  \norms{T_n} \norms{T^*}
 +
\norm{ 2 \epsilon T_n}
\norm{ \frac{1}{2 \epsilon} f_T}
\\ &\leq
 \frac{g_A^T}{2}  \norms{T_n}^2
+ \frac{g_A^T}{2}  \norms{T^*}^2
 +
\epsilon \norm{T_n}^2
+
\frac{1}{4 \epsilon} \norm{f_T}^2
\end{aligned}
\end{equation*}
and after reordering:
\begin{equation}
\begin{aligned}
\frac{1}{2}\frac{d}{dt} \norm{ T_n}^2
 + \frac{g_A^T}{2} \norms{T_n}^2
 + \norm{\tensor{\kappa}^{1/2} \boldsymbol{ \nabla} T_n}^2
\leq
\epsilon \norm{T_n}^2
+
\frac{g_A^T}{2}  \norms{T^*}^2
 +
\frac{1}{4 \epsilon} \norm{f_T}^2
\label{eq:boundwithgrad}
\end{aligned}
\end{equation}
for any $\epsilon>0$. Recall that all of the parameters are rescaled to render the entire problem, and each of its constituents, dimensionless, therefore $\epsilon$ is also dimensionless.

Apply Lemma {C.}\ref{lemma:3} to the left-hand side, to deduce:
\begin{equation}
\frac{d}{dt} \norm{T_n}^2
 + 2(\lambda-\epsilon) \norm{T_n}^2
 \leq
 g_A^T \norms{T^*}^2
 +
\frac{1}{2\epsilon} \norm{f_T}^2
\equiv
\zeta.
\label{eq:C1ineq}
\end{equation}
By choosing $0<\epsilon<\lambda$, and defining $\eta\equiv2(\lambda-\epsilon)>0$, we obtain a bound on the growth of $T_n$,
$
\frac{d}{dt} \norm{T_n}^2
 + \eta \norm{T_n}^2
 \leq
\zeta
$,
and using Gronwall's inequality,
\begin{equation*}
\norm{T_n}^2
\leq
e^{-\eta t} \norm{T_0}^2
+ \frac{\zeta}{\eta} (1-e^{-\eta t})
.
\end{equation*}
This means that $\norm{T_n}$ is bounded, and the bound is independent of $n$, and for all $t\in[0,{\mathcal{T}}]$:
\begin{equation}
\norm{T_n}^2(t)
\leq
C_1^T
\equiv 
\norm{T_0}^2
+ \frac{g_A^T}{2\lambda(1 - a)} \norms{T^*}^2
 +
\frac{1}{4\lambda^2 a (1 -  a)} \norm{f_T}^2
\label{eq:L2Tbound}
\end{equation}
where $a \in (0,1)$,
{Observe that $C_1^T$ obtains its minimal value for
$a = \frac{1}{1 + \sqrt{1 + 2 g_A^T \frac{\norms{T^*}^2}{\norm{f_T}^2}}}$.}

 {(ii)
 \underline{Uniform in $n$, $L^2(0,\mathcal{T}; H^1(\Omega))$ bounds on $T_n$:}}
 
Combining equations \eqref{eq:boundwithgrad} and \eqref{eq:L2Tbound} yields:
\begin{equation*}
\frac{1}{2} \frac{d}{dt} \norm{T_n}^2
 + \norm{\tensor{\kappa}^{1/2} \boldsymbol{ \nabla} T_n}^2
 \leq
a_2 \lambda C_1^T + \frac{g_A^T}{2} \norms{T^*}^2 + \frac{1}{4 {a_2} \lambda} \norm{f_T}^2
\end{equation*}
for any $a_2 \in (0,1)$.
Integrate over $[0,\mathcal{T}]$, and set $a_2 = a$ :
\begin{equation*}
\frac{1}{2} \norm{T}^2 ({\mathcal{T}}) + 
\int_0^{\mathcal{T}} dt \norm{\tensor{\kappa}^{1/2} \boldsymbol{ \nabla} T}^2
\leq 
 \frac{1}{2} \norm{T_0}^2
 +
 {\mathcal{T}} \left(
 a \lambda \norm{T_0}^2
+ \frac{g_A^T}{2(1-a)} \norms{T^*}^2
+ \frac{1}{4 (1-a) a \lambda} \norm{f_T}^2
 \right).
\end{equation*}
This is true for any value of $a\in(0,1)$, and the value of $a$ can be tweaked to obtain an optimal bound. However, we do not care about the exact value of this bound, so to simplify notation, in this subsection we set $a=1/2$ to obtain:
\begin{equation*}
\frac{1}{2} \norm{T}^2 ({\mathcal{T}}) + 
\int_0^{\mathcal{T}} dt \norm{\tensor{\kappa}^{1/2} \boldsymbol{ \nabla} T}^2
\leq 
C_2^T + {\mathcal{T}} C_3^T
\end{equation*}
where
\begin{equation*}
\begin{aligned}
C_2^T &\equiv \frac{1}{2}\norm{T_0}^2; \; \;
C_3^T &\equiv \frac{\lambda}{2} \norm{T_0}^2
+ g_A^T \norms{T^*}^2
+ \frac{1}{\lambda} \norm{f_T}^2.
\end{aligned}
\end{equation*}
\newline
 {(iii)
\underline{Uniform in $n$, $L^\infty([0,\mathcal{T}]; L^2(\Omega))$ bounds on $S_n$:}}
 
Taking the same approach for $S_{n}$ here as we did for $T_{n}$, we note that the only difference between the two functions is in the boundary conditions, therefore we can immediately write:
\begin{equation*}
\frac{1}{2} \frac{d}{dt} \norm{S_n}^2
- g_A^S \int_{{\sigma_1}} S_n(x,y,z=1,t) S^*(x,y)d\sigma
+ \norm{\tensor{\kappa}^{1/2}\boldsymbol{ \nabla} S_n }^2 
= \int_\Omega S_n f_S dV
\end{equation*}
for almost all times $[0,{\mathcal{T}}]$.
Rearranging the terms and using Cauchy-Schwarz and Young inequalities on the boundary and source terms, we obtain:
\begin{equation*}
\frac{1}{2} \frac{d}{dt} \norm{S_n}^2
+
\norm{\tensor{\kappa}^{1/2}\boldsymbol{ \nabla} S_n }^2 
\leq 
\frac{g_A^S}{2 \epsilon_1} \norms{S^*}^2 + \frac{1}{2 \epsilon_2} \norm{f_S}^2+ 
\frac{g_A^S}{2} \epsilon_1 \norms{S_n}^2 + \frac{\epsilon_2}{2} \norm{S_n}^2,
\end{equation*}
for any $\epsilon_1, \epsilon_2 > 0$.
Using Lemma {C.}\ref{lemma:4},
we are left with:
\begin{equation}
\frac{1}{2} \frac{d}{dt} \norm{ S_n}^2
+
\left(1-\frac{g_A^S}{2 \kappa_{min}} \epsilon_1\right)
\norm{\tensor{\kappa}^{1/2}\boldsymbol{ \nabla} S_n }^2 
\leq
\frac{g_A^S}{2 \epsilon_1} \norms{S^*}^2 + \frac{1}{2 \epsilon_2} \norm{f_S}^2+ 
\left(\frac{\epsilon_2}{2} + g_A^S \epsilon_1\right) \norm{S_n}^2,
\label{eq:boundwithgradS}
\end{equation}
where $\epsilon_1 < \frac{2 \kappa_{min}}{g_A^S}$.
Since we set $\left<{S}_n\right>_{\Omega} = 0$, Lemma {C.}\ref{lemma:5} gives us
\begin{equation*}
\frac{d}{dt} \norm{S_n}^2
+
\nu \norm{S_n}^2
\leq
\frac{g_A^S}{ \epsilon_1} \norms{S^*}^2 + \frac{1}{\epsilon_2} \norm{f_S}^2,
\end{equation*}
where
$
\nu \equiv {2 \pi \kappa_{min}} - {(\pi+2)} g_A^S \epsilon_1 - \epsilon_2.
$
Choosing $\epsilon_2$ small enough such that $\nu > 0$, we can use the Gronwall inequality to establish a bound on $\norm{S_n}(t)$ for all $t \in [0,{\mathcal{T}}]$:
\begin{equation}
\label{eq:snepsilonbound}
\norm{S_n}^2(t)
\leq
 \norm{S_0}^2
+ 
\frac{g_A^S}{ \epsilon_1 \nu} \norms{S^*}^2 + \frac{1}{\epsilon_2 \nu} \norm{f_S}^2.
\end{equation}
Taking $\epsilon_1 = 2 \frac{\kappa_{min}}{g_A^S} \frac{b}{1+2/\pi}$, $\epsilon_2 = 2 {\kappa_{min}}\pi d$ for $b,d>0$ satisfying $b+d < 1$, the requirements on $\epsilon_1, \epsilon_2$ are guaranteed.
One can check that the optimal values for $b$ and $d$ are given by
$
b = \left({2
+
\frac{2}{
g_A^S \sqrt{2 + \pi}
}
\frac{\norm{f_S}}{\norms{S^*}}
}\right)^{-1}
\; , \; \; d = \frac{1}{2} - b
$,
and the bound simplifies to:
\begin{equation*}
\norm{S_n}^2(t)
\leq
C_1^S \equiv
\norm{S_0}^2 + \frac{1}{2 \pi \kappa_{min}^2}
\left(
 \frac{1+\frac{2}{\pi}}{b} (g_A^S)^2  \norms{S^*}^2
+
\frac{1}{\pi (\frac{1}{2} - b)} \norm{f_S}^2
\right).
\end{equation*}
\newline
 {(iv) \underline{Uniform in $n$, $L^2(0,{\mathcal{T}}; \dot{H}^1(\Omega))$ bounds on $S_n$:}}

Using the bound on $\norm{S_n}^2$ obtained above with equation \eqref{eq:boundwithgradS}, we deduce:
\begin{equation*}
\begin{aligned}
&\frac{1}{2} \frac{d}{dt} \norm{S_n}^2 + 
(1 - \frac{g_A^S}{2\kappa_{min}} \epsilon_1)\norm{\tensor{\kappa}^{1/2} \boldsymbol{ \nabla} S_n}^2
\leq
&\frac{g_A^S}{2 \epsilon_1} \norms{S^*}^2 + \frac{1}{2 \epsilon_2} \norm{f_S}^2 + (\frac{\epsilon_2}{2} + g_A^S \epsilon_1)
\norm{S_n}^2
\end{aligned}
\end{equation*}
for any $\epsilon_1$, $\epsilon_2$.
Integrate over the time interval $[0,\mathcal{T}]$, and use equation \eqref{eq:snepsilonbound} to deduce
\begin{equation*}
\begin{aligned}
&\int_0^{\mathcal{T}} dt \norm{\tensor{\kappa}^{1/2} \boldsymbol{ \nabla} S}^2 \leq 
\frac{\kappa_{min}}{2 \kappa_{min} - g_A^S  \epsilon_1} \norm{S_0}^2
+\\
&+{\mathcal{T}} 
\frac{(2 g_A^S \epsilon_1 + \epsilon_2)\kappa_{min}}{2 \kappa_{min} - g_A^S \epsilon_1} \norm{S_0}^2
+
\frac{\kappa_{min}}{\epsilon_1 (2 \kappa_{min} - g_A^S \epsilon_1 (1+2/\pi) - \epsilon_2/\pi)}
(\frac{1}{\epsilon_2} \norm{f_S}^2
+
 g_A^S\norms{S^*}^2).
\end{aligned}
\end{equation*}
One can enlarge the bounds and choose values for $\epsilon_1$ and $\epsilon_2$ small enough to obtain 
\begin{equation*}
\begin{aligned}
\int_0^{\mathcal{T}} \norm{\tensor{\kappa}^{1/2} \boldsymbol{ \nabla} S}^2 dt  \leq C_2^S + {\mathcal{T}} C_3^S,
\end{aligned}
\end{equation*}
where
\begin{equation*}
\begin{aligned}
C_2^S &\equiv \frac{\pi \kappa_{min}}{2}  \norm{S_0}^2;
\; \;
C_3^S \equiv 
\frac{\pi \kappa_{min}}{2}  \norm{S_0}^2
+
6 \frac{g_A^S}{\kappa_{min}}
\left(
g_A^S \norms{S^*}^2
+
\frac{2}{\pi \kappa_{min}} \norm{f_S}^2
\right).
\end{aligned}
\end{equation*}
{$\blacksquare$}


\subsubsection{Convergence of the iterative approximate sequences}

In this section we prove that the sequence $\{ (T_n, S_n) \}_{n=1}^\infty$ of strong solutions to the corresponding iteratively defined problems $\left\{ \left(P^{linear}_{T_0, S_0, T_{n-1}, S_{n-1}} \right)\right\}_{n=1}^\infty$ established in section \ref{sec:C21} converges to limit functions, that we denote $(T_{\infty}, S_{\infty})$. 
Eventually, we will prove that these limit functions are solutions to the nonlinear problem $\left(P_{T_{0}, S_{0}, T^*, S^*, f_{T}, f_{S}}\right)$.

\begin{app-proposition}\label{prop:3}
Let $\mathcal{T}>0$ be given, and let $\{(T_n, S_n)\}_{n=1}^\infty$ be strong solutions on $[0,\mathcal{T}]$ of the iteratively defined sequence of problems $\left\{ \left(P^{linear}_{T_0, S_0, T_{n-1}, S_{n-1}} \right)\right\}_{n=1}^\infty$, as established in section \ref{sec:C21}.
There exists $\tau\in(0,\mathcal{T}]$ such that 
$\{ T_n\}_{n=1}^\infty$, $\{S_n \}_{n=1}^\infty$ converge to limit functions, respectively $T_\infty$ and $S_\infty$, strongly in $L^\infty([0,\tau];L^2(\Omega))$.
\end{app-proposition}

{Proof:}
It is enough to show that there exists $\tau\in(0,\mathcal{T}]$
 for which $\{ (T_n, S_n) \}_{n=1}^\infty$ is a Cauchy sequence in the Banach space $(L^\infty([0,\tau];L^2(\Omega)))^2$,
hence it has a limit $(T_\infty, S_\infty) \in (L^\infty([0,\tau];L^2(\Omega)))^2$.

Define $\delta T_n \equiv T_{n} - T_{n-1}$, $\delta S_n \equiv S_{n} - S_{n-1}$, $\eta_n(t) \equiv \norm{\delta T_n}^2(t) + \norm{\delta S_n}^2(t)$, $n \in \mathbb{N}$. 
Let $m\in\mathbb{N}$. 
Then $\delta T_{m+1}$ can be used as a test function for weak solutions in equation \eqref{eq:weakT2} according to Lemma {C.}\ref{lemma:2}.
Thus, we subtract equation \eqref{eq:weakT2} for $T_{m+1}$ in place of $T_n$ and $\delta T_{m+1}$ in place of $\psi$,  from equation  \eqref{eq:weakT2} for $T_m$ in place of $T_n$ and $\delta T_{m+1}$ in place of $\psi$, deducing:
\begin{equation*}
\begin{aligned}
&\int_0^\mathcal{T} dt
\left(
\frac{1}{2}\frac{d}{dt} \norm{\delta T_{m+1}}^2
+ \norm{\tensor{\kappa}^{1/2} \boldsymbol{\nabla} \delta T_{m+1}}
+ g_A^T \norms{\delta T_{m+1}} + \int_{\Omega} \delta T_{m+1} \delta \boldsymbol{u}_m \boldsymbol{\cdot} (\boldsymbol{\nabla} T_m)
\right)
= 0,
\end{aligned}
\end{equation*}
where we define $\delta \boldsymbol{u}_m \equiv \boldsymbol{u}_{m} - \boldsymbol{u}_{m-1}$, add and subtract $T_m \boldsymbol{u}_m \boldsymbol{\cdot} (\boldsymbol{\nabla} T_m)$, and use the generalized divergence theorem \citep{constantin1988navier} and assumptions $(a4), (a5)$ in \eqref{eq:a}.
Therefore,
\begin{equation}
\begin{aligned}
\frac{1}{2}\frac{d}{dt} \norm{\delta T_{m+1}}^2(s)
&\leq
\frac{1}{2}\frac{d}{dt}\norm{\delta T_{m+1}}^2(s)
+\norm{\tensor{\kappa}^{1/2} \boldsymbol{ \nabla} \delta T_{m+1}}^2(s)
+ g_A^T \norms{ \delta T_{m+1}}^2(s)
\\
&\leq
\int_\Omega |\delta T_{m+1} | |\delta \boldsymbol{u}_m | |\boldsymbol{ \nabla} T_m|(t) dV,
\label{eq:2mainineq}
\end{aligned}
\end{equation}
where as before, we find bounds for a general time $s\in[0,\mathcal{T}]$, and will eventually integrate over time.
We note that $\delta \boldsymbol{u}_m = (a_I^m - a_I^{m-1})\boldsymbol{u}_I$. Since $\boldsymbol{u}_I$ is a known, bounded function by construction, we can bound it with its $L^\infty$ norm $u_I^{\hbox{max}} = \max_{\boldsymbol{r}\in\Omega}{|\boldsymbol{u}_I(\boldsymbol{r})|} < \infty$. 
Thus,
\begin{equation}
\begin{aligned}
\norm{\delta \boldsymbol{u}_m}_{L^\infty(\Omega)} &\leq u_I^{\hbox{max}} \Gamma
\left(
\frac{1}{|D_1|}  \int_{D_1} (\alpha |\delta T_m| + \beta |\delta S_m|) dV
+ \frac{1}{|D_2|} \int_{D_2} (\alpha |\delta T_m| + \beta |\delta S_m|) dV
\right)
\\
&\leq
c  \int_{\Omega} ( |\delta T_m| +  |\delta S_m|),
 \label{eq:deltaubound}
\end{aligned}
\end{equation}
where
$c = u_{I}^{\hbox{max}} \Gamma \frac{ \max \{\alpha, \beta \}}{\min \{ |D_1|, |D_2| \}}$.
By \eqref{eq:2mainineq},  \eqref{eq:deltaubound}, and Cauchy-Schwarz, we obtain
\begin{equation*}
\frac{d}{dt} \norm{\delta T_{m+1}}^2 
 = 2  \norm{\delta T_{m+1}} 
 \frac{d}{dt} \norm{\delta T_{m+1}}
\leq
2 c \norm{\delta T_{m+1}}
\norm{\boldsymbol{ \nabla} T_m}
\left( \int_{\Omega} ( |\delta T_m| +  |\delta S_m|) dV\right),
\end{equation*}
therefore
\begin{equation*}
 \frac{d}{dt} \norm{\delta T_{m+1}}
\leq
c
\norm{\boldsymbol{ \nabla} T_m}
\left(\int_\Omega (|\delta T_m| + |\delta S_m|) dV \right)
.
\end{equation*}
Integrate over the time interval $[0,t]$ for $t\in[0,\mathcal{T}]$ to obtain:
\begin{equation*}
\norm{\delta T_{m+1}}(t)
\leq
c \sup_{0\leq s \leq t} \left\{ \int_\Omega 
\left(
|\delta T_m|(s) + |\delta S_m|(s) 
\right)
dV
\right\}
 \int_0^t\norm{\boldsymbol{ \nabla} T_m} dt,
\end{equation*}
while observing that $\norm{\delta T_{m+1}}(0)=0$.
By the Cauchy-Schwarz inequality,
\begin{equation*}
\norm{\delta T_{m+1}}(t) \leq
c \sup_{0\leq s \leq t} \left\{ \int_\Omega 
\left(|\delta T_m| + |\delta S_m| \right) dV
\right\}
\left(\int_0^t\norm{\boldsymbol{ \nabla} T_m}^2 dt\right)^{1/2}
 t^{1/2}. 
\end{equation*}
By virtue of bound $(ii)$ in Proposition {C.}\ref{prop:2},
\begin{equation*}
\norm{\delta T_{m+1}}(t)
\leq
b_T
\sup_{0\leq s \leq t} \left\{ \int_\Omega 
\left(
|\delta T_m| + |\delta S_m| \right) dV
\right\}
 (1 + t )^{1/2} t^{1/2}
\end{equation*}
where $b_T = \frac{c \max \{ C_2^T, C_3^T\}^{1/2}}{
\min \{ \kappa_x,\kappa_y,\kappa_z\}} 
$.
Observe that
\begin{equation*}
\begin{aligned}
\left(\int_\Omega (|\delta T_m| + |\delta S_m|)dV\right)^2
\leq
2 \left( \int_\Omega |\delta T_m|dV \right)^2 + 
2 \left( \int_\Omega |\delta S_m| dV \right)^2
\leq
2 |\Omega | \eta_m,
\end{aligned}
\end{equation*}
therefore
\begin{equation}
\begin{aligned}
\label{eq:deltaTm}
\norm{\delta T_{m+1}}^2 (t) 
\leq
2 |\Omega |
b_T^2
\sup_{0\leq s \leq t} \{ \eta_m  (s)\}
(1+ t) t.
\end{aligned}
\end{equation}
Similar calculations may be performed for
 $\delta S_{m+1}$, yielding
\begin{equation}
\begin{aligned}
\label{eq:deltaSm}
\norm{\delta S_{m+1}}^2 (t) 
\leq
2 |\Omega |
b_S^2
\sup_{0\leq s \leq t} \{ \eta_m  (s)\}
(1+t) t,
\end{aligned}
\end{equation}
where $b_S = \frac{c \max \{ C_2^S, C_3^S\}^{1/2}}{2 
\min \{ \kappa_x,\kappa_y,\kappa_z\}} 
$.
Combining \eqref{eq:deltaTm} and \eqref{eq:deltaSm} yields, for any $t \in [0,\mathcal{T}]$ and $\tau\in[t,\mathcal{T}]$,
\begin{equation}
\begin{aligned}
\eta_{m+1}(t)
&\leq
2 |\Omega | \max \{b_T^2, b_S^2\} (1+t) t 
\sup_{0\leq s \leq t} \{\eta_m(s)\}
\\
&\leq
2 |\Omega | \max \{b_T^2, b_S^2\} (1+\tau) \tau
\sup_{0\leq s \leq \tau} \{\eta_m(s)\}.
\label{eq:C56}
\end{aligned}
\end{equation}
Hence,  \eqref{eq:C56} implies
\begin{equation*}
\sup_{0\leq s \leq \tau} \{\eta_{m+1}(s) \}
\leq
\theta 
\sup_{0\leq s \leq \tau} \{\eta_m(s)\}
\end{equation*}
where 
\begin{equation*}
\theta = b (1+\tau)\tau, \; with \; b =  2 |\Omega |  \max \{b_T^2, b_S^2 \}.
\end{equation*}
Let $\tau$ be chosen small enough such that $\theta = 1/2$. 
This yields
\begin{equation*}
\sup_{0\leq s \leq \tau} \{\eta_{m+1} \}
\leq
2^{-m} 
\sup_{0\leq s \leq \tau} \{\eta_1\},
\end{equation*}
which implies that the sequence $\{ (T_n, S_n) \}_{n=1}^\infty$ is a Cauchy sequence in $(L^\infty([0,\tau];L^2(\Omega)))^2$.
{$\blacksquare$}
%
%
%
%
\subsubsection{$(T_\infty, S_\infty)$ is a global solution to the nonlinear model $\left(P_{T_{0}, S_{0}, T^*, S^*, f_{T}, f_{S}}\right)$}

\begin{app-proposition}
\label{prop:4}
Let $(T_\infty, S_\infty)$ and $\tau>0$ be as in Proposition {C.}\ref{prop:3}. Then $(T_\infty, S_\infty)$ is a strong solution to 
$\left(P_{T_{0}, S_{0}, T^*, S^*, f_{T}, f_{S}}\right)$
 on $[0,\tau]$.
 Furthermore, $T_\infty$ and $S_\infty$  satisfy bounds $(i),(ii)$ and $(iii), (iv)$, respectively, of Proposition {C.}\ref{prop:2}, where $n$ is replaced by $\infty$ and $\mathcal{T}$ is replaced by $\tau$.
 \end{app-proposition}

{Proof:}
In this proof, we  focus on the time interval $[0,\tau]$.
By Proposition {C.}\ref{prop:1},  $\{ T_n \}_{n=1}^\infty, \{ S_n \}_{n=1}^\infty \subset C([0,\tau]; L^2(\Omega))$.
Moreover, by the proof of Proposition {C.}\ref{prop:3} the sequences 
are Cauchy in $L^\infty([0,\tau]; L^2(\Omega))$.
%
Therefore, since the convergence is uniform in time we conclude that
$T_\infty, S_\infty \in C([0,\tau]; L^2(\Omega))$.

By Proposition {C.}\ref{prop:2}, the sequence $\{ T_n \}_{n=1}^\infty$ is bounded in the Hilbert space $L^2(0,\tau; H^1(\Omega))$; hence it has a subsequence $\{T_{n_k}\}_{k=1}^\infty$ that converges weakly to  $\bar{T}$ in  $L^2(0,\tau;H^1(\Omega))$ and in $L^2(0,\tau;L^2(\Omega))$  \citep{yosida5p1}. On the other hand, from Proposition {C.}\ref{prop:3}, $T_\infty$ is a strong limit of $\{T_{n_k}\}_{k=1}^\infty$ in $L^\infty([0,\tau]; L^2(\Omega))$. Since the $L^\infty$ norm is stronger than the $L^2$ norm on a bounded interval, and from uniqueness of the weak limit, it follows that $\bar{T} = T_\infty$, therefore $T_\infty \in L^2(0,\tau; H^1(\Omega)) \cap C([0,\tau]; L^2(\Omega))$. The same argument holds for $S_\infty$.
Therefore the limit functions $T_\infty, S_\infty$ satisfy the regularity conditions of weak solutions.

%
%
%
%
%

Given a test function $\psi \in H^1([0,\tau]; H^1(\Omega))$, we want to show that $T_\infty$, $S_\infty$ satisfy equation \eqref{eq:weakTSnonlinear}.
For all $n\in\mathbb{N}$, $T_n, S_n$ satisfy equations \eqref{eq:weakT}, \eqref{eq:weakS} respectively, with $T_n$ replacing $T$, $S_n$ replacing $S$, and $\boldsymbol{u}_{n-1}$ in place of $\boldsymbol{u}_m$.
Due to the strong convergence of $\{ T_n \}_{n=1}^\infty, \{ S_n \}_{n=1}^\infty$ in $C([0,\tau]; L^2(\Omega))$, the first term in each of the equations \eqref{eq:weakT}, \eqref{eq:weakS} converges to, respectively, 
$\int_0^\tau \int_\Omega T_\infty(t) \partial_t \psi(t) dt dV,$
$\int_0^\tau \int_\Omega S_\infty(t) \partial_t \psi(t) dt dV$.

Let us move into the subsequence $\{T_{n_k}\}_{k=1}^\infty$. Then 
\begin{equation*}
\int_0^\tau
\int_\Omega 
\left(
\tensor{\kappa}  (\boldsymbol{\nabla} T_{n_k} - \boldsymbol{\nabla} T_{\infty} ) \boldsymbol{\cdot} (\boldsymbol{\nabla} \psi)
\right) dV dt
\underset{k\rightarrow\infty}{\rightarrow} 0
\end{equation*}
due to the weak convergence in $ L^2(0,\tau;H^1(\Omega))$.
The equivalent argument shows that the same thing is true for $S_\infty$ as well.
Regarding the velocity term, observe that
\begin{equation*}
\begin{aligned}
|\boldsymbol{u}_{n_k-1}| 
\leq \mathcal{U} \equiv
|a_E| u_{E}^{\hbox{max}} +
u_{I}^{\hbox{max}} \frac{\Gamma |\Omega|^{1/2}}{\min\{|D_1|, |D_2|\}}(\alpha (C_1^T)^{1/2}+ \beta (C_1^S) ^{1/2}),
\end{aligned}
\end{equation*}
where $u_{E}^{\hbox{max}} = \sup_{\boldsymbol{r} \in \Omega} |\boldsymbol{u}_E|$. Therefore, by the strong convergence of $T_{n}$ to $T_\infty$ in $C([0,\tau]; L^2(\Omega))$ and the Cauchy-Schwarz inequality,
\begin{equation*}
\begin{aligned}
\left| \int_0^\tau \int_\Omega \boldsymbol{u}_{n_k-1} (T_{n_k} - T_\infty) \boldsymbol{\cdot}\boldsymbol{ \nabla} \psi dV dt\right|
\leq
\mathcal{U} \int_0^\tau \int_\Omega |T_{n_k} - T_\infty| |\boldsymbol{ \nabla} \psi| dV dt
\underset{k\rightarrow\infty}{\rightarrow} 0.
\end{aligned}
\end{equation*}
From equation \eqref{eq:deltaubound}, 
\begin{equation*}
\begin{aligned}
&\left|
\int_0^\tau \int_\Omega
T_\infty (\boldsymbol{u}_{n_k-1} - \boldsymbol{u}_\infty)  \boldsymbol{\cdot}(\boldsymbol{ \nabla} \psi)  dV dt \right|
\\
&\leq
c \int_0^\tau \left(
\norm{T_\infty} \norm{\boldsymbol{\nabla} \psi}
\int_{\Omega} \left( |T_{n_k-1} - T_\infty| +  |S_{n_k-1} - S_\infty|\right) dV
\right)dt \\
&\leq
2 c d |\Omega|^{1/2} 
\left(\int_0^\tau 
\norm{\boldsymbol{\nabla} \psi}
\left(
\int_{\Omega} \left( |T_{n_k-1} - T_\infty|^2 +  |S_{n_k-1} - S_\infty|^2\right) dV
\right)^{1/2} dt
 \right)\\
&\leq
2 c d g |\Omega|^{1/2} 
\left(\int_0^\tau 
\int_{\Omega} \left( |T_{n_k-1} - T_\infty|^2 +  |S_{n_k-1} - S_\infty|^2\right) dV
dt
 \right)^{1/2}\underset{k\rightarrow\infty}{\rightarrow} 0,
\end{aligned}
\end{equation*}
where $c,d,g$ are some positive constants, and we used Cauchy-Schwarz and the regularity of $T_\infty$ and $\psi$. The same calculation holds for $S_\infty$.

The last term we need to take care of is the boundary term, $\int_0^\tau \int_{{\sigma_1}} g_A^T T_n(t) \psi(t) dx dy dt$ in equation \eqref{eq:weakT}, relevant only to the temperature equation. We define an auxilliary function
 $\beta_T(q) = \begin{cases}
 g_A^T \; \hbox{if}\; q \in \sigma_1
 \\ 0 \;\; \,\hbox{if}\; q \in \partial \Omega \setminus \sigma_1
 \end{cases}$. 
 Thus,
 \begin{equation*}
 \int_0^\tau \int_{\partial\Omega} \beta_T (T_{n_k}-T_\infty) \psi d\sigma dt
 \underset{k\rightarrow\infty}{\rightarrow} 0,
 \end{equation*}
since $T_{n_k}$ converges weakly in $L^2([0,\tau]; H^1(\Omega))$ to $T_\infty$.
Therefore, $(T_\infty, S_\infty)$ is a weak solution of 
$\left( P_{T_{0}, S_{0}, T^*, S^*, f_{T}, f_{S}}\right)$ on $[0,\tau]$.


Next, we prove that the weak solution $(T_\infty, S_\infty)$ established above is indeed a strong solution of $\left( P_{T_{0}, S_{0}, T^*, S^*, f_{T}, f_{S}}\right)$
on $[0,\tau]$. 
Let us define $\boldsymbol{u}_\infty \equiv \boldsymbol{u}_\infty(\boldsymbol{r}; \left< T_\infty \right>, \left< S_\infty \right>)$ following the notation of 
$(P4^l)$ in \eqref{eq:Plinear}.
 Then $\boldsymbol{u}_\infty \in L^\infty((0,\tau); L^\infty(\Omega))^3$ is given, and can be used to define the linear problem
$\left(P^{linear}_{T_0, S_0, T_\infty, S_\infty}\right)$ on $[0,\tau]$.
Due to Remark 2.15 in \cite{Nittka2014}, this problem has a strong solution $(T,S)$. 
Observe that $(T_\infty, S_\infty)$ is also a weak solution of the linear problem $\left(P^{linear}_{T_0, S_0, T_\infty, S_\infty}\right)$ on $[0,\tau]$, since it solves the nonlinear problem.
Since a strong solution to the linear problem is also a weak solution, and weak solutions of the linear problem are unique, the strong solution $(T, S)$ must equal $(T_\infty, S_\infty)$ on the time interval $[0,\tau]$.

Regarding the bounds of Proposition {C.}\ref{prop:2}, 
since $T_\infty$, $S_\infty$ are strong limits of $T_{n_k}$, $S_{n_k}$ in $C([0,\tau]; L^2(\Omega))$ and  weak limits in $L^2(0,\tau; H^1(\Omega))$, respectively, then $T_\infty$, $S_\infty$ enjoy the same bounds as the sequence itself, as established in Proposition {C.}\ref{prop:2}.
{$\blacksquare$}

Next, we prove that the solution $\left(T_\infty, S_\infty \right)$ as established in Proposition {C.}\ref{prop:4} is unique, and that it has a continuous dependence on the data of the system, namely the initial and boundary conditions.

\begin{app-proposition}\label{prop:5}
There exists ${\bar{\tau}}>0$ such that 
 $\left( P_{T_{0}, S_{0}, T^*, S^*, f_{T}, f_{S}}\right)$
 has a unique strong solution on $[0,{\bar{\tau}}]$, with a continuous dependence on the problem's data - the initial conditions, the boundary conditions, and the sources.
\end{app-proposition}


{Proof:}
Let $( T_1^{\hbox{nl}}, S_1^{\hbox{nl}})$ and $( T_2^{\hbox{nl}}, S_2^{\hbox{nl}} )$ be strong solutions of the nonlinear problems
$\left( P_{T_{0,i}, S_{0,i}, T^*_i, S^*_i, f_{T,i}, f_{S,i}}\right)$ on 
$[0,\tau_i]$, $\tau_i>0$,
for $i = 1$ and $i=2$, respectively, as established in Proposition {C.}\ref{prop:4}.
Define ${\bar{\tau}} \equiv \min\{\tau_1, \tau_2\}$, $\delta T \equiv T_2^{\hbox{nl}} - T_1^{\hbox{nl}}$, $\delta S \equiv S_2^{\hbox{nl}} - S_1^{\hbox{nl}}$.
Then $\delta T$, $\delta S$ can be used as test functions for weak solutions on $[0,\bar{\tau}]$ according to Lemma {C.}\ref{lemma:2}. 

Let us start with considering equation \eqref{eq:weakTSnonlinear} for the temperature, with $T_i^{\hbox{nl}}$, $S_i^{\hbox{nl}}$ in place of $T$, $S$ for $i=1,2$ and $\delta T$ as a test function in place of $\psi$. 
The difference between the equations satisfies
\begin{equation*}
\begin{aligned}
&
\int_0^{\bar{\tau}} \left(
\frac{1}{2} \frac{d}{dt} \norm{\delta T}^2
+ \norm{{\tensor{\kappa}}^{1/2} \boldsymbol{ \nabla} \delta T}^2
-
\int_\Omega (
\boldsymbol{u}_2 T_2^{\hbox{nl}} - \boldsymbol{u}_1 T_1^{\hbox{nl}}) \boldsymbol{\cdot}(\boldsymbol{ \nabla} \delta T) dV 
+
g_A^T \norms{ \delta T}^2 
\right) dt =
\\
&
\int_0^{\bar{\tau}} \left(
\int_\Omega \delta f_{T} \delta T dV
+ 
\int_{{\sigma_1}} g_A^T \delta T^* \delta T dx dy  \right) dt
\end{aligned}
\end{equation*}
where $\boldsymbol{u}_i= \boldsymbol{u}\left(\boldsymbol{r}; \left< T_i^{\hbox{nl}} \right>, \left< S^{\hbox{nl}}_i\right>\right)$ for $i = 1,2$, 
$\delta f_T = f_{T,2} - f_{T,1}$
and $\delta T^* = T^*_2 - T^*_1$.
Inside the time integrand we add and subtract $\int_\Omega (\boldsymbol{u}_2 T_1^{\hbox{nl}} \boldsymbol{\cdot}(\boldsymbol{ \nabla} \delta T))$. We also use (a4), (a5)  in \eqref{eq:a}
to simplify the velocity terms and deduce
\begin{equation*}
\begin{aligned}
&
\frac{1}{2} \frac{d}{dt} \norm{\delta T}^2
+ \norm{{\tensor{\kappa}}^{1/2} \boldsymbol{ \nabla} \delta T}^2
+
g_A^T \norms{ \delta T}^2 
\leq
\int_\Omega
\left(
 |\delta T| |\delta\boldsymbol{ {u} }| |\boldsymbol{ \nabla} T_1^{\hbox{nl}}|
+
 |\delta f_{T}| |\delta T| \right) dV
+ 
g_A^T\int_{{\sigma_1}} | \delta T^*| |\delta T| d\sigma
\end{aligned}
\end{equation*}
where $\delta\boldsymbol{ u} \equiv \boldsymbol{u}_2 - \boldsymbol{u}_1$.
Employing Cauchy-Schwarz and using Lemma {C.}\ref{lemma:3}, we deduce:
\begin{equation*}
\begin{aligned}
\frac{1}{2} \frac{d}{dt} \norm{\delta T}^2
\leq
\int_\Omega |\delta T| |\delta\boldsymbol{ {u} }| |\boldsymbol{ \nabla} T_1^{\hbox{nl}}|dV
+
\frac{1}{2\lambda} \norm{\delta f_T}^2 
+
\frac{g_A^T}{2}  \norms{\delta T^*}^2.
\end{aligned}
\end{equation*}
Use the bound on $\delta\boldsymbol{ u}$ from \eqref{eq:deltaubound}, and employ Cauchy-Schwarz again, to obtain:
\begin{equation}
\begin{aligned}
\label{eq:C69}
\frac{d}{dt} \norm{\delta T}^2
\leq
h (\norm{\delta T} + \norm{\delta S}) 
\norm{\delta T} \norm{\boldsymbol{ \nabla} T_1^{\hbox{nl}}}
+
\frac{1}{\lambda} \norm{\delta f_T}^2 
+
g_A^T  \norms{\delta T^*}^2,
\end{aligned}
\end{equation}
where $h = 2 u_{I}^{\hbox{max}} \Gamma \frac{\max \{ \alpha, \beta \}}{\min \{ |D_1|, |D_2| \}} |\Omega|^{1/2}$.
Applying similar arguments on $S$ and using Lemmas {C.}\ref{lemma:4} and {C.}\ref{lemma:5} , there is some constant $\epsilon>0$ small enough for which we have
\begin{equation}
\begin{aligned}
\label{eq:C70}
\frac{d}{dt} \norm{\delta S}^2
\leq
h (\norm{\delta T} + \norm{\delta S}) \norm{\delta S}  \norm{\boldsymbol{ \nabla} S_1^{\hbox{nl} }}
+
\frac{1}{2\epsilon} \norm{\delta f_S}^2 
+
\frac{g_A^S}{2\epsilon}  \norms{\delta S^*}^2.
\end{aligned}
\end{equation}
Defining $\eta \equiv \norm{\delta T}^2 + \norm{\delta S}^2$, from \eqref{eq:C69} and \eqref{eq:C70} we deduce
\begin{equation*}
\begin{aligned}
\frac{d}{dt} \eta
-
\zeta(t) \eta
\leq
\mu
\end{aligned}
\end{equation*}
where $\zeta(t) = h \left( \norm{\boldsymbol{\nabla} T_1^{\hbox{nl}} }+\norm{\boldsymbol{\nabla} S_1^{\hbox{nl}} }\right)$, and 
$\mu = \frac{1}{\lambda} \norm{\delta f_T}^2 
+
g_A^T  \norms{\delta T^*}^2 + 
\frac{1}{2\epsilon} \norm{\delta f_S}^2 
+
\frac{g_A^S}{2\epsilon}  \norms{\delta S^*}^2
$.
Then, by Gronwall's inequality, for $t \in [0,\bar{\tau}]$,
\begin{equation*}
\eta(t) \leq \left( \eta(0) + \mu t \right) \exp\left(\int_0^t \zeta(t') dt' \right) 
\leq \left( \eta(0) + \mu t \right) \exp\left(\left( \tilde{a} t + \tilde{b} t^2 \right)^{1/2}\right),
\end{equation*}
where we employ the bounds (ii) and (iv) from Proposition {C.}\ref{prop:2}, and $\tilde{a}$ and $\tilde{b}$ are positive constants that depend on the readily established estimates on solutions, equations \eqref{eq:Cs}.
Hence,
\begin{equation}
\begin{aligned}
\label{eq:continuousdependence}
& \sup_{t\in[0,\bar{\tau}]} \eta \leq \left( \eta(0) + \mu \bar{\tau} \right) \exp\left(\left( \tilde{a} \bar{\tau}+ \tilde{b} \bar{\tau}^2 \right)^{1/2}\right),
\end{aligned}
\end{equation} 
implying uniqueness and continuous dependence on data on the time interval $[0,\bar{\tau}]$.
{$\blacksquare$}

\begin{app-proposition}\label{prop:bounds}
Let $(T,S)$ be a strong solution of $\left(P_{T_{0}, S_{0}, T^*, S^*, f_{T}, f_{S}}\right)$ on $[0,\mathcal{T}]$ for $\mathcal{T}>0$.
Then, bounds \eqref{eq:bounds} are satisfied on the time interval $[0,\mathcal{T}]$.
\end{app-proposition}
Proof:
By Remark {C.}\ref{remark:lemma2}, 
 a strong solution can be used as a test function in the analogous equations to \eqref{eq:weakT2} and \eqref{eq:weakS2}, as described in the remark.
Then, following similar arguments as in Proposition {C.}\ref{prop:2}, one can conclude the proof of our proposition.
{$\blacksquare$}

Next, we establish global existence and uniqueness.

\begin{app-proposition}\label{prop:6}
Let $ \bar{\mathcal{T}}\in(0,\infty)$ be given and let \eqref{eq:a} hold.
Then:
\begin{enumerate}
\item 
Problem $\left(P_{T_{0}, S_{0}, T^*, S^*, f_{T}, f_{S}}\right)$ has a unique strong solution
$(T, S)$ on $[0,\bar{\mathcal{T}}]$, which depends continuously on the  initial conditions, boundary functions and sources in the sense described in equation \eqref{eq:continuousdependence},
and satisfies the bounds from equations \eqref{eq:bounds} and \eqref{eq:Cs}.
\item
If the initial conditions additionally satisfy assumptions \eqref{eq:b}
then $(T, S)$ is also a classical $L^2$-solution of 
$\left(
P_{T_{0}, S_{0}, T^*, S^*, f_{T}, f_{S}}
\right)$
on $[0,\bar{\mathcal{T}}]$.
\end{enumerate}
\end{app-proposition}

{Proof:}
\begin{enumerate}
\item
Let  $\tau>0$ such that $(T, S)$ is a unique strong solution of 
$\left(P_{T_{0}, S_{0}, T^*, S^*, f_{T}, f_{S}}\right)$
 on $[0,\tau]$, according to Propositions {C.}\ref{prop:4} and {C.}\ref{prop:5}.
 Let ${\mathcal{T}} 
 \geq \tau$ be the maximal interval of existence of the strong solution,
 ${\mathcal{T}} = \sup\left\{{s\in[\tau,\infty]}:\;  \forall t < s, \; (T,S)\hbox{ is a unique strong solution on }[0,t]\right\}$.

If ${\mathcal{T}}=\infty$, the solution is global in time and we are done.

Assume by contradiction that ${\mathcal{T}}<\infty$.
Then necessarily $\limsup_{t\rightarrow\mathcal{T}^-}\norm{T}(t) = \infty$ or  $\limsup_{t\rightarrow\mathcal{T}^-}\norm{S}(t) = \infty$. Else, 
using Proposition {C.}\ref{prop:3}, the solution can be extended beyond the maximal interval of existence $\mathcal{T}$, which is a contradiction.
However, 
by Proposition {C.}\ref{prop:bounds}
 $\limsup_{t\rightarrow\mathcal{T}^-}\norm{T}(t) < \infty$ and  $\limsup_{t\rightarrow\mathcal{T}^-}\norm{S}(t) < \infty$, which contradicts the fact that ${\mathcal{T}}<\infty$. Hence, $\mathcal{T}$ must be infinite.

Uniqueness and continuous dependence on initial conditions are established in the same manner as in Proposition {C.}\ref{prop:5}.

\item
Since $(T, S)$ is a strong solution
of $\left( P_{T_{0}, S_{0}, T^*, S^*, f_{T}, f_{S}}\right)$ on $[0,\bar{\mathcal{T}}]$,
 we can define $\boldsymbol{u} \equiv \boldsymbol{u}(\boldsymbol{r}; \left< T \right>, \left< S \right>)$ from 
$(P4^l)$.
 Then $\boldsymbol{u} \in L^\infty(0,\bar{\mathcal{T}}; L^\infty(\Omega))^3$, and can be considered as a given function and used to define the linear problem
$\left(P^{linear}_{T_0, S_0, T, S}\right)$.
Due to Proposition 2.7 in \cite{Nittka2014}, under the additional assumptions \eqref{eq:b} this problem has a classical $L^2$-solution $(T^{\hbox{c}},S^{\hbox{c}})$ on $[0,\bar{\mathcal{T}}]$. Since a classical solution is also a strong solution, and strong solutions of the linear problem are unique, then $(T^{\hbox{c}},S^{\hbox{c}}) = (T, S)$
on $[0,\bar{\mathcal{T}}]$.
\end{enumerate}
{$\blacksquare$}

%
%
%

To summarize, we have proved the main theorem, that the nonlinear problem
$\left( P_{T_{0}, S_{0}, T^*, S^*, f_{T}, f_{S}}\right)$
is well-posed in the sense of Hadamard: the model has 
\begin{enumerate}
\item a unique, global, strong solution,
\item with a smooth dependence on the problem's data: initial conditions, boundary conditions and sources.
\item This solution satisfies the bounds described in equation \eqref{eq:bounds}. 
\item With the additional regularity conditions on the initial conditions of \eqref{eq:b}, the solution is also a classical $L^2$-solution. 
\end{enumerate}

{$\blacksquare$}

%
%
%
%
%
%


\begin{app-corollaryprf}\label{corproof:1}
Let $\tensor{\kappa}$, $\Gamma$, $\alpha$, $\beta$, $a_E$, $\boldsymbol{u}_E$, $\boldsymbol{u}_I$, $f_T$, $f_S$, $g_A^T$, $g_A^S$, $T^*$, $S^*$, $T_0$, and $S_0$ be as in \eqref{eq:a}, and 
let $(T,S)$ be the global strong solution of $\left( P_{T_{0}, S_{0}, T^*, S^*, f_{T}, f_{S}}\right)$, as established in Theorem {C.}\ref{theoreminproof}.
The dynamic weight of the velocity function, $a_I(\left<T\right>,\left<S\right>)$, as defined in equation (P3) in \eqref{eq:P}, is bounded at all times:
\begin{equation*}
|a_I(\left<T\right>,\left<S\right>)| \leq \frac{\Gamma |\Omega|^{1/2}}{\min\{ |D_1|, |D_2|\} } \left(\alpha (C_1^T )^{1/2} + \beta (C_1^S)^{1/2}\right),
\end{equation*}
where the values of $C_1^T$ and $C_1^S$ are given in equation \eqref{eq:Cs}.
\end{app-corollaryprf}

{Proof:}
Using the Cauchy-Schwarz and the Young inequalities, and the bounds \eqref{eq:bounds} from Theorem {C.}\ref{theoreminproof}, we obtain a bound on $a_I$:
\begin{equation*}
\begin{aligned}
|a_I(\left<T\right>,\left<S\right>)|
&\leq \Gamma\left(\frac{1}{|D_1|} \int_{D_1} |-\alpha T + \beta S| dV + \frac{1}{|D_2|} \int_{D_2} |-\alpha T + \beta S| dV \right) \\
&\leq
\frac{\Gamma |\Omega|^{1/2}}{\min \{|D_1|, |D_2|\}}\left(\alpha \norm{T} + \beta \norm{S}\right)
\leq
\frac{\Gamma |\Omega|^{1/2}}{\min \{|D_1|, |D_2|\}}\left(\alpha (C_1^T)^{1/2}+ \beta (C_1^S)^{1/2}\right).
\end{aligned}
\end{equation*}
{$\blacksquare$}

%
%
%
%
%
%
%
%
%
%
%
%
%
%
%
%

\subsection{The steady-state problem}\label{sec:steadystate}
This section investigates steady-state solutions to the nonlinear problem defined above.
Consider the nonlinear time-independent problem
\begin{equation}
\begin{aligned}
\label{eq:Pss}
\left(P^{\hbox{ss}}\right)
\begin{cases}
(P1^{\hbox{ss}}) \; \; - \boldsymbol{ \nabla}\boldsymbol{\cdot}(\tensor{\kappa}\boldsymbol{ \nabla} T(\boldsymbol{r}) )
+ (\boldsymbol{u}(\boldsymbol{r}; \left<T\right>, \left<S\right>)\boldsymbol{\cdot\nabla})T(\boldsymbol{r}) 
= f_T(\boldsymbol{r}), & \boldsymbol{r}\in\Omega
\\
(P2^{\hbox{ss}}) \; \;  - \boldsymbol{ \nabla}\boldsymbol{\cdot}(\tensor{\kappa}\boldsymbol{ \nabla} S(\boldsymbol{r})) +
 (\boldsymbol{u}(\boldsymbol{r}; \left<T\right>, \left<S\right>)\boldsymbol{\cdot\nabla})S(\boldsymbol{r}) = f_S(\boldsymbol{r}),&  \boldsymbol{r}\in\Omega
\\
\\
(P3^{\hbox{ss}}) \; \; a_I(\left<T\right>, \left<S\right>) = \Gamma  \left(-\alpha(
\left<T\right>_{2} - \left<T\right>_{1})
+
\beta(\left<S\right>_{2} - \left<S\right>_{1})
\right),
&
\\
(P4^{\hbox{ss}}) \; \; \boldsymbol{u}(\boldsymbol{r};
 \left<T\right>, \left<S\right>) = a_E \boldsymbol{u}_E(\boldsymbol{r}) 
+ a_I(\left<T\right>, \left<S\right>) \boldsymbol{u}_I(\boldsymbol{r}), & \boldsymbol{r}\in\Omega
\\
\\
(P5^{\hbox{ss}}) \; \; (\tensor{\kappa} \boldsymbol{ \nabla} T(q)) \boldsymbol{\cdot}\boldsymbol{\hat{n}}(q)
=
\begin{cases}
g_A^T (T^*(x,y) - T(q)) & \hbox{if}\; \; q \in \sigma_1\\
 0 & \hbox{else}
\end{cases}
,
&  q\in\partial\Omega
\\
(P6^{\hbox{ss}}) \; \; (\tensor{\kappa} \boldsymbol{ \nabla} S(q)) \boldsymbol{\cdot}\boldsymbol{\hat{n}}(q)
=
\begin{cases}
g_A^S S^*(x,y) & \hbox{if} \; \; q \in \sigma_1\\
 0 & \hbox{else}
\end{cases}
,
&  q\in\partial\Omega.
\end{cases}
\end{aligned}
\end{equation}
where the parameters satisfy \eqref{eq:a}.
Since if $S$ is a solution then so is $S+a$ for any constant $a$, then we restrict ourselves to the class where the average of $S$ over $\Omega$ is zero as we did for the time-dependent problem. We thus define a weak solution to $P^{\hbox{ss}}$ in the same spirit as of weak solutions of the nonlinear time-dependent problem:
\begin{app-definition}
$( T, S) \in H^1(\Omega)\times \dot{H}^1(\Omega)$ is called a {weak solution to 
$\left(P^{\hbox{ss}}\right)$}
 if, for all test functions $\varphi \in H^1(\Omega)$, $\psi \in \dot{H}^1(\Omega)$, the following holds:
\begin{equation}
\label{eq:weakTSnonlinearss}
\begin{aligned}
&
\int_\Omega \left( (\tensor{\kappa} \boldsymbol{ \nabla} T) \boldsymbol{\cdot}(\boldsymbol{ \nabla} \varphi) +((\boldsymbol{u} \boldsymbol{\cdot}\nabla)T ) \varphi
\right)
dV +
\int_{{\sigma_1}} g_A^T T \varphi dx dy
=
\int_\Omega f_T \varphi dV
+ 
\int_{{\sigma_1}} g_A^T T^*(x,y) \varphi dx dy;
\\
\\
&
\int_\Omega \left( (\tensor{\kappa} \boldsymbol{ \nabla} S) \boldsymbol{\cdot}(\boldsymbol{ \nabla} \psi) +((\boldsymbol{u} \boldsymbol{\cdot}\nabla)S ) \psi
\right)
dV 
=
\int_\Omega f_S \psi dV
+ 
\int_{{\sigma_1}} g_A^S S^*(x,y) \psi dx dy,
\end{aligned}
\end{equation}
where $\boldsymbol{u} = \boldsymbol{u}(\boldsymbol{r}; \left< T\right>, \left< S\right>)$ is defined according to equation $(P4)$ in \eqref{eq:P}.
\end{app-definition}

\begin{app-remark}
A weak solution $(T,S)$ to the steady state problem $\left(P^{\hbox{ss}}\right)$ is a strong solution of the nonlinear time-dependent problem $\left(P_{T, S, T^*, S^*, f_{T}, f_{S}}\right)$ that is independent of time.
\end{app-remark}

%
%
%

To conclude this appendix, in Theorem {C.}\ref{theoreminproof2} we show that for any set of parameters, boundary functions and source functions satisfying \eqref{eq:a}, the nonlinear steady-state problem $\left(P^{\hbox{ss}}\right)$ has a weak solution with bounded norms.
Note that we do not show uniqueness for the general problem; indeed for a given set of parameters, uniqueness of a weak solution to the nonlinear steady state problem $\left(P^{\hbox{ss}}\right)$ is not guaranteed nor expected in general.
However, we show that
given some restriction on the size of the parameters under which the system is not vigorously forced with respect to its dissipation, the steady-state solution \textit{is} unique; furthermore, all solutions to the time-dependent problem converge to this unique steady state solution as $t\rightarrow\infty$, as could be expected from a dissipative dynamical system induced by advection-diffusion-type problems.
In fact, one should be able to show that the infinite-dimensional dynamical system induced by the evolution of the nonlinear problem $\left( P_{T_{0}, S_{0}, T^*, S^*, f_{T}, f_{S}}\right)$ possess a finite-dimensional global attractor, a subject which is outside the scope of this article \citep{temam2012infinite}.

\begin{app-theoremprf}\label{theoreminproof2}
Let $\tensor{\kappa}$, $\Gamma$, $\alpha$, $\beta$, $a_E$, $\boldsymbol{u}_E$, $\boldsymbol{u}_I$, $f_T$, $f_S$, $g_A^T$, $g_A^S$, $T^*$, and $S^*$ be as in
\eqref{eq:a}.
Then:
\begin{enumerate}
\item The nonlinear steady-state problem $\left( P^{\hbox{ss}}\right)$ has a weak solution.
\item A solution $(T,S)$ to $\left( P^{\hbox{ss}}\right)$ satisfies the following bounds:
\begin{equation}
\begin{aligned}
\label{eq:ssH1bounds}
\norm{T}^2 \leq C_4^T&, 
\; \; 
\norm{\tensor{\kappa}^{1/2} \boldsymbol{ \nabla} T}^2 \leq C_5^T,
\; \; 
\norm{S}^2 \leq C_4^S,
\; \; 
\norm{\tensor{\kappa}^{1/2} \boldsymbol{ \nabla} S}^2 \leq C_5^S;
\end{aligned}
\end{equation}
where
\begin{equation}
\begin{aligned}
\label{eq:ssH1Cs}
&C_4^T = 
\frac{1}{\lambda^2} \norm{f_T}^2 
+
 \frac{g_A^T}{\lambda} \norms{T^*}^2,
\; \;
C_5^T =
\frac{1}{\lambda} \norm{f_T}^2 
+
 \frac{2 g_A^T}{3} \norms{T^*}^2,
\\
&C_4^S =
\frac{1}{\epsilon_1} \norms{S^*}^2
+
\frac{1}{\epsilon_2} \norm{f_S}^2,
\; \;
C_5^S = 
\frac{1}
{\epsilon_1} \norms{S^*}^2
+
\frac{1}{\epsilon_3} \norm{f_S}^2.
\end{aligned}
\end{equation}
The constants are given by $\epsilon_1 = 2\kappa_{min}/\left(g_A^S\right)^2$,
$\epsilon_2 = 4 \kappa_{min} \left(3 \pi/8 -1\right)$,
$\epsilon_3 = 2 \left(\kappa_{min}(\pi-2)+1 \right)$.
\item
Let $T_0, S_0$ be as \eqref{eq:a}, and let $(\tilde{T},\tilde{S})$ be a global strong solution to $\left( P_{T_{0}, S_{0}, T^*, S^*, f_{T}, f_{S}}\right)$, as established in Theorem {C.}\ref{theoreminproof}.
If the following condition is satisfied by the problem parameters:
\begin{equation*}
u_{I}^{\hbox{max}} \Gamma
<
\frac
{\min \{ |D_1|, |D_2| \}}
{\max \{ \alpha, \beta \} 
|\Omega| ^{1/2}  }
\frac
{
\min\{\kappa_x, \kappa_y, \kappa_z\}
\min \{ 2 \pi \kappa_x, 2 \pi \kappa_y, \frac{g_A^T}{2}, \frac{\kappa_z}{2} \}
}{
4 \max \{ C_5^T, C_5^S \}
}
;
\; \; u_{I}^{\hbox{max}} = \max_{\boldsymbol{r}} |\boldsymbol{u}_I(\boldsymbol{r})|,
\end{equation*}
{
where $C_5^T$, $C_5^S$ are given in equation \eqref{eq:ssH1Cs},
then $(T,S)$ converges to a unique steady-state solution as $t\rightarrow \infty$.}
\end{enumerate}
\end{app-theoremprf}

Proof:

(ii)
We start by proving the second point, establishing the {a priori} bounds \eqref{eq:ssH1bounds}. To this end, assume $(T,S)$ is a weak solution to $\left(P^{\hbox{ss}}\right)$.
Since $(T,S) \in H^1(\Omega) \times \dot{H}_1(\Omega)$, then each can be used as a test function in equation \eqref{eq:weakTSnonlinearss}.
Thus, by using the divergence theorem, Cauchy-Schwarz, Young's inequality and Lemma {C.}\ref{lemma:3} one can immediately conclude the equations for $T$ in \eqref{eq:ssH1bounds} and \eqref{eq:ssH1Cs}.
Similarly, using Lemmas {C.}\ref{lemma:4} and {C.}\ref{lemma:5} instead of Lemma {C.}\ref{lemma:3}, one obtains the equations for $S$ in  \eqref{eq:ssH1bounds} and \eqref{eq:ssH1Cs}.

(i)
Next, we prove that the nonlinear steady-state problem $\left( P^{\hbox{ss}}\right)$ has a weak solution. 

\underline{Step 1.} Induced linear problem.

Let $\bar{T}, \bar{S} \in L^2(\Omega)$ be given. 
We introduce the following linear stationary problem induced by $\left(P^{\hbox{ss}}\right)$:
\begin{equation}
\begin{aligned}
\label{eq:Psslinear}
\left(P^{\hbox{ss,lin}}_{\bar{T}, \bar{S}}\right)
\begin{cases}
(P1^{\hbox{ss,lin}}) \; \; - \boldsymbol{ \nabla}\boldsymbol{\cdot}(\tensor{\kappa}\boldsymbol{ \nabla} T(\boldsymbol{r}) )
+ (\bar{\boldsymbol{u}}(\boldsymbol{r}; \left<\bar{T}\right>, \left<\bar{S}\right>)\boldsymbol{\cdot\nabla})T(\boldsymbol{r}) 
= f_T(\boldsymbol{r}), & \boldsymbol{r}\in\Omega
\\
(P2^{\hbox{ss,lin}}) \; \;  - \boldsymbol{ \nabla}\boldsymbol{\cdot}(\tensor{\kappa}\boldsymbol{ \nabla} S(\boldsymbol{r})) +
 ({\bar{\boldsymbol{u}}}(\boldsymbol{r}; \left<{\bar{T}}\right>, \left<{\bar{S}}\right>)\boldsymbol{\cdot\nabla})S(\boldsymbol{r}) = f_S(\boldsymbol{r}),&  \boldsymbol{r}\in\Omega
\\
\\
(P3^{\hbox{ss,lin}}) \; \; a_I(\left<{\bar{T}}\right>, \left<{\bar{S}}\right>) = \Gamma  \left(-\alpha(
\left<{\bar{T}}\right>_{2} - \left<{\bar{T}}\right>_{1})
+
\beta(\left<{\bar{S}}\right>_{2} - \left<{\bar{S}}\right>_{1})
\right),
&
\\
(P4^{\hbox{ss,lin}}) \; \; {\bar{\boldsymbol{u}}}(\boldsymbol{r};
 \left<{\bar{T}}\right>, \left<{\bar{S}}\right>) = a_E \boldsymbol{u}_E(\boldsymbol{r}) 
+ a_I(\left<{\bar{T}}\right>, \left<{\bar{S}}\right>) \boldsymbol{u}_I(\boldsymbol{r}), & \boldsymbol{r}\in\Omega
\\
\\
(P5^{\hbox{ss,lin}}) \; \; (\tensor{\kappa} \boldsymbol{ \nabla} T(q)) \boldsymbol{\cdot}\boldsymbol{\hat{n}}(q)
=
\begin{cases}
g_A^T (T^*(x,y) - T(q)) & \hbox{if} \; \; q \in \sigma_1\\
 0 & \hbox{else}
\end{cases}
,
&  q\in\partial\Omega
\\
(P6^{\hbox{ss,lin}}) \; \; (\tensor{\kappa} \boldsymbol{ \nabla} S(q)) \boldsymbol{\cdot}\boldsymbol{\hat{n}}(q)
=
\begin{cases}
g_A^S S^*(x,y) & \hbox{if} \; \; q \in \sigma_1\\
 0 & \hbox{else}
\end{cases}
,
&  q\in\partial\Omega.
\end{cases}
\end{aligned}
\end{equation}
We define the bilinear forms:
\begin{equation*}
\begin{aligned}
B_T: H^1(\Omega) \times H^1(\Omega)\rightarrow \mathbb{R},
\; \;&B_T(T,\varphi) = \int_\Omega 
\left(
(\tensor{\kappa} \boldsymbol{ \nabla} T) \boldsymbol{\cdot}(\boldsymbol{ \nabla} \varphi) + (({\bar{\boldsymbol{u}}} \boldsymbol{\cdot}\boldsymbol{ \nabla})T)  \varphi
\right) dV
+ \int_{{\sigma_1}} g_A^T T \varphi dx dy,
\\
B_S: \dot{H}^1(\Omega) \times \dot{H}^1(\Omega)\rightarrow \mathbb{R},
\; \;&B_S(T,\varphi) =
\int_\Omega 
\left(
(\tensor{\kappa} \boldsymbol{ \nabla} S) \boldsymbol{\cdot}(\boldsymbol{ \nabla} \varphi) + (({\bar{\boldsymbol{u}}} \boldsymbol{\cdot}\boldsymbol{ \nabla})S)  \varphi
\right) dV,
\end{aligned}
\end{equation*}
and the linear functionals: 
\begin{equation*}
\begin{aligned}
l_T:H^1(\Omega)\rightarrow\mathbb{R},
\; \; &l_T(\varphi) = \int_\Omega f_T \varphi dV + \int_{{\sigma_1}} g_A^T T^* \varphi dx dy,
\\
l_S:\dot{H}^1(\Omega)\rightarrow\mathbb{R},
\; \; &l_S(\varphi) = \int_\Omega f_S \varphi dV + \int_{{\sigma_1}} g_A^S S^* \varphi dx dy.
\end{aligned}
\end{equation*}
We define a weak solution to $\left(P^{\hbox{ss,lin}}_{\bar{T}, \bar{S}} \right)$ 
to be $( T, S) \in H^1(\Omega)\times \dot{H}^1(\Omega) $
such that
 \begin{equation}
 \label{eq:weakTSlinearss}
B_T(T,\varphi) = l_T(\varphi),
\; \;
B_S(S,\psi) = l_S(\psi)
\end{equation}
for every test function $\varphi \in H^1(\Omega), \psi \in \dot{H}^1(\Omega)$.
%
%

Next, we show that the linear problem has a unique weak solution,
employing the Lax-Milgram theorem, see, e.g., \citet{EvansPDE}, chapter 6.2.
%
%
To this end, we need to check that the conditions of the Lax-Milgram theorem are valid, namely the boundedness of the bilinear forms $B_T$ and $B_S$ and their coercivity, as well as the boundedness of the linear functionals $l_T$ and $l_S$.

Using the Cauchy-Schwarz inequality and the trace theorem, we have
\begin{equation*}
\begin{aligned}
|B_T(\rho, \varphi)| &= 
\left| \int_\Omega (\tensor{\kappa} \boldsymbol{ \nabla} \rho - {\bar{\boldsymbol{u}}} \rho) \boldsymbol{\cdot}(\boldsymbol{ \nabla} \varphi) dV
+ \int_{{\sigma_1}} g_A^T \rho \varphi dx dy
\right|
\\
&\leq
\left(
\left(
\int_\Omega |\tensor{\kappa} \boldsymbol{ \nabla} \rho|^2
\right)^{1/2} + \int_\Omega\left(|{\bar{\boldsymbol{u}}} \rho|^2\right)^{1/2}\right)
\left(
\int_\Omega |\boldsymbol{ \nabla}\varphi|^2
\right)^{1/2}
+
g_A^T 
\left(\int_{{\sigma_1}}  \rho^2
\right)^{1/2} 
\left(\int_{{\sigma_1}} \varphi^2\right)^{1/2}
\\
&\leq
M_T \norm{\rho}_{H^1(\Omega)} \norm{\varphi}_{L^2(\Omega)} + g_A^T \norm{\rho}_{L^2(\partial\Omega)} \norm{\varphi}_{L^2(\partial\Omega)}
\leq
\alpha_T\norm{\rho}_{H^1(\Omega)} \norm{\varphi}_{H^1(\Omega)},
\end{aligned}
\end{equation*}
where $M_T>0$ exists because $\tensor{\kappa}$ is constant and ${\bar{\boldsymbol{u}}}$ is bounded by construction (equation \eqref{eq:a}),
and $\alpha_T>0$. 
Using similar arguments,
\begin{equation*}
\begin{aligned}
|B_S(\rho, \varphi)| &= 
\left|
\int_\Omega (\tensor{\kappa} \boldsymbol{ \nabla} \rho - {\bar{\boldsymbol{u}}} \rho) \boldsymbol{\cdot}(\boldsymbol{ \nabla} \varphi) dV
\right|
\leq
\left(
\left(
\int_\Omega |\tensor{\kappa} \boldsymbol{ \nabla} \rho|^2
\right)^{1/2} + \int_\Omega
\left(\left|{\bar{\boldsymbol{u}}} \rho\right|^2\right)^{1/2}\right)
\left(\int_\Omega 
\left|
\nabla\varphi
\right|^2
\right)^{1/2}
\\
&\leq
M_S \norm{\rho}_{H^1(\Omega)} \norm{\varphi}_{L^2(\Omega)}
\leq
\alpha_S \norm{\rho}_{H^1(\Omega)} \norm{\varphi}_{H^1(\Omega)}.
\end{aligned}
\end{equation*}
Next, to show coercivity, we use the divergence theorem and Lemma {C.}\ref{lemma:3} to deduce
\begin{equation*}
\begin{aligned}
B_T(\varphi, \varphi) = 
\norm{\tensor{\kappa}^{1/2} \boldsymbol{ \nabla} \varphi}^2_{L_2(\Omega)}
+
g_A^T \norms{\varphi}^2
\geq
\beta_T \norm{\varphi}_{H^1(\Omega)}^2
\end{aligned}
\end{equation*}
where $\beta_T = \frac{\min\{\lambda, \kappa_{min} \}}{2}$.
Moreover, using Lemma  {C.}\ref{lemma:5}, we conclude
\begin{equation}
\begin{aligned}
B_S(\varphi, \varphi) = 
\norm{\tensor{\kappa}^{1/2} \boldsymbol{ \nabla} \varphi}^2_{L_2(\Omega)}
\geq \beta_S \norm{\varphi}_{H^1(\Omega)}^2
\end{aligned}
\end{equation}
where $\beta_S = \frac{\min\{1,\kappa_{min}\}}{1+1/(\pi \kappa_{min})}$.
Finally, to show boundedness of $l_T$ and $l_S$, one uses Cauchy-Schwarz and Lemmas {C.}\ref{lemma:3} and {C.}\ref{lemma:5} as above.

\underline{Step 2.} Existence of a steady state solution as a fixed point of a continuous operator.

We denote the function $Q:L^2\times \dot{L}^2  \rightarrow L^2 \times \dot{L}^2 $ as follows:
$Q((\bar{T},\bar{S})) = (T,  S)$ where $(T, S)$ is the solution of $\left(P^{\hbox{ss,lin}}_{\bar{T},\bar{S}}\right)$.
Notice that a fixed point of the mapping $Q$ is exactly a steady-state solution of the nonlinear stationary problem $\left(P^{\hbox{ss}}\right)$.
Next, we show that the mapping $Q$ indeed has a fixed point, employing the Schauder-Tychonof fixed point theorem (see \cite{zeidler2013nonlinearI}, Corollary 2.13).
Thus, in order to conclude the proof we have left only to check that the conditions of the Schauder-Tychonof fixed point theorem are valid,  namely that $Q$ is a continuous operator from a compact, convex subset of $L^2$ to itself.

Indeed, let $(\bar{T}, \bar{S})$, $( T, S)$ satisfy $Q((\bar{T}, \bar{S})) = ( T, S)$. Then $( T, S)$ satisfy the bounds from equation \eqref{eq:ssH1bounds}. Therefore, $( T, S)$ is inside the closed ball 
in $H^1(\Omega)\times \dot{H}^1(\Omega)$ defined by these bounds, and this closed ball is compact in $L^2(\Omega)\times \dot{L}^2(\Omega)$ by the Rellich-Kondrachov Lemma \citep{EdmundsEvansSpectral}. 

To prove continuity of $Q$ in the sense of $L^2$, 
%
%
%
%
%
%
%
let $\bar{T}_i, \bar{S}_i, T_i, S_i$ satisfy $Q((\bar{T}_i, \bar{S}_i)) = (T_i, S_i)$ for $i = 1,2$. Then $T_i, S_i$ satisfy
$B_T(T_i,\varphi) = l_T(\varphi)$, $B_S(S_i,\psi) = l_S(\psi)$, respectively. Using assumptions (a4), (a5)  in \eqref{eq:a}, and subtracting equation \eqref{eq:weakTSlinearss} for $T_1$ from equation  \eqref{eq:weakTSlinearss} for $T_2$, both with the same test function $\varphi = T_2 - T_1 \equiv \delta T$, we deduce
\begin{equation*}
\norm{\tensor{\kappa}^{1/2} \boldsymbol{ \nabla} \delta T}_{L^2(\Omega)}
+ \int_\Omega ((\delta{\boldsymbol{ \bar{u}}} \boldsymbol{\cdot}\nabla)T_1) \delta T dV
+ g_A^T \norms{\delta T}^2 = 0
\end{equation*}
where 
$\bar{\boldsymbol{u}}_i = \bar{\boldsymbol{u}}(\boldsymbol{r}; \left< \bar{T}_{i}\right>, \left< \bar{S}_{i} \right>)$ is given by equation $(P4^{\hbox{ss,lin}})$ in for $i = 1,2$ in \eqref{eq:Psslinear},
$\delta{\boldsymbol{ \bar u}} = \bar{\boldsymbol{u}}_{2} - \bar{\boldsymbol{u}}_{2}$.
Therefore, using Lemma {C.}\ref{lemma:3} and equation \eqref{eq:deltaubound},  
and defining $\delta \bar{T} \equiv \bar{T}_2 - \bar{T}_1$, 
$\delta \bar{S} \equiv \bar{S}_2 - \bar{S}_1$, we obtain
\begin{equation*}
\lambda \norm{\delta T}^2_{L^2(\Omega)} 
\leq
b \int_\Omega (|\delta \bar{T}| + |\delta\bar{ S}|) dV \int_\Omega |\boldsymbol{ \nabla} T_1| |\delta T| dV
\end{equation*}
where $b>0$ is some constant. Using Cauchy-Schwarz, equation \eqref{eq:ssH1bounds} and Young's inequality, we deduce
\begin{equation}
\norm{\delta T}^2_{L^2(\Omega)} 
\leq
\tilde{b} \left(\norm{\delta \bar{T}}_{L^2(\Omega)} + 
\norm{\delta \bar{S}}_{L^2(\Omega)}\right)^2
\label{eq:continuityTss}
\end{equation}
where  $\tilde{b}>0$ is some constant.
We repeat similar steps for $S$, using Lemma {C.}\ref{lemma:5} and defining $\delta S \equiv S_2 - S_1$, to obtain
\begin{equation}
\norm{\delta S}^2_{L^2(\Omega)}\leq
\tilde{c} \left(\norm{\delta \bar{T}}_{L^2(\Omega)} + 
\norm{\delta \bar{S}}_{L^2(\Omega)}\right)^2
\label{eq:continuitySss}
\end{equation}
where $\tilde{c}>0$ is some constant.
Equations \eqref{eq:continuityTss} and \eqref{eq:continuitySss} prove continuity of $Q$ in the sense of $L^2(\Omega)$
since small 
$\norm{\delta \bar{T}}_{L^2(\Omega)}$,
$\norm{\delta \bar{S}}_{L^2(\Omega)}$
imply small 
$\norm{\delta T}_{L^2(\Omega)}$,
$\norm{\delta S}_{L^2(\Omega)}$.

%
%



(iii)
To conclude the proof, we derive a condition on the parameters under which solutions to the time-dependent problem converge to a unique steady state solution.
Let $T_0, S_0$ be as \eqref{eq:a}, and let $(\tilde{T},\tilde{S})$ be a global strong solution to $\left( P_{T_{0}, S_{0}, T^*, S^*, f_{T}, f_{S}}\right)$, as established in Theorem {C.}\ref{theoreminproof}.
Further, let $( {{T}}, {{S}} )$ be a steady state solution to
$\left( P_{{{T}}, {{S}}, T^*, S^*, f_{T}, f_{S}}\right)$
as established above.
By Remark {C.}\ref{remark:lemma2}, using the analogous equation to \eqref{eq:weakT2} for $T$ and $\tilde{T}$ both with $\delta T \equiv \tilde{T} - {{T}}$ as a test function, we deduce
%
\begin{equation*}
\frac{1}{2}\frac{d}{dt}\norm{\delta T}^2
 + g_A^T \norms{\delta T}^2
+ \norm{\tensor{\kappa}^{1/2} \boldsymbol{ \nabla} \delta T}^2
\leq 
\int_\Omega |\delta T | |\delta u| |\boldsymbol{ \nabla} {{T}}|.
\end{equation*}
By Lemma {C.}\ref{lemma:3}, equation \eqref{eq:deltaubound} and the Cauchy-Schwarz inequality, we obtain
\begin{equation*}
\frac{1}{2}\frac{d}{dt}\norm{\delta T}^2
 + \lambda \norm{\delta T}^2
 \leq 
 b 
 \frac{C_5^T}{\kappa_{min}} 
 (\norm{\delta T} + \norm{\delta S}) \norm{\delta T} 
\end{equation*}
where $\delta S \equiv \tilde{S} - {{S}}$ and $b \equiv u_{I}^{\hbox{max}} \Gamma 
  \frac{\max \{ \alpha, \beta \}}{\min \{ |D_1|, |D_2| \} }
  |\Omega| ^{1/2}$.
Following similar arguments for $S$, 
\begin{equation*}
\frac{1}{2} \frac{d}{dt} \norm{\delta S}^2 + {\pi \kappa_{min}}\norm{\delta S}^2 \leq 
b\frac{C_5^S}{\kappa_{min}} 
 (\norm{\delta T} + \norm{\delta S}) \norm{\delta S}.
\end{equation*}
Hence, by defining $\tilde{\lambda} = \min \{ {2 \pi \kappa_{min}}, \frac{g_A^T}{2}, \frac{\kappa_z}{2} \} $, 
$c = 4 \frac{b}{\kappa_{min}} 
  \max \{ C_5^T, C_5^S \} $,
  and
  $\eta(t) = \norm{\delta T}^2(t) + \norm{\delta S}^2(t)$, we obtain
\begin{equation*}
\begin{aligned}
\frac{d}{dt} 
\eta + 
\tilde{\lambda} \eta
\leq
\frac{c}{2} (\norm{\delta T} + \norm{\delta S})^2
 \leq
 c
\eta.
 \end{aligned}
\end{equation*}
This allows employing Gronwall's inequality to obtain the following bound:
\begin{equation*}
\eta(t) \leq \eta(0) e^{-(\tilde\lambda - c)t}.
\end{equation*}
Thus, if $\tilde{\lambda} > c$, i.e.
\begin{equation*}
u_{I}^{\hbox{max}} \Gamma
<
\frac
{\min \{ |D_1|, |D_2| \}}
{\max \{ \alpha, \beta \} 
 |\Omega| ^{1/2}  }
\frac
{
\kappa_{min}
\min \{ 
4 \pi \kappa_{min}, g_A^T, \kappa_z \}
}{
4\max \{ C_5^T, C_5^S \}
},
\end{equation*}
then $(\tilde{T}, \tilde{S})$ converges to the steady-state solution $({{T}}, {{S}})$ at a rate of at least $1/(\tilde{\lambda} - c)$, independent of the initial conditions $T_0, S_0$.
Thus, in this case, the steady-state solution is unique.
{$\blacksquare$}

\subsection{Proof of useful Lemmas}
\label{sec:lemmaproofs}
\subsubsection{Proof of Lemma {C.}\ref{lemma:3} (Poincar{\'e} inequality)}
Using the Cauchy-Schwarz inequality and Young's inequality,
\begin{equation*}
\begin{aligned}
T^2(x,y,z)
&= T^2(x,y,1) + \int_1^z \partial_zT^2(x,y,z')dz'= T^2(x,y,1) + 2\int_1^z T\partial_zT(x,y,z')dz'\\
&\leq
T^2(x,y,1) + 2\int_1^z |T| \left| \frac{\partial T}{\partial z}\right|dz' 
\leq
T^2(x,y,1) +
\left( \int_1^z |T|^2dz' \right) ^{1/2}
\left( \int_1^z 4 \left| \frac{\partial T}{\partial z} \right| ^2dz' \right) ^{1/2} \\
&
\leq
T^2(x,y,1) +
\frac{1}{2} \int_1^0 |T|^2dz +
2  \int_1^0 \left| \frac{\partial T}{\partial z} \right|^2dz'.
\end{aligned}
\end{equation*}
By integrating over the domain $\Omega$, we obtain
$
\norm{T}^2 \leq
2 \norms{ T}^2 + 4 \norm{\partial_z T}^2
$, therefore
\begin{equation*}
\begin{aligned}
\lambda \norm{T}^2&\leq
2 \lambda \norms{T}^2+
4 \lambda \norm{\partial_z T}^2
\leq 
\frac{g_A^T}{2} \norms{T}^2 +
\kappa_z \norm{\partial_z T}^2 
\leq
\frac{ g_A^T}{2} \norms{T}^2 +
\norm{\tensor{\kappa}^{1/2} \boldsymbol{ \nabla} T}.
 \end{aligned}
\end{equation*}
{$\blacksquare$}

\subsubsection{Proof of Lemma {C.}\ref{lemma:4}}
Using the Cauchy-Schwarz inequality and Young's inequality,
\begin{equation*}
\begin{aligned}
 S^2(x,y,1)
&= S^2(x,y,z) - 2 \int_1^z S \partial_zS(x,y,z')dz' 
\leq
S^2(x,y,z) + 2 \int_1^z |S| \left|\partial_z S\right|dz' \\
&\leq
S^2(x,y,z) +  \frac{1}{\epsilon} \int_1^0 |S|^2dz' + \epsilon \int_1^0|\partial_z S|^2 dz'.
\end{aligned}
\end{equation*}
Integrating over the  domain $\Omega$, we deduce
\begin{equation*}
\begin{aligned}
\norms{S}^2
\leq
\left(
1+\frac{1}{\epsilon}
\right)
 \norm{S}^2 
+ \frac{\epsilon}{\kappa_{min} } 
\norm{\tensor{\kappa}^{1/2} \boldsymbol{ \nabla} S}^2.
\end{aligned}
\end{equation*}
{$\blacksquare$}

\subsubsection{Proof of Lemma {C.}\ref{lemma:5}}
%
%
Note that the inequality $\norm{S - \left<S\right>_{\Omega}}^2 \leq c \norm{\boldsymbol{ \nabla} S}^2 $ is the Poincar{\'e}-Wirtinger inequality for $p=2$, where $c$ is a constant determined only by the domain and $p$.
In fact, since $\Omega$ is a smooth bounded cube with side lengths $1$, we can calculate the constant: it equals $1/\lambda_1$, where $\lambda_1$ is the smallest eigenvalue of minus the Laplacian, solving $- \boldsymbol{ \nabla}^2 S = \lambda_1 S$, and equals $\pi/\max\{L_x, L_y, L_z\}$. In this simple case,  $\lambda_1 = \pi$, therefore $c = 1/\pi$, and
\begin{equation}
\norm{S - \left<S\right>_{\Omega}}^2  \leq \frac{1}{\pi \kappa_{min}} \norm{\tensor{\kappa}^{1/2}\boldsymbol{ \nabla} S}^2.
\end{equation}
{$\blacksquare$}

\bibliographystyle{jfm}

\bibliography{ref-nl}

\end{document}